\documentclass[prd,a4paper,nofootinbib,
showpacs, 
]{revtex4}
\usepackage{graphicx}
\usepackage{amsfonts}
\usepackage{slashed} 
\usepackage[section]{placeins}
\usepackage{mathrsfs}
\usepackage{amsmath}
\def\eq#1{{eq.~(\ref{#1})}}

\def\Im{\mbox{Im}\,}

\def\Tr{\mbox{Tr}\,}

\def\hbar{\hspace{0pt}\raisebox{1pt}{$-$} \hspace{-7pt} h}

\def\5{\overline 5}

\newcommand{\be}{\begin{equation}}
\newcommand{\ee}{\end{equation}}
\newcommand{\bea}{\begin{eqnarray}}
\newcommand{\eea}{\end{eqnarray}}
\newcommand{\nn}{\nonumber}

{% for the vertical padding in tables
\usepackage{color}
\usepackage{hyperref}
\def\hhref#1{\href{http://arxiv.org/abs/#1}{#1}} % in bibliography

\definecolor{oucrimsonred}{rgb}{0.6, 0.0, 0.0}
\definecolor{persianblue}{rgb}{0.11, 0.22, 0.73}
\definecolor{forestgreen}{rgb}{0.13,0.35,0.13}
 \hypersetup{colorlinks, citecolor=oucrimsonred, linkcolor=persianblue, urlcolor=oucrimsonred}
\begin{document}
%%%%%%%%%%%%%%%%%%%%%%%%%%%%%%%%%%%%%%%%%%%%%%%%%%%%%%%%%%%  FRONT PAGE
\title[]{Vector boson scattering at the LHC\\
{\small A study of the  $WW\rightarrow WW$ channels with the Warsaw cut}}
%%%%%%%%%%%%%%%%%%%%%%%%
\date{\today}
\author{M. Fabbrichesi$^{\dag}$}
\author{M. Pinamonti$^{\ddag\ast}$}
\author{A. Tonero$^{\circ}$}
\author{A. Urbano$^{\dag\ast}$}
\affiliation{$^{\dag}$INFN, Sezione di Trieste}
\affiliation{$^{\ddag}$INFN, Sezione di Trieste, Gruppo collegato di Udine}
\affiliation{$^\ast$SISSA, via Bonomea 265, 34136 Trieste, Italy}
\affiliation{$^{\circ}$ICTP-SAIFR, Rua Dr.\ Bento Teobaldo Ferraz 271, 
01140-070 S\~ao Paulo, Brazil }
\begin{abstract}
\noindent  We study  $W$ boson  scattering in the same- and opposite-sign   channels under the assumption that no resonances are present in the collider processes 
$pp\rightarrow l^\pm \nu_{ l} l^\pm  \nu_{ l} j j$ and 
$pp\rightarrow l^\pm \nu_{ l} l^\mp  \nu_{ l} j j$, respectively. 
Basic selection cuts together with a restriction on the combination of the final lepton and jet momenta (the Warsaw cut)  makes it possible to argue that at the LHC a luminosity of 100 fb$^{-1}$ and a center-of-mass energy of $\sqrt{s}=$  13 TeV will  allow to constrain the leading effective lagrangian coefficients at the permil level. We also discuss 
limits on the other coefficients of the effective lagrangian as well as stronger  constraints provided by  higher  energy and luminosity. We show that the same-sign $WW \rightarrow WW$  channel suffices in providing the most stringent  constraints.
\end{abstract}
%\keywords{}
%\preprint{}
%%%%%%%%%%%%%%%%%%%%%%%%%%%%%%%%%%%%%%%%%%%%%%%%%%%%%%%%%%%%%%%%%%%
\maketitle
%%%%%%%%%%%%%%%%%%%%%%%%%%%%%%%%%%%%%%%%%%%%%%%%%%%%%%%%%%%%%%%%%%%
%
%%%%%%%%%%%%%%%%%%%%%%%%%%
\vskip1.5em

%%%%%%%%%%%%%%%%%%%%%%%%%%%%%%%%
\section{Introduction} 
%%%%%%%%%%%%%%%%%%%%%%%%%%%%%%%%

Vector boson scattering (VBS) at the LHC provides a direct window on the mechanism responsabile for the breaking of the electroweak (EW) symmetry. 
The  tree-level amplitude for VBS  is  the combination of seven subprocesses in which gauge  and Higgs bosons are exchanged. In the standard model (SM) the terms leading in energy cancel leaving an amplitude and a cross section  consistent with unitary. If any or all  among the trilinear and quartic gauge couplings and the Higgs boson coupling to the vector bosons are modified these delicate cancellations fail and tree-level unitarity is lost. In particular, if either the trilinear or the quartic gauge couplings are changed, terms proportional to the fourth power of the center-of-mass (CM) energy will be present. 

After the existence of the Higgs boson has been confirmed~\cite{Chatrchyan:2012ufa}, we know that this particle plays a role in EW symmetry breaking but the details may  differ from the basic scenario in which the Higgs boson is linearly and minimally coupled.  If the gauge couplings  are left unchanged but the Higgs boson couplings to the vector bosons  are modified, terms proportional to the square of the CM energy will be present in the amplitude for VBS.

All these potential departures from the SM represent signals for new physics. Since there are many possibilities---ranging from an extended Higgs sector to strong dynamics---they are best described by means of an effective field theory.  

 Terms in the amplitude growing with the CM energy arise when considering the scattering among the longitudinal components of the vector bosons. Using the equivalence theorem~\cite{Chanowitz:1985hj}, these components  can be identified with the Goldsone bosons of the EW symmetry breaking and behave as  scalar particles with derivative couplings: their  scattering amplitudes are similar to those for $\pi\pi$-scattering in QCD and the same techniques can be used. The transverse components give rise to terms in the amplitude that are bounded in the CM energy and subleading---for all practical purposes, they are part of the background.
The natural language for computing the relevant amplitudes is that of the effective nonlinear (chiral) EW lagrangian first introduced in \cite{Appelquist:1980vg}. 

 Depending on the symmetry group used, there exist different  effective lagrangians which are equivalent but differ in the order-by-order  terms and therefore in the  dimension and  field content of the operators.
Compared to other effective lagrangian based on the linear theory and the full symmetry group, the chiral EW lagrangian has the advantage of being optimised for  VBS. 

The loss of tree-level unitarity suggests the presence of a strongly interacting sector. We expect unitarity to be restored by the presence of resonances.  Barring the spectacular case of the LHC actually seeing one or more of these resonances, this loss and its eventual restoration can be studied by the effective EW lagrangian in terms of bounds of its coefficients. Because we now know that the theory also contains a Higgs boson, such a lagrangian must be  completed by the introduction of  this field~\cite{Espriu:2012ih,Delgado:2013loa}---the effect of which is parameterised in terms of additional coefficients. 

 The same-sign $W^\pm W^\pm\rightarrow W^\pm W^\pm$ channel stands out  in this search because of the suppressed QCD background and the reduced contribution from  channels where transverse and longitudinal gauge bosons are mixed. It is a channel in which is easier to single out the scattering of the longitudinal components of the gauge bosons and the most likely place to look for possible deviations from the SM.  
 
 Possible  resonances in this channel are expected to be either present  in the $t$-channel (and therefore leading to only a decrease of the cross section) or carrying isospin 2 and doubly charged and therefore heavier than those in other decay channels. Under the assumption that no resonance has been seen in this or other channels, it is reasonable to unitarize the amplitude by the simplest and model-independent means without worrying about  the value of the resonances' masses and widths. Experimental cross sections for the process    $pp\rightarrow l^\pm \nu_{ l} l^\pm  \nu_{ l} j j$ can then be compared with the SM and provide the means to constrain the coefficients of the effective lagrangian and  the physics behind the EW symmetry breaking.
 
 Even in the same-sign $WW$ channel, the extraction of the coefficients is challenging. Appropriated  selection cuts are required to isolate the VBS process from other, often larger backgrounds. In addition, we want to isolate the longitudinally from the transversally polarised vector boson. The former is mostly produced together with a final quark which is more forward than in the case in which the $W$ is transversally polarised. These requirements provide a standard set of selection rules to which we add a final requirement (the Warsaw cut~\cite{Doroba:2012pd}) on the size and direction of the final transverse momenta of jets and leptons which has been shown to be  effective in disentangling longitudinal and transverse vector boson polarizations.
 
 The opposite-sign $W^\pm W^\mp \rightarrow W^\pm W^\mp$ channel is less clean mainly because of the large background generated by the production of $t\bar t$ pairs. It would be best to do without it and we find that indeed it is possible to establish the  most stringent constraints by means of  only the same-sign channel. 
 
% Even though the presence of low-laying resonances in the opposite-sign channel cannot be argued away like in the same-sign channel, we will assume again that none have been found and unitarize the amplitude by the same means as in the same-sign channel. A set of selection cuts similar to that used in the same-sign channel can be applied here too.
 
 The study of the cross sections $\sigma(pp\rightarrow l^\pm \nu_{ l} l^\pm  \nu_{ l} j j)$ and $\sigma(pp\rightarrow l^\pm \nu_{ l} l^\mp  \nu_{ l} j j)$ at the LHC can lead to either the discovery or the exclusion of the terms in the effective lagrangian at the permil level. This is the size of these coefficients expected on dimensional grounds. For the first time we will be able to study the breaking of the EW symmetry at its fundamental level.
 
 In this introduction we recall the relevant literature in section \ref{sec:Story}, introduce the notation in  section \ref{sec:Notation}, discuss coefficients size and higher-order terms in section \ref{sec:coef}, compare the nonlinear (chiral) lagrangian with the linear and anomalous couplings formulations to provide  a dictionary for the relevant coefficients in section \ref{sec:Others} . We collect the existing limits and estimates in section \ref{sec:Bounds}.

%%%%%%%%%%%%%%%%%%%%%%%%%%%%%%%%
\subsection{The story so far} \label{sec:Story}
%%%%%%%%%%%%%%%%%%%%%%%%%%%%%%%%

The importance of VBS in the study of the EW symmetry breaking was recognised early on~\cite{Chanowitz:1985hj,duncan}. The unique role played by the same-sign channel was singled out in~\cite{Barger:1990py} and the identification of the central jet veto to distinguish the EW signal from the QCD background was first introduced in~\cite{Bagger:1995mk}
where the purely leptonic ``gold-plated'' decay channels were also identified. In~\cite{Butterworth:2002tt} the study was extended to semi-leptonic decay modes. 

More recently, with the coming of the LHC, many different groups and authors have discussed VBS from different points of view. Of relevance for the present work, the papers in
\cite{Englert:2008tn} and ~\cite{Accomando:2005hz}  have provided new insights on both the gold-plated and the semi-leptonic decay channel as well as the determination of resonances and the coefficients of the effective lagrangian.
In a parallel development, the extraction of bounds on anomalous triple and quartic gauge couplings from the LHC data was discussed in~\cite{Belyaev:1998ih}.

The parameterisation of the experimental results in terms of the effective chiral lagrangians was begun in  \cite{Dobado:1990jy} and further discussed in~\cite{Fabbrichesi:2007ad,Brivio:2013pma,Espriu:2012ih,Delgado:2013loa}. The analysis in \cite{Eboli:2006wa} provides an estimate of the possible limits at the LHC on the effective lagrangian coefficients---of which  our work can be considered an improved and updated version.

For a more comprehensive review of the literature, the interested reader is referred to \cite{Szleper:2014xxa}.

%%%%%%%%%%%%%%%%%%%%%%%%%%%%%%%%
\subsection{Notation}\label{sec:Notation}
%%%%%%%%%%%%%%%%%%%%%%%%%%%%%%%%

In this work we choose to adopt the non-linear parametrization for the EW symmetry breaking sector. This choice is particularly suitable for our purposes, since the non-linear formulation puts the longitudinal degrees of freedom of the EW gauge bosons---dominant in the VBS processes we are interested in---in foreground position.

The effective non-linear lagrangian that describes the dynamics of the Goldstone bosons associated to the $SU(2)_L\times U(1)_Y \rightarrow U(1)_{em}$ symmetry breaking pattern is organized as an expansion in powers of Goldstone bosons momenta and the number of possible operators is restricted by Lorentz, gauge, charge and parity symmetry. The leading term is of $O(p^2)$  and---in the presence of a light Higgs particle $h$---it can be written as
\be
{\cal L}_0 = \frac{v^2}{4}\left[ 1 +  2 a \frac{h}{v} + b \left(\frac{h}{v}\right)^2  \right] \Tr \left[ (D_\mu U)^\dag (D^\mu U) \right] +\frac{1}{2} \partial_\mu h \partial^\mu h  -V(h) \label{L0} \, ,
\ee
where $a$ and $b$ are coefficients parametrizing the Higgs interactions with the gauge bosons. The Goldstone bosons $\pi^a$ $(a=1,2,3)$ are encoded into the matrix
\be
U = \exp ( i \pi^a \sigma_a/v) \, ,
\ee
where $\sigma_a$ are the Pauli matrices and $v=246$ GeV is the EW vacuum. The Goldstone matrix $U$ has well-defined transformation properties under $SU(2)_L\times U(1)_Y$: $U\to \mathcal{G}_L U \mathcal{G}_R^{\dag}$
with $ \mathcal{G}_L = \exp(i\alpha^j\sigma_j/2)\in SU(2)_L$ and $ \mathcal{G}_R = \exp(i\alpha_Y\sigma_3/2)\in U(1)_Y$. It constitutes
the building-block for the effective lagrangian with broken (non-linearly realized) EW symmetry. In \eq{L0} the covariant derivative is given by
\be
D_\mu U = \partial_\mu U + i g \hat{W}_{\mu} U - i g^\prime U \hat B_\mu \, ,
\ee
where $\hat{W}_{\mu}\equiv \sigma_a W_{\mu}^a/2$ and $\hat B_\mu\equiv \sigma^3B_{\mu}/2$. The fields $W_\mu^{a}$ and $B_\mu$ are the $SU(2)_L\times U(1)_Y$ gauge fields with standard kinetic terms
\be\label{Lg}
{\cal L}_{gauge} =-\frac{1}{2} \Tr \hat{W}_{\mu\nu}\hat{W}^{\mu\nu} -\frac{1}{2} \Tr \hat{B}_{\mu\nu}\hat{B}^{\mu\nu}\, ,
\ee
where $\hat{W}_{\mu\nu} =  \partial_\mu \hat{W}_\nu  - \partial_\nu \hat{W}_\mu + i g[\hat{W}_\mu,\hat{W}_\nu]$ and  $ \hat{B}_{\mu\nu} = \partial_{\mu}\hat{B}_{\nu} -  \partial_{\nu}\hat{B}_{\mu}$.

In \eq{L0} the quantity $V(h)$ is the Higgs boson potential with the generic structure $V(h)= \frac{1}{2} m_h^2 h^2+ d_3(m_h^2/2v) h^3 + d_4 (m_h^2/8v^2) h^4$, where the parameters $d_3$ and $d_4$ are related to the triple and quartic Higgs self-interactions, respectively. 

We extend the lagrangian in \eq{L0} by adding a set of higher dimensional operators parametrizing the following $O(p^4)$ lagrangian
\bea \label{L1}
{\cal L}_1 & = & \frac{1}{2} a_1 g g^{\prime} B_{\mu\nu}   \Tr (T \hat{W}^{\mu\nu})+\frac{i}{2} a_2 g^{\prime}   B_{\mu\nu} \Tr ( T [V^\mu,V^\nu]) +2  i a_3 g \Tr ( \hat{W}_{\mu\nu} [ V^\mu,V^\nu])  \nn \\
& + & a_4 \left[ \Tr (V_\mu V_\nu)\right]^2
+ a_5 \left[ \Tr (V_\mu V^\mu)\right]^2 \,,
\eea
where $V_\mu = (D_\mu U) U^\dag$ and $T \equiv U\sigma^3 U^{\dag}$. The complete list of operators entering in the chiral lagrangian at $O(p^4)$ can be found in \cite{Appelquist:1980vg}. Here we restrict to a subset of those given by \eq{L1} because we are interested only in operators that modify triple and quartic gauge boson couplings and are relevant for VBS processes. In particular, the coefficient $a_1$ modifies the vertices with both two and three gauge bosons, $a_2$ and $a_3$ those with three gauge bosons while $a_4$ and $a_5$ only vertices with four gauge bosons. 
%Notice that the operators associated to $a_3$, $a_4$ and $a_5$ respect $SU(2)$-custodial invariance.
In principle, being the Higgs boson a singlet, we can add a multiplicative function of $h$ in front of all the operators of \eq{L1}; a function similar to the one between squared brackets of \eq{L0} but with different coefficients, as shown in~\cite{Brivio:2013pma}. Here we assume these corrections to be sub-leading and neglect them.

In the framework we have introduced, the SM corresponds to the choice $a = b = d_3 = d_4 = 1$ and $a_1 = a_2 = a_3 = a_4 = a_5 = 0$. Any departure from these values  can be interpreted as  presence of new physics.

%Notice that these two operators do not formally break custodial invariance since they vanish in the limit $g^{\prime} \to 0$.

%%%%%%%%%%%%%%%%%%%%%%%%%%%%%%%%
\subsection{Coefficients size and higher-order terms} \label{sec:coef}
%%%%%%%%%%%%%%%%%%%%%%%%%%%%%%%%

The effective field theory approach to physics beyond the SM is made into an even more powerful tool after few assumptions on the ultraviolet (UV) physics are made.
Without such, admittedly, speculative arguments, it  remains a mere classification of effective operators without offering any particular physical insight.

The use of a non-linear realization of the electroweak symmetry naturally emerges by assuming the existence of a new strongly-interacting sector responsible for its breaking.
The new sector can be characterised by two parameters:  a coupling, $g_*$, and a mass scale, $\Lambda$. The latter identifies
 the mass of 
the heavy states populating the new sector.
Furthermore---in the spirit of the non-linear $\sigma$ model used in eq.~(\ref{L0})---it is natural to assume that  the Goldstone bosons originate from the spontaneous breaking of
a global symmetry of the strong sector; in this regard, the $\sigma$-model scale $v$ is linked to the parameter of the strong sector via the relation
 $g_*v\approx \Lambda$. Having in mind a cut-off scale $\Lambda$ of a few  TeVs, the relation $g_*v\approx \Lambda$
  points towards a maximally strongly coupled sector in which one expects $g_*\approx 4\pi$.
 In this picture the Higgs boson emerges as a light resonance of the strong sector.
 
 The size of the effective operators generated integrating out the heavy resonances of the strong sector 
 can be estimated by means of the so-called na\"{\i}ve dimensional analysis (NDA)~\cite{Giudice:2007fh}. 
 Integrating out heavy fields at the tree level in the strong sector, the effective Lagrangian takes the following general form
  \begin{equation}\label{eq:NDA}
 \mathcal{L}_{\rm eff} = \frac{\Lambda^4}{g_*^2}\hat{\mathcal{L}}\left[
 \frac{\partial_{\mu}}{\Lambda},\frac{g_* h}{\Lambda}, \frac{g_* \pi^a}{\Lambda},
 \frac{g A_{\mu}}{\Lambda},\frac{g A_{\mu\nu}}{\Lambda^2}
 \right]~,
 \end{equation}
where $A_{\mu}$ ($A_{\mu\nu}$) denotes a generic gauge field (field strength) while $\hat{\mathcal{L}}$ is a dimensionless functional. 
For simplicity, we neglect fermionic contributions since they are not
important in our setup. 
 The most relevant information in eq.~(\ref{eq:NDA}) is that the Goldstone bosons and the Higgs are always 
accompanied by an insertion of $g_*$ since they are directly coupled to the strong sector they belong to. 

We can now 
analyze by power counting the effective operators, written in eq.~(\ref{L1}), relevant for the $WW$ scattering process we are interested in:

\begin{itemize}

\item[$\circ$]  The effective operators $a_4 \left[ \Tr (V_\mu V_\nu)\right]^2$ and $a_5 \left[ \Tr (V_\mu V^\mu)\right]^2$ generate the quadrilinear vertex 
involving four Goldstone boson derivatives. Using the rules of NDA we find the corresponding $WW$ scattering amplitude to be proportional to $g_*^2(E/\Lambda)^4$, where $E$ is the characteristic
center-of-mass energy of the process (for the sake of simplicity we do not distinguish here between different $WW$ channels, since we are simply interested in an order-of-magnitude estimate of the amplitude);

\item[$\circ$] The operator 
$a_3 \Tr ( \hat{W}_{\mu\nu} [ V^\mu,V^\nu])$ generates the trilinear coupling 
\be
\epsilon_{kAB}(\partial_{\mu}W_{\nu}^k - \partial_{\nu}W_{\mu}^k)(\partial^{\mu}\pi^A
\partial^{\nu}\pi^B - \partial^{\nu}\pi^A
\partial^{\mu}\pi^B)\, .
\ee
 The corresponding $WW$ scattering amplitude involves the $s$-, $t$-, and $u$-channel exchange of the EW gauge bosons $W^{k=1,2,3}$, and from NDA we obtain  an amplitude proportional to
 $g^2(E/\Lambda)^4$;

\item[$\circ$] The operator 
$a_2 B_{\mu\nu}\Tr(T[ V^\mu,V^\nu])$ generates the trilinear coupling 
\be
\epsilon_{3AB}(\partial_{\mu}B_{\nu} - \partial_{\nu}B_{\mu})(\partial^{\mu}\pi^A
\partial^{\nu}\pi^B - \partial^{\nu}\pi^A
\partial^{\mu}\pi^B)\, . 
\ee
The corresponding $WW$ scattering amplitude involves the $s$-, $t$-, and $u$-channel exchange of the EW gauge boson $B$, and from NDA we obtain an amplitude proportional to
 $g^{\prime 2}(E/\Lambda)^4$;

\item[$\circ$] Finally, 
the $\sigma$-model operator 
$\Tr\left[ (D_\mu U)^\dag (D^\mu U) \right]$ generates the trilinear structures
\be
\epsilon_{kAB}W_{\mu}^k[(\partial^{\mu}\pi^A)\pi^B - (\partial^{\mu}\pi^B)\pi^A]\quad \mbox{and} \quad \epsilon_{3AB}B_{\mu}[(\partial^{\mu}\pi^A)\pi^B - (\partial^{\mu}\pi^B)\pi^A]\, . 
\ee
By combining these vertices with the trilinear interactions extracted before from $a_3 \Tr ( \hat{W}_{\mu\nu} [ V^\mu,V^\nu])$ and $a_2 B_{\mu\nu}\Tr(T[ V^\mu,V^\nu])$, 
we find an amplitude proportional to, respectively, $g^2(E/\Lambda)^2$ and $g^{\prime 2}(E/\Lambda)^2$. 

\end{itemize}
Notice that the energy dependence of these amplitudes---obtained here  by dimensional analysis---will be confirmed by means of a direct computation in section~\ref{sec:Unitarity}.

We can now compare the amplitude proportional to $a_{4,5}$ against that proportional to  $a_2$. Both these amplitudes 
grow with $E^4$; however, the contribution coming from the operators $a_4 \left[ \Tr (V_\mu V_\nu)\right]^2$ and $a_5 \left[ \Tr (V_\mu V^\mu)\right]^2$
is parametrically enhanced since proportional to $g_*^2$.
Similarly, we can compare the same amplitude  against  that proportional to  $a_3$. 
The former dominates if the condition $g_*(E/\Lambda) > g$ is satisfied. 
Since $g_*v\approx \Lambda$, it implies $E > gv$, a condition easily satisfied at typical LHC energies.

It therefore seems natural to expect that in the presence of  a genuinely strongly coupled new sector the most relevant contribution to the $WW$ scattering arises from the pure Goldstone operators $a_4 \left[ \Tr (V_\mu V_\nu)\right]^2$ and $a_5 \left[ \Tr (V_\mu V^\mu)\right]^2$. For this reason in section~\ref{sec:methods} 
we will focus our Monte Carlo analysis on the two coefficients $a_4$ and $a_5$, setting $a_2=a_3 = 0$.

Finally, notice that the same NDA argument  can be  used in order to estimate the contribution of $O(p^6)$ (or higher) operators.
For definiteness, let us consider the $O(p^6)$ operator $\epsilon_{ABC}(W^A)_{\mu}^{\nu}(W^B)_{\nu\rho}(W^C)^{\rho\mu}$. It generates the quadrilinear vertex 
\be
\epsilon_{ABC}\epsilon_{AB'C'}(\partial_{\mu}W^{B,\nu}-\partial^{\nu}W^B_{\mu})(
\partial_{\nu}W^C_{\rho} - \partial_{\rho}W^C_{\nu}
)W^{B',\rho}W^{C',\mu}
\ee
 which contributes to the $WW$ (transverse) scattering according to $g^2(g^2/g_*^2)(E/\Lambda)^2$.\footnote{A further loop-suppression $g_*/(4\pi)^2$ is expected, since this operator can not be generated by integrating out at the tree level any resonance with spin less than 2. However, since we have in mind the limit $g_* \approx 4\pi$, the presence of this extra factor does not change our estimate.} As evident from the previous discussion, the maximally strongly coupled limit $g_* \approx 4\pi$ suppresses this contribution
that in principle could interfere with the perturbative expansion.

%%%%%%%%%%%%%%%%%%%%%%%%%%%%%%%%
\subsection{Mapping to other formulations} \label{sec:Others}
%%%%%%%%%%%%%%%%%%%%%%%%%%%%%%%%

It is  useful to map the non-linear formalism into other popular parameterizations---thus providing a dictionary
through which to translate all the available bounds. In the following, we briefly discuss the relations with
${\it i)}$ the phenomenological lagrangian commonly used to parametrize triple and quartic anomalous gauge boson couplings and  ${\it ii)}$ the higher dimensional effective lagrangian obtained by imposing the additional assumption  that the Higgs field $h$ is part of a $SU(2)_L$ doublet  that  breaks  the  EW  symmetry.

%%%%%%%%%%%%%%5
\subsubsection{Anomalous triple and quartic gauge couplings}
%%%%%%%%%%%%%%
Traditionally bounds on triple gauge boson couplings (TGC) have been expressed in terms of anomalous coefficients~\cite{Hagiwara:1992eh}, according to the following phenomenological lagrangian
\begin{eqnarray}
\mathcal{L}_{\rm TGC} &=& ie\left[
g_{1}^{\gamma}A_{\mu}\left(
W_{\nu}^{-}W^{+\mu\nu} - W_{\nu}^+W^{-\mu\nu}
\right) + \kappa^{\gamma}W_{\mu}^-W_{\nu}^+A^{\mu\nu} + \frac{\lambda^{\gamma}}{m_W^2} W^{-\nu}_{\mu}W^+_{\nu\rho}A^{\rho\mu}
\right]\nonumber \\
&+&\frac{iec_W}{s_W}\left[
g_{1}^{Z}Z_{\mu}\left(
W_{\nu}^{-}W^{+\mu\nu} - W_{\nu}^+W^{-\mu\nu}
\right) + \kappa^{Z}W_{\mu}^-W_{\nu}^+Z^{\mu\nu} + \frac{\lambda^{Z}}{m_W^2} W^{-\nu}_{\mu}W^+_{\nu\rho}Z^{\rho\mu}
\right]~,
\end{eqnarray}
where $W^{\pm}_{\mu\nu}\equiv \partial_{\mu}W^{\pm}_{\nu}-\partial_{\nu}W^{\pm}_{\mu}$, $V_{\mu\nu}\equiv \partial_{\mu}V_{\nu} - \partial_{\nu}V_{\mu}$, with $V=A,Z$.
The SM corresponds to $g_1^{\gamma,Z}=\kappa^{\gamma,Z}=1$, $\lambda^{\gamma,Z} = 0$.

In our case $\kappa_Z$, $\kappa_\gamma$ and $g_1^Z$ ($g_1^\gamma$ is fixed to be 1 by gauge invariance) are modified by the presence of the operators in \eq{L1}. By inspection, we can identify the following identities:
\be
\Delta g_1^Z =  \frac{g^{\prime 2}}{c_W^2 - s_W^2}a_1  + \frac{2 g^2}{c_W^2}a_3~,~~~~
\Delta\kappa^{\gamma} = g^2(a_2 - a_1) + 2 g^2 a_3~,~~~~\Delta\kappa^Z =  \frac{g^{\prime 2}}{c_W^2 - s_W^2}a_1  - g^{\prime 2}(a_2 - a_1) + 2 g^2 a_3~.
\ee
Furthermore, it follows that $\Delta\kappa^Z = \Delta g_1^Z - (g^{\prime 2}/g^2)\Delta\kappa^{\gamma}$.
For illustrative purposes we can take $a_1 = 0$, as suggested by the stringent fit of LEP data of~\cite{Falkowski:2013dza}. In this case the previous relations
simplify to
\begin{equation}\label{ac23}
\Delta g_1^Z = \frac{2 g^2}{c_W^2}a_3~,~~~~\Delta\kappa^{\gamma} -\Delta\kappa^{Z} =  \left(g^2 + g^{\prime 2} \right)a_2~.
\end{equation}
As far as the anomalous quartic gauge couplings (QGC) are concerned, they are usually parametrized as follows
\begin{eqnarray}
\mathcal{L}_{QGC} &=&  e^2 g_{WWVV} \left[
g_{1}^{VV}V^{\mu}V^{\nu}W_{\mu}^-W_{\nu}^+ - g_2^{VV}V^{\mu}V_{\mu}W^{-\nu}W^+_{\nu}
\right]\nonumber\\
&+& \frac{e^2c_W}{s_W}\left[
g_1^{\gamma Z}A^{\mu}Z^{\nu}(W_{\mu}^-W_{\nu}^+ + W_{\mu}^+W_{\nu}^-) - 2g_{2}^{\gamma Z}A^{\mu}Z_{\mu}W^{-\nu}W_{\nu}^+
\right]\nonumber \\
&+& \frac{e^2}{2 s_W^2}
\left[
g_1^{WW} W^{-\mu}W^{+\nu}W_{\mu}^-W_{\nu}^+  - g_2^{WW}(W^{-\mu}W_{\mu}^+)^2
\right] + \frac{e^2}{4s_W^2 c_W^4}h^{ZZ}(Z_{\mu}Z^{\mu})^2~,
\end{eqnarray}
with $g_{WW\gamma\gamma} = 1$, $g_{WWZZ} = c_W^2/s_W^2$. The SM corresponds to $g_{1/2}^{VV^{\prime}}=1$, $h^{ZZ} = 0$.
The effective operators of \eq{L1} produce the following corrections
\begin{align}
\Delta g_{1}^{\gamma Z} &=  \Delta g_{2}^{\gamma Z} =  \frac{g^{\prime 2}}{c_W^2 - s_W^2}a_1  + \frac{2 g^2}{c_W^2}a_3~,    &   \Delta g_2^{ZZ} & = 2\Delta g_1^{\gamma Z} - \frac{g^2}{c_W^4}a_5~,   \\
\Delta g_{1}^{ZZ} &= 2\Delta g_{1}^{\gamma Z}  + \frac{g^2}{c_W^4}a_4~,   & \Delta g_1^{WW}  & = 2c_W^2\Delta g_1^{\gamma Z} + g^2 a_4 ~,   \\
 h^{ZZ} &= g^2(a_4 + a_5) ~,   & \Delta g_2^{WW}  & = 2c_W^2\Delta g_1^{\gamma Z} - g^2 (a_4  + 2 a_5)~.
\end{align}

\subsubsection{Comparison with the linear realization}

At dimension 6, the bosonic operators relevant for our discussion are~\cite{Falkowski:2014tna}
\begin{align}
\mathcal{O}_{WB} & = \frac{g\kappa_{WB}}{4m_W^2}B^{\mu\nu}W_{\mu\nu}^kH^{\dag}\sigma^k H~, & 
\mathcal{O}_{3W} & = \frac{g\kappa_{3W}}{6 m_W^2}\epsilon^{ijk}W_{\mu\nu}^iW_{\rho}^{j \nu}W^{k \rho\mu}~, &  
\mathcal{O}_{H} & = \frac{\kappa_H}{v^2}\partial^{\mu}(H^{\dag}H)\partial_{\mu}(H^{\dag}H)~, \nn    \\ 
\mathcal{O}_{HW} & = \frac{ig\kappa_{HW}}{m_W^2}(D^{\mu}H)^{\dag}\sigma^k (D^{\nu}H)W_{\mu\nu}^k~, & 
\mathcal{O}_{W} & = \frac{ig\kappa_W}{2m_W^2}H^{\dag}\sigma^k\overleftrightarrow{D}_{\mu}H(D_{\nu}W^{k \mu\nu})~,  &  
\mathcal{O}_{WW} & = \frac{g^2\kappa_{WW}}{4m_W^2}(H^{\dag}H)W_{\mu\nu}^k W^{k\mu\nu}~, \nn \\
\mathcal{O}_{HB} & = \frac{ig^{\prime}\kappa_{HB}}{m_W^2}(D_{\mu}H)^{\dag}(D_{\nu}H)B^{\mu\nu}~, & 
\mathcal{O}_{B} &= \frac{ig^{\prime}\kappa_B}{2m_W^2}H^{\dag}\overleftrightarrow{D}_{\mu}H (\partial_{\nu}B^{\mu\nu})~,  & 
\mathcal{O}_{BB} &= \frac{g^{\prime 2}\kappa_{BB}}{4m_W^2}(H^{\dag}H)B_{\mu\nu}B^{\mu\nu}~, \nn \\
\mathcal{O}_{2W} & = \frac{g^2\kappa_{2W}}{16m_W^2}(D_{\rho}W_{\mu\nu}^k)^2~, & 
\mathcal{O}_{2B} & = \frac{g^{\prime 2}\kappa_{2B}}{16m_W^2}(D_{\rho}B_{\mu\nu})^2~,\label{eq:O3}
\end{align}
with $H^{\dag}\overleftrightarrow{D}_{\mu}H = H^{\dag}(D_{\mu}H) - (D_{\mu}H)^{\dag}H$. $H$ is the Higgs doublet of the SM with hypercharge $Y_H = 1/2$. The operators $\mathcal{O}_{H,WW,BB}$ in the last column affect only Higgs physics 
while the remaining ones affect the electroweak precision observables. Three operators $\mathcal{O}_{3W}$, $\mathcal{O}_{WW}$ and $\mathcal{O}_{W}$ enter in $WW$ vector boson scattering.

Notice that there is a redundancy in this list, since it is possible to remove some of these operators using the equation of motion of the gauge fields and 
the operator identities $\mathcal{O}_{HB} = \mathcal{O}_B - \mathcal{O}_{WB} - \mathcal{O}_{BB}$, and $\mathcal{O}_{HW} = \mathcal{O}_W - \mathcal{O}_{WB} - \mathcal{O}_{WW}$.\footnote{
In~\cite{Baak:2013fwa} (using the notation of \eq{eq:O3}) 
the subset $\{\mathcal{O}_H, \mathcal{O}_{3W}, \mathcal{O}_{HW},  \mathcal{O}_{HB}, \mathcal{O}_{WW}, \mathcal{O}_{BB} \}$
was considered. 
} 
For instance in the SILH basis used in~\cite{Elias-Miro:2013gya} the operators $\mathcal{O}_{2W,2B,WB,WW}$ are dropped while in the so-called Warsaw basis~\cite{Grzadkowski:2010es} the operators 
$\mathcal{O}_{2W,2B,W,B,HW,HB}$ are dropped.
By comparing the anomalous TGC, we find 
\begin{equation}
\Delta g_1^Z = g^2\left(
\frac{s_W^2}{c_W^2 - s_W^2}a_1 + \frac{2 a_3}{c_W^2}
\right) = \kappa_W + \kappa_{HW}~,~~~~~\Delta \kappa^{\gamma} = g^2(a_2 - a_1 + 2 a_3) = - \kappa_{WB} + \kappa_{HW} + \kappa_{HB}~.
\end{equation}

There are  18 operators of dimension 8 but only two
\begin{equation}
\mathcal{O}_{S,0} = \frac{f_{S,0}}{\Lambda^4}[(D_{\mu}H)^{\dag}D_{\nu}H][(D^{\mu}H)^{\dag}D^{\nu}H]~,~~~~~\mathcal{O}_{S,1} = \frac{f_{S,1}}{\Lambda^4}[(D_{\mu}H)^{\dag}D^{\mu}H][(D^{\nu}H)^{\dag}D_{\nu}H]~,
\end{equation}
are relevant for us. The other 16 operators of dimension 8---five of which enter $WW$ scattering---have derivative terms in the vector bosons in addition to those with the Higgs field, and would have to be matched  to higher order terms in the chiral lagrangian. For the $WW$ channel we are interested in, we find~\cite{Baak:2013fwa}:
\be
a_4 = \frac{f_{S,0}}{\Lambda^4} \frac{v^4}{8} \, , \quad \mbox{and} \quad a_4+2 a_5 = \frac{f_{S,1}}{\Lambda^4} \frac{v^4}{8}\, . \label{f01}
\ee

%%%%%%%%%%%%%%%%%%%%%%%%%%%%%%%%
\subsection{Current and estimated bounds} 
\label{sec:Bounds}
%%%%%%%%%%%%%%%%%%%%%%%%%%%%%%%%

Current bounds on the coefficients of the operators in \eq{L1} come from EW precision measurements performed at LEP-I and LEP-II and from data collected at LHC run 1. Estimated bounds are meant to be for LHC run 2. 
 
%%%%%%%%%%%%%%
\subsubsection{Electroweak precision tests}
\label{sec:EWPT}
%%%%%%%%%
The coefficient $a_1$ is strongly constrained by LEP-I and LEP-II data because it contributes at tree-level to the $S$ parameter
\begin{equation}\label{eq:Sparamater}
\Delta S = -16\pi a_1\,.
\end{equation}
A simple fit of LEP data~\cite{Falkowski:2013dza} performed including the correction in eq.~(\ref{eq:Sparamater}) shows that  
\be
a_1 = (1.0  \pm 0.7) \times 10^{-3} \, .
\ee
On the other hand, the other coefficients $a_2$, $a_3$, $a_4$ and $a_5$ contribute to the $S,T,U$ parameters only at at one-loop. 
%%%%%%%%%%%%%%%%%%%
\begin{figure}[ht!]
\begin{center}
\includegraphics[width=3in]{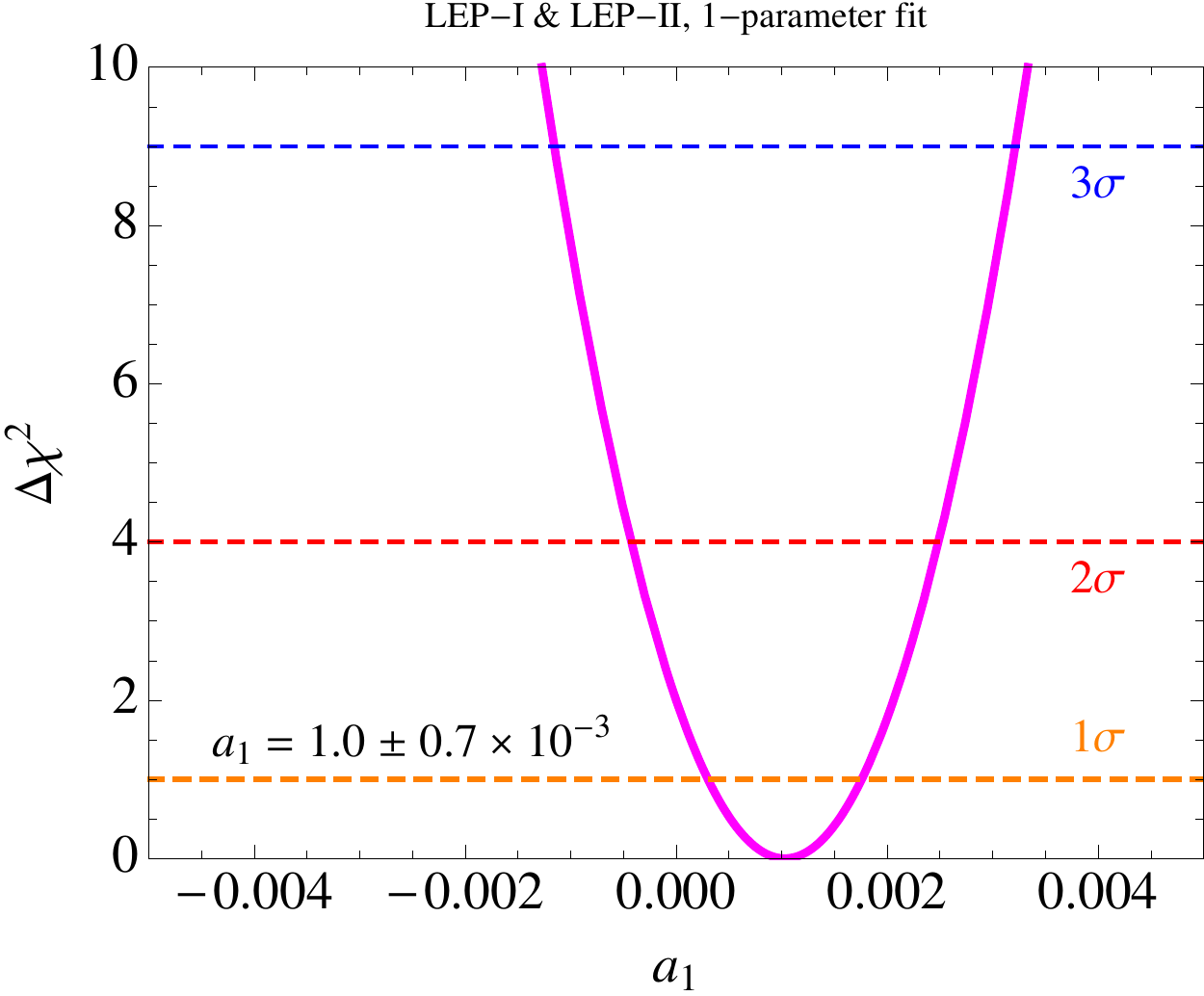}
\caption{\small  \textit{$\Delta \chi^2$ plot for the limit on the  coeficient $a_1$ from LEP I and II precision  tests.
\label{fig_a1}}}
\end{center}
\end{figure}
%%%%%%%%%%%%%%%%%%%%%%%%%% 
In particular, the one-loop contributions of $a_4$ and $a_5$ to EW precision measurements lead to the following (rather weak) bounds on these coefficients at 90\% CL~\cite{Brivio:2013pma}
\be
-0.094< a_4 < 0.10 \quad \mbox{and} \quad -0.23 < a_5 < 0.26\, . \label{45}
\ee

The combined LEP bounds on TGC~\cite{Lombardo:2013daa} are
\be
- 0.054 < \Delta g_1^Z  <0.021\qquad  \qquad - 0.07< \Delta \kappa_Z < 0.051 \qquad  \qquad- 0.099<\Delta \kappa_\gamma < 0.066 \quad (95\%\, \mbox{CL})\,.
\ee
By means of the relations in \eq{ac23} we can translate the above bounds into limits on the coefficients $a_2$ and $a_3$
\be
-0.26 < a_2 < 0.26 \quad \mbox{and} \quad  -0.10 < a_3 <  0.04 \, ,
\ee
which are in agreement with the ones found in \cite{Brivio:2013pma}.

%%%%%%%%%%%%%%
\subsubsection{LHC run 1 and run 2}
%%%%%%%%%%%%%%%
Current experimental limits on $a_4$ and $a_5$ based on LHC run 1 are still rather weak and comparable to those in \eq{45} coming from EW precision measurements. ATLAS and CMS find~\cite{Aad:2014zda}
\be
-0.14 < a_4 < 0.16 \quad \mbox{and}  \quad -0.23 < a_5 < 0.24  \label{limitsLHC}
\ee
at the 95\% CL and with a luminosity of 20.3 fb$^{-1}$ (CM energy of 8 TeV). These bounds are obtained by studying the double charged channel (after unitarization by means of the $K$-matrix method).

Estimated bounds on $a_4$ at the LHC run 2 presented in~\cite{Degrande:2013yda} represent a substantial improvement with respect to the current LHC limits, namely
\be
a_4 \leq 0.066\, .
\ee
This limit is obtained at 95\% CL and for a luminosity of 300 fb$^{-1}$ (CM energy of 14 TeV).  

The best existing estimated limit is obtained in~\cite{Eboli:2006wa} where they combine same- and opposite-sign channels. They find
\be
-22 < \frac{f_0}{\Lambda^4} (\mbox{TeV}^{-4}) < 24 \quad \mbox{and} \quad -25 < \frac{f_1}{\Lambda^4} (\mbox{TeV}^{-4}) < 25
\ee
at the 99\% CL and for a luminosity of 100 fb$^{-1}$ (CM energy of 14 TeV). These bounds are
 equivalent by means of \eq{f01} to
\be
-0.01 < a_4 < 0.01 \quad \mbox{and} \quad -0.01 < a_5 < 0.01\, .
\ee

Recent data on the Higgs boson decays indicate a value for the Higgs coupling to the gauge bosons very close to the SM value, namely~\cite{Ellis:2013lra}
\be
a = 1.03 \pm 0.06 \, . \label{a}
\ee
No dramatic improvement on this limit is expected from future LHC runs due to systematic errors~\cite{CMS:2013xfa}.

%%%%%%%%%%%%%%%%%%%%%%
\subsubsection{Analyticity and causality}
%%%%%%%%%%%%%%%%%
The causal and analytic structure of the amplitudes leads to theoretical bounds on the possible values the  two coefficients $a_4$ and $a_5$  can assume~\cite{Pham:1985cr,Fabbrichesi:2007ad}. The most stringent of these comes from the requirement that the underlying theory respects causality:
\be
a_4(\mu) \geq \frac{1}{12} \frac{1}{(4\pi)^2} \log \frac{\Lambda^2}{\mu^2}
\quad \mbox{and} \quad a_4(\mu) + a_5(\mu) \geq \frac{1}{8} \frac{1}{(4\pi)^2} \log \frac{\Lambda^2}{\mu^2} \, ,
\ee
where $\Lambda$ represents the cutoff of the effective theory and $\mu<\Lambda$ the scale at which the amplitude is evaluated. For most practical proposes, we can neglect the logarithms and take 
\be
a_4>0 \quad \mbox{and} \quad a_4+a_5>0
\ee
as our causality bounds.   In our limits, we will assume them to be satisfied. Even though a violation of the above constraints would imply a (hard to entertain) breach in the causal structure of the theory, it is useful to bear in mind that this possibility cannot be ruled out \textit{a priori} and that an observation of a negative value of $a_4$ or of the combination $a_4+ a_5$ would be a really striking discovery in as much as it would  challenge the very foundations of quantum field theory.

%%%%%%%%%%%%%%%%%%%%%%%%%%%%%%%%%%%%%%
\section{Methods}
\label{sec:methods}
%%%%%%%%%%%%%%%%%%%%%%%%%
In section \ref{sec:mc} we present some details about the Monte Carlo simulation we have implemented in order to generate the VBS processes we are interested in. In section \ref{sec:Cuts} we describe the selection cuts we have employed. The statistical framework and the estimation of the effects of systematic errors are presented in section \ref{sec:Statistics}. Finally, in section \ref{sec:Unitarity} we discuss the violation of unitarity that can potentially arise and explain the unitarization procedure we have applied.

%%%%%%%%%%%%%%%%%%%%%%%%%%%%%%%%
\subsection{Monte Carlo simulation} \label{sec:mc}
%%%%%%%%%%%%%%%%%%%%%%%%%%%%%%%%

We have modeled the effective lagrangian consisting of the sum of the terms in \eq{L0}, \eqref{Lg} and \eqref{L1} by means of {\sc FeynRules}~\cite{Christensen:2008py} {\sc v2.0.28} in order to create the  {\sc UFO} module that is used in {\sc MadGraph5}~\cite{madgraph} {\sc v2.2.3} to simulate signal and background events related to the VBS processes we are interested in. 

Pure EW same-sign (SS)  $WW$ events in $pp\rightarrow W^\pm W^\pm\,jj\rightarrow l^\pm \nu_{ l} l^\pm  \nu_{ l}\, j j$ 
and  EW opposite-sign (OS) $WW$ events in $pp\rightarrow W^\pm W^\mp\,jj\rightarrow l^\pm \nu_{ l} l^\mp  \nu_{ l}\, j j$ are   $O(\alpha_W^6)$. Mixed QCD/EW SS and OS $WW$ events   are  $O(\alpha_W^4\alpha_s^2)$. 

The  relevant diagrams for probing the symmetry breaking dynamics  must contain direct $W$ boson interactions. They  are only a small fraction of the whole set   in pure EW events---which are dominated by diagrams in which the $W$ bosons are radiated from the incoming quarks, do not interact and  have predominantly a transverse polarisation. 
Mixed QCD/EW events---in which the vector bosons are produced from strongly scattered quarks---only contain diagrams in which the $W$ bosons do not interact. These two processes constitute the main irreducible background for our analysis.

%at $O(\alpha_W^6)$ by means of the following {\sc MadGraph5} syntax:\\[0.3em]
%\verb|generate p p > w+ w+ j j QCD=0 QED=6, w+ > l+ vl|\\
%\verb|add process p p > w- w- j j QCD=0 QED=6, w- > l- vl~|\,.\\[0.3em]
%We generate mixed QCD/EW same-sign $WW$ events $pp\rightarrow W^\pm W^\pm\,jj\rightarrow l^\pm \nu_{ l} l^\pm  \nu_{ l}\, j j$ at $O(\alpha_W^2\alpha_s^2)$ by means of the following {\sc MadGraph5} syntax:\\[0.3em]
%\verb|generate p p > w+ w+ j j QCD=2 QED=2, w+ > l+ vl|\\
%\verb|add process p p > w- w- j j QCD=2 QED=2, w- > l- vl~|.\\[0.3em]
%Pure EW opposite-sign (OS) $WW$ events $pp\rightarrow W^\pm W^\mp\,jj\rightarrow l^\pm \nu_{ l} l^\mp  \nu_{ l}\, j j$ have been generated at $O(\alpha_W^6)$ by using the following {\sc MadGraph5} syntax:\\[0.3em]
%\verb|generate p p > w+ w- j j QCD=0 QED=6, w+ > l+ vl, w- > l- vl~|\,.\\[0.3em]

%by using the following {\sc MadGraph5} syntax:\\[0.3em]
%\verb|generate p p > w+ w- j j QCD=2 QED=2, w+ > l+ vl, w- > l- vl~|.\\[0.3em]

Other background processes that contribute to SS and OS $WW$ channels are the following:
\begin{itemize}
\item $Z$+jets: events from this process can easily enter the OS channel and even the  SS channel if the sign of one lepton is mis-identified; 
\item $t\bar t$: the same considerations apply as for $Z$+jets, but this kind of events are expected to be harder to suppress due to the higher probability of having more energetic jet and lepton pairs with large angular separation (and therefore higher invariant masses);
\item $WZ$+jets, $t\bar tW$, $t\bar tZ$ and $t\bar tH$: events from these processes can originate high energy jets together with two or more charged leptons, 
which can  even enter the SS leptons selection, in case of three or more leptons or one lepton from the $t \bar t$ decay and another one from the associated boson decay;
\item single-lepton+jet ({\it e.g.} from $W$+jets): these events can enter any of the two channels if a jet is mis-identified as an additional isolated lepton.
\end{itemize}

Among the processes listed above, we have included  the $WZ$+jets  background   in the study of the SS  channel and the $t \bar t$ background   in that of the OS   channel. 
%The $t\bar t$ background  has been generated using the following {\sc MadGraph5} syntax:\\[0.3em]
% \verb|generate p p > t t~, (t > w+ b, w+ > l+ vl), (t~ > w- b~, w- > l- vl~)|\\
 %\verb|add process p p > t t~ j, (t > w+ b, w+ > l+ vl), (t~ > w- b~, w- > l- vl~)|\\
 %\verb|add process p p > t t~ j j, (t > w+ b, w+ > l+ vl), (t~ > w- b~, w- > l- vl~)|.\\[0.3em]
%  The $WZ$+jets background  has been generated using the following {\sc MadGraph5} syntax:\\[0.3em]
% \verb|generate p p > w z, (w > l+ vl, z >  l- l+)|\\
% \verb|add process p p > w z  j, (w > l+ vl, z >  l- l+)|\\
 %\verb|add process p p > w z j j, (w > l+ vl, z >  l- l+)|.\\[0.3em]
 The other processes are highly suppressed by the selection cuts, resulting in negligible effects in the analysis. We are, however, aware that this suppression depends on our Monte Carlo simulation which does not  predict correctly the effects of lepton charge mis-identification and jets reconstructed as leptons in the detector. 

The simulated events have been showered using {\tt Pythia 6.4}~\cite{Sjostrand:2006za} and subsequently processed through {\tt Delphes}~\cite{deFavereau:2013fsa} in order to simulate the response of a generic LHC detector. 
All the settings for both {\tt Pythia} and {\tt Delphes} have been kept as default 
({\it i.e.}, leaving the default options when installing the software through the {\tt Madgraph5} interface).

The number of events from each process has been then rescaled according to the LO cross-section and the expected integrated luminosity in each of the considered cases, 
to obtain an expected yield after the event selection.

%In this section we present the methods we used, starting in \ref{sec:Cuts} with the selection cuts employed. We discuss the statistical framework and estimate the effect of systematic errors in section \ref{sec:Statistics}. In section \ref{sec:Unitarity}  we show the potential violation of unitarity that arises and explain the unitarization procedure we have applied.

%%%%%%%%%%%%%%%%%%%%%%%%%%%%%%%%
\subsection{Selection cuts} \label{sec:Cuts}
%%%%%%%%%%%%%%%%%%%%%%%%%%%%%%%%
As already discussed, the pure EW production of $WW$ pairs in association with two jets at the LHC is dominated by events that have no direct relevance for the mechanism of electroweak symmetry breaking. Typically these events come from soft collisions involving incoming partons which lead to soft accompanying parton jets in the final state and can be rejected by appropriate cuts on their rapidity. In order to suppress this irreducible background and select events with hard $WW$ interactions we apply the following selection criteria 
%apply a set of basic cuts that retain events with
\begin{itemize}
\item[-]  small pseudo-rapidity and large transverse momentum for the $W$ gauge bosons;
\item[-]  two opposite tagging jets at large pseudo-rapidities and relatively small transverse momentum.
\end{itemize}
%These features, which are preserved in going from the partonic $WW$ production to the full process including decays and hadronization, have suggested the so called  ``central jet veto'' on which most of past analyses have been based. ,
Beside reducing the irreducible EW background, these cuts also suppress the mixed EW/QCD one.

Subsequently we have to impose additional cuts in order to wean out the transversally polarised vector bosons---which  accounts for more than 90\% of the total produced $W$ pairs---and select the longitudinally polarised ones. At the parton level, the production of longitudinally polarised $W$ is characterised by the final quark which is emitted more forward than in the case of the production of transversally polarised $W$. Moreover, after being produced by bremsstrahlung, the $W_L$ (mostly) conserve their polarisation---as long as we stay above the on-shell production threshold. 

The complete set of cuts applied in the case of SS and OS $WW$ channels are summarized below.

%%%%%%%%%%%%%%%%%%%%%%%%%%%%%%%%
\subsubsection{Same-sign $WW$ channel}
%%%%%%%%%%%%%%%%%%%%%%%%%%%%%%%%

We select events by applying the following set of cuts:
\begin{itemize}
\item two same-sign leptons with $p_T^{l^\pm}>20$ GeV and $|\eta_{l^\pm}| <  2.5$;
\item at least two jets ($p_T^j>25$ GeV and $|\eta_j|<4.5$) with relative rapidity $|\Delta y_{jj}| > 2.4$;
\item the two highest $p_T$ jets with an invariant mass $m_{jj} > 500$ GeV;
\item missing transverse energy $E_T^{\, miss} >$ 25 GeV.
\end{itemize}
This combined set of cuts has been optimized for VBS at the energy of 14 TeV, condering an integrated luminosity of 300 fb$^{-1}$
and are rather close to those already in use by the LHC experimental collaborations.

%%%%%%%%%%%%%%%%%%%
\begin{figure}[!htb!]
\minipage{0.5\textwidth}
  \includegraphics[width=.8\linewidth]{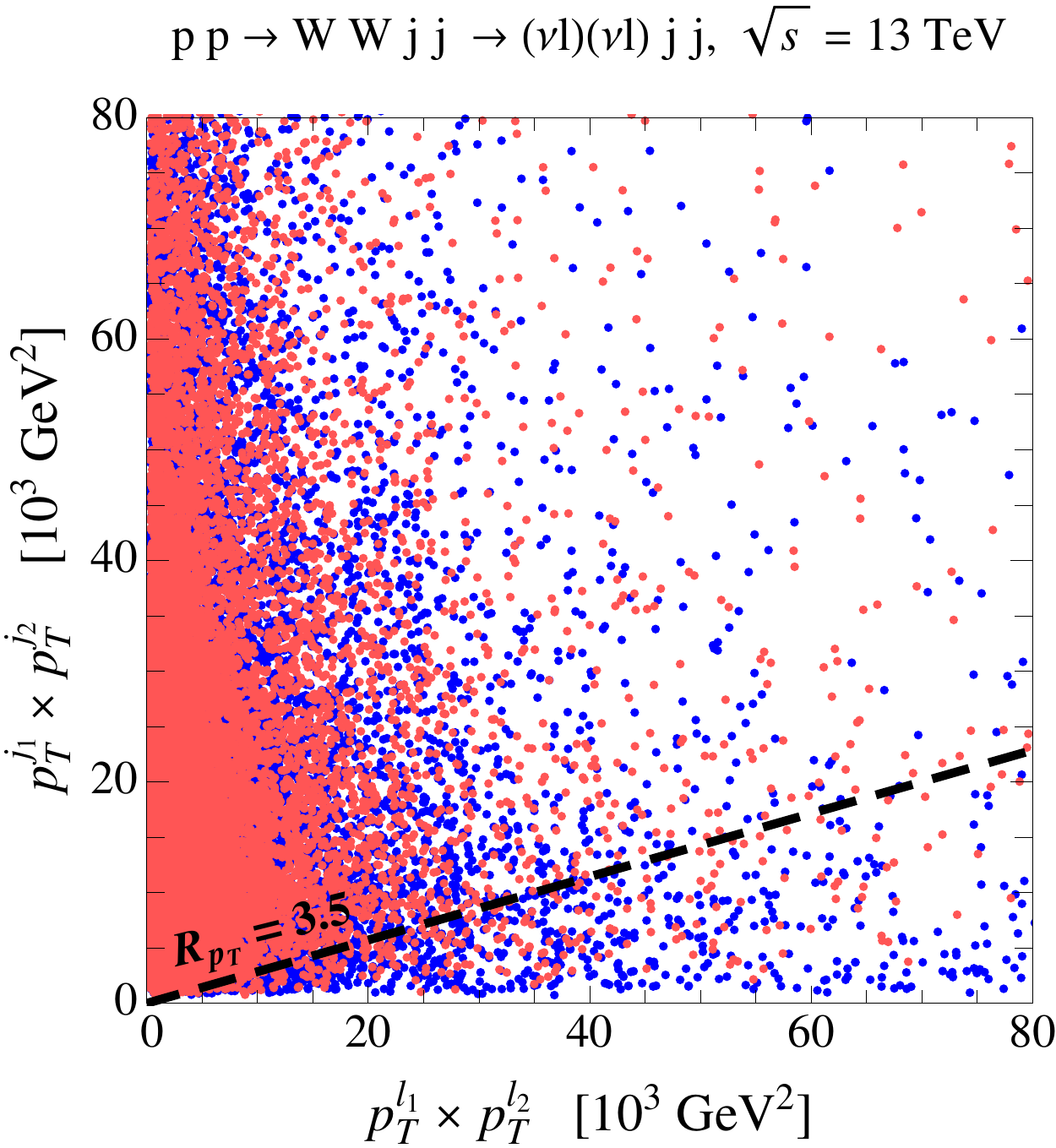}
\endminipage\hfill
\minipage{0.5\textwidth}
  \includegraphics[width=.8\linewidth]{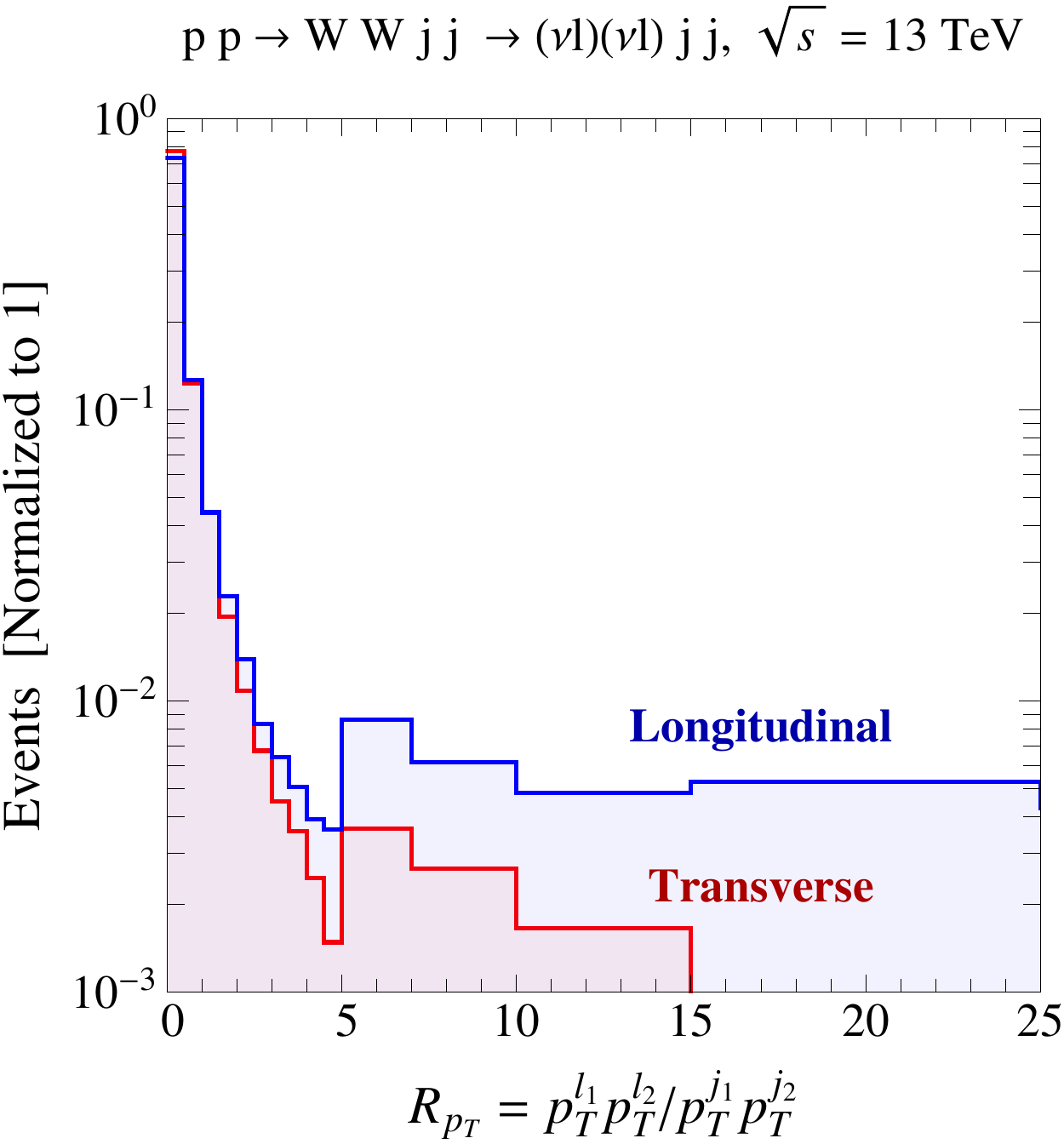}
%  \caption{A really Awesome Image}\label{fig:awesome_image2}
\endminipage\\
\caption{\small \textit{
Distribution of final state events obtained generating the processes $pp \to W^{\pm}W^{\pm}jj \to (l^{\pm}\nu_l)(l^{\pm}\nu_l)jj$ at the LHC with $\sqrt{s} = 13$ TeV.
We show in blue (red) the events with leptons coming from the decay of longitudinal (transverse) polarized W bosons.
}}\label{fig:Distributions}
\end{figure}
%%%%%%%%%%%%%%%%%%%

%%%%%%%%%%%%%%%%%%%
\begin{figure}[!htb!]
\minipage{0.5\textwidth}
  \includegraphics[width=.9\linewidth]{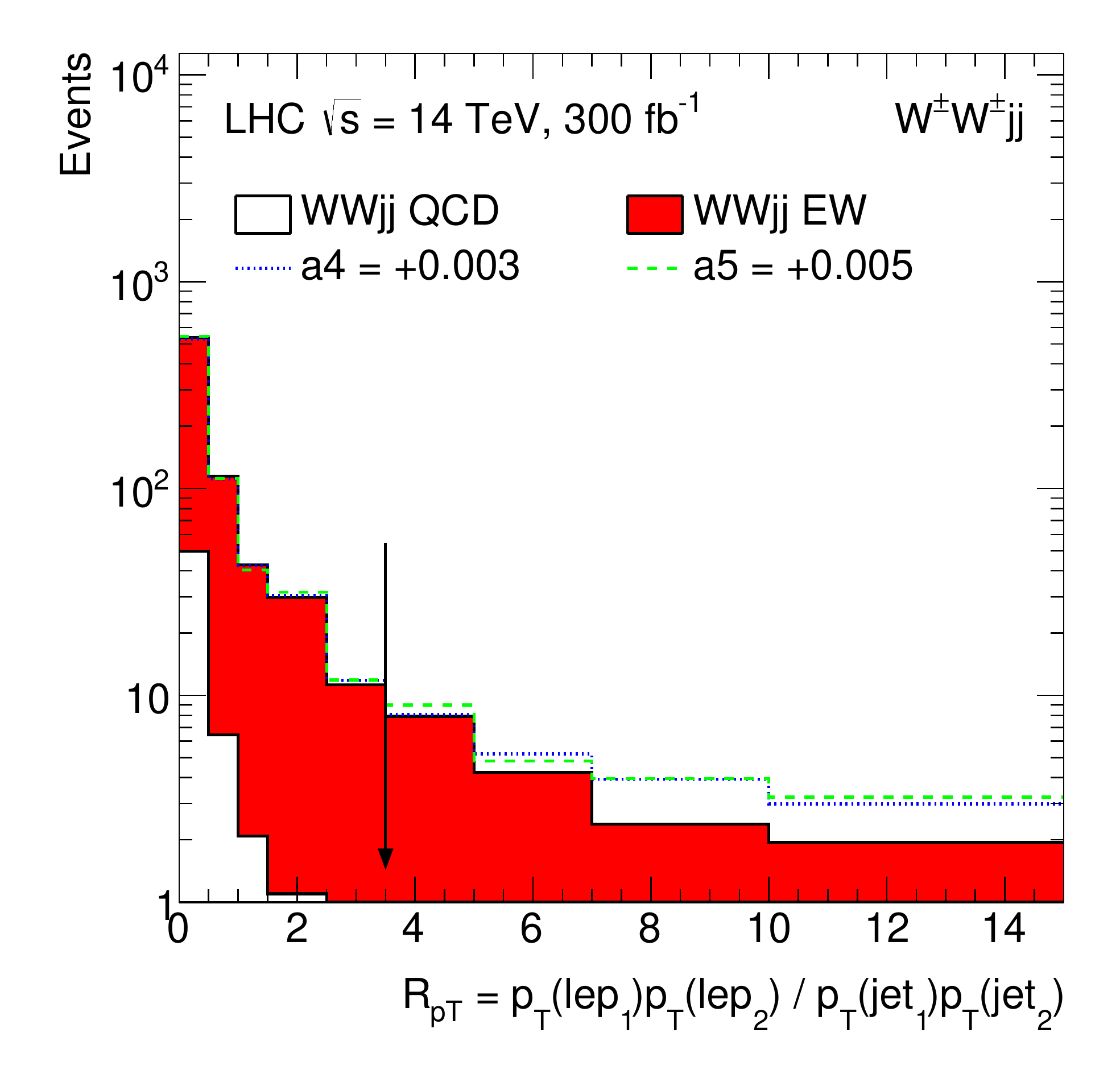}
\endminipage\hfill
\minipage{0.5\textwidth}
  \includegraphics[width=.9\linewidth]{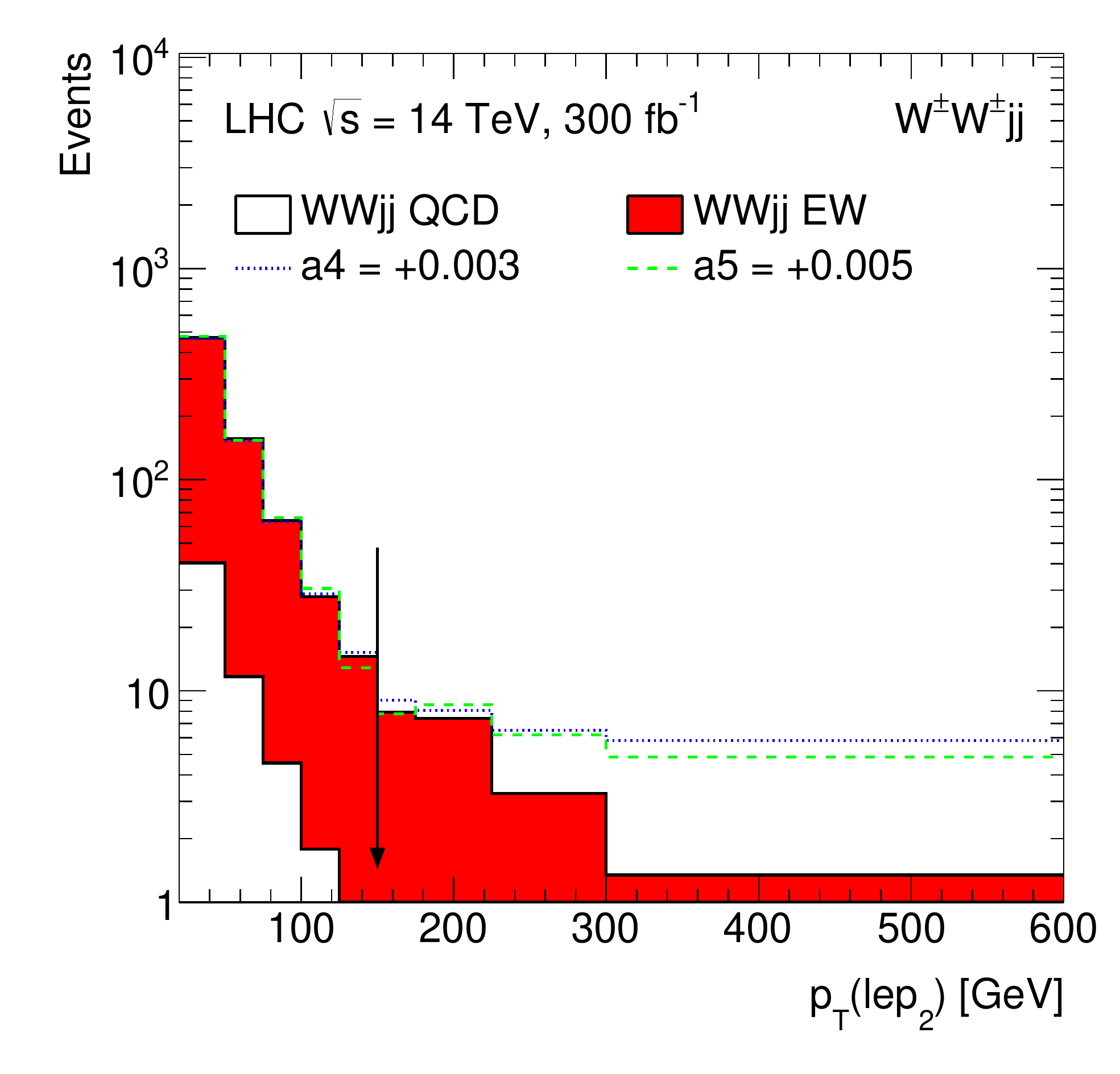}
%  \caption{A really Awesome Image}\label{fig:awesome_image2}
\endminipage\\
\caption{\small \textit{Comparison of  selection cuts: $R_{p_T}>3.5$ vs.\ $p_T^{lep}>150$ GeV. In red (white)  the EW (QCD) contribution. The dashed lines mark the number of events in the presence of non-vanishing coefficients of the effective lagrangian ($a_4=0.003$ and $a_5=0.005$)} }
\label{cuts}
\end{figure}
%%%%%%%%%%%%%%%%%%%%%%%%%% 

The cuts above only partially succeed in singling out the longitudinal $W$ bosons and a rather large pollution from the transversally polarized ones  is still present. To improve further the selection efficiency of the longitudinal modes we add the Warsaw cut~\cite{Doroba:2012pd} defined as follows 
\be\label{eq:WarsawCut}
R_{p_T} = \frac{p_T^{l_1} \; p_T^{l_2}}{p_T^{j_1} \; p_T^{j_2}}>3.5\,.
\ee
The $R_{p_T}$ variable contains the information about  the momenta of the final leptons and is very effective in separating the transverse from the longitudinal modes.

The discriminating power of this cut is illustrated in the left plot of Fig.~\ref{fig:Distributions}. The red (blue) points represent the distribution in the $[p_T^{l_1} p_T^{l_2}, p_T^{j_1} p_T^{j_2} ]$ plane of $pp \to W^{\pm}W^{\pm}jj \to l^{\pm}\nu_l l^{\pm}\nu_l\,jj$ events at the LHC ($\sqrt{s} =$ 13 TeV) containing transverse (longitudinally) polarized $WW$ pairs. By inspection we see that the cut $R_{p_T} > 3.5$ is very useful in discriminating longitudinal from transverse polarized $W$ bosons. The power of this selection is even more evident from the histogram shown in the right panel of Fig~\ref{fig:Distributions}, where the same distribution of events is plotted as a function of the ratio $R_{p_T}$.

%%%%%%%%%%%%%%%%%%%%%%%
\begin{table}[ht!]
\begin{center}
\caption{\textit{Comparison of upper exclusion limits (at 95\% and 99\% CL) and discovery significance (at $3$ and $5 \sigma$)  for the effective lagrangian coefficients $a_4$ and $a_5$, for CM energy $\sqrt{s}= 14$ TeV and luminosity 300 fb$^{-1}$, using the selection cut on $R_{p_T}$ and $p_T$. Values for both coefficients obtained by only using the same-sign  $WW$ channel.}}  
\label{comp}
\vspace{0.2cm}
%\begin{ruledtabular}
\begin{tabular}{c|c|c|c|c|}
\cline{2-5} 
 &\multicolumn{4}{|c|}{ \quad  $\sqrt{s}=14$ TeV, 300 fb$^{-1}$ \quad }  \cr
 \cline{2-5}
 &\multicolumn{2}{|c|}{\quad $R_{p_T} > 3.5$  \quad }  & \multicolumn{2}{|c|}{ \quad $p_T^{lep} > 150$ GeV\quad }  \cr
 \cline{2-5}
  \cline{2-5}
 & \qquad  95\% (99\%) \qquad  & \qquad  3$\sigma$ (5$\sigma$) \qquad  & \qquad  95\% (99\%) \qquad  & \qquad  3$\sigma$ (5$\sigma$) \qquad   \cr
\cline{2-5}
\hline
\multicolumn{1}{|c|}{\quad $a_4$ \quad} &  0.0027 (0.0034) & 0.0032 (0.0041) &  0.0031 (0.0038) & 0.0036 (0.0047)\cr
\hline
\multicolumn{1}{|c|}{\quad $a_5$ \quad} &  0.0055 (0.0068) & 0.0064 (0.0084) &  0.0063 (0.0078) & 0.0074 (0.0097)\cr
\hline
\end{tabular}
%\end{ruledtabular}
\end{center}
\end{table}
%%%%%%%%%%%%%%%%%%%%%%%%%

In \cite{Eboli:2006wa} the selection on the $W$ polarisation is carried out by means of a selection on the lepton momentum instead of the Warsaw cut. Fig.~\ref{cuts} compares the  two choices and Table~\ref{comp} shows the upper limits for  the coefficients of the effective lagrangian obtained by means of the two possible selection cuts. We find the Warsaw cut   to be better in weaning out the transverse polarizations.  In any case, the similarity in the selection choice is reflected in our final limits that turn out to be rather close to those of \cite{Eboli:2006wa} for comparable energies and luminosities.

\begin{figure}[!htb!]
 \includegraphics[width=0.32\linewidth]{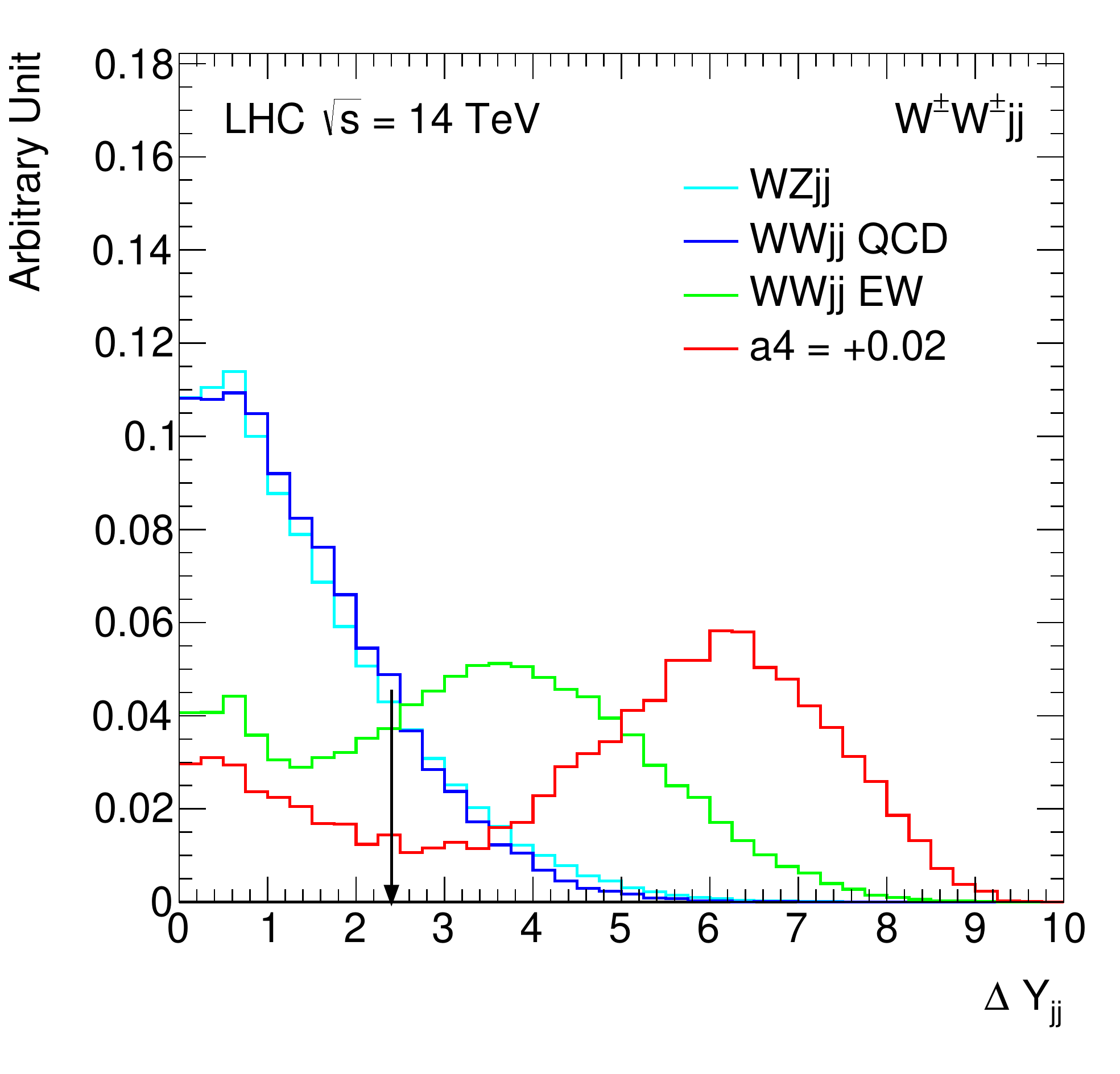}
  \includegraphics[width=0.32\linewidth]{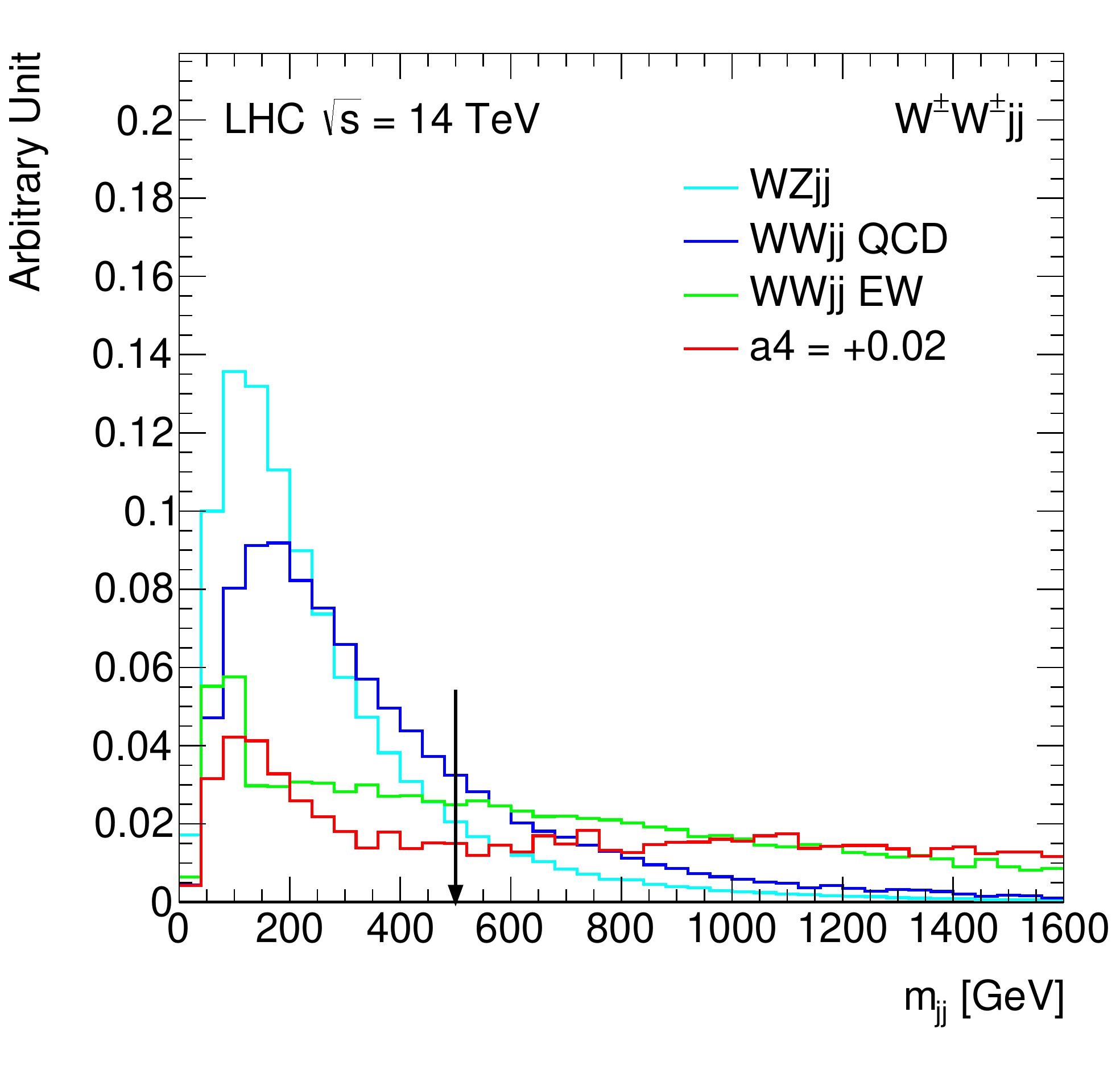}
  \includegraphics[width=0.32\linewidth]{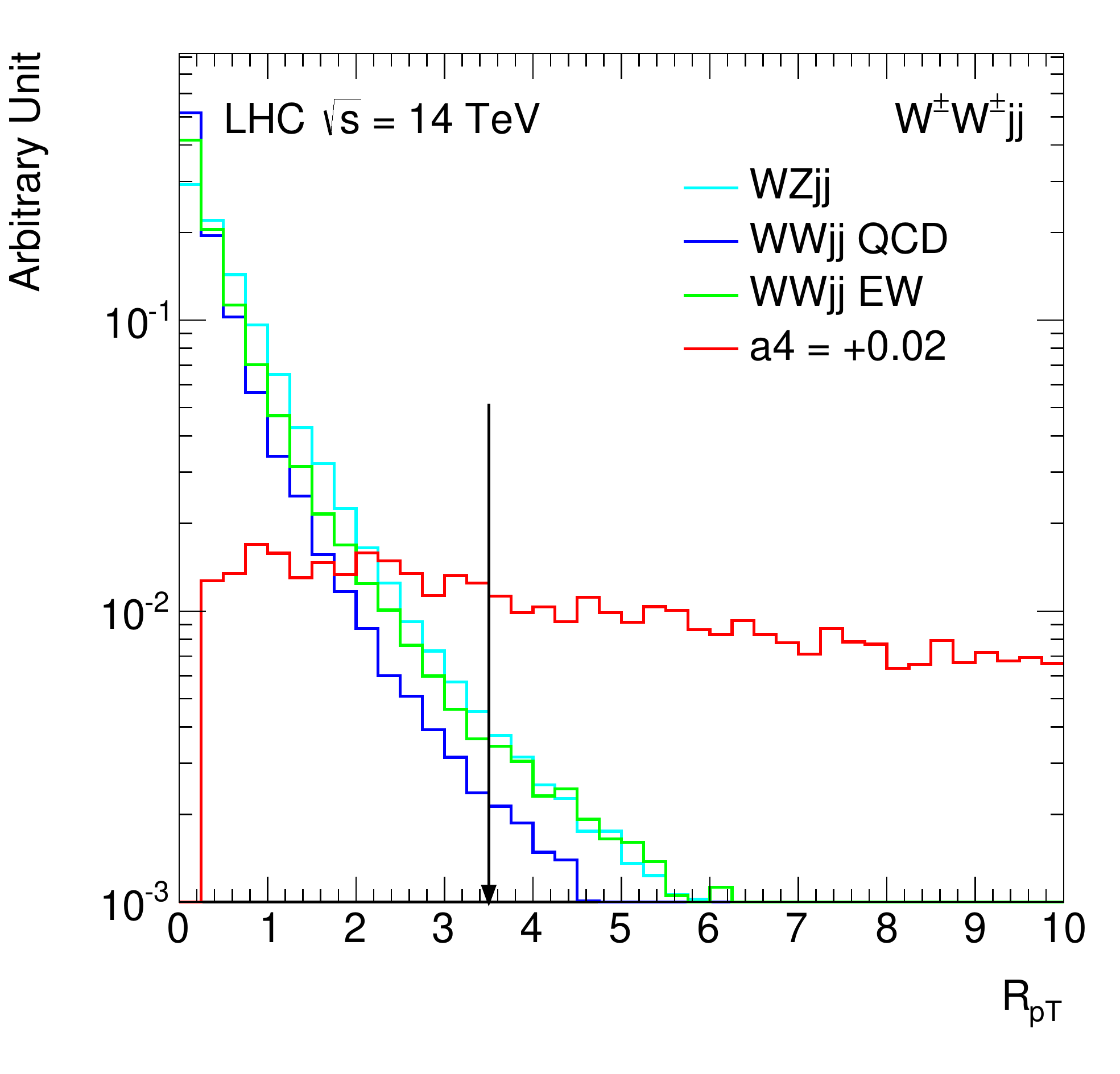}
\caption{\small \textit{Position of the cut selection for the three variables $\Delta y_{jj}$, $m_{jj}$ and $R_{p_T}$ in the SS channel.}}\label{fig:shapeSS}
\end{figure}

Table~\ref{cutflowSS} shows the effect of the various selection cuts on the number of surviving events in the SS channel. Fig.~\ref{fig:shapeSS} shows the position of the cut selection for the variables  $\Delta y_{jj}$, $m_{jj}$ and $R_{p_T}$ for this channel.

%%%%%%%%%%%%%%%%%%%%%%%
\begin{table}[ht!]
\begin{center}
\caption{\textit{Cutflow (number of events for each process cut by cut)  for the SS channel for CM energy $\sqrt{s}= 14$ TeV and luminosity 300 fb$^{-1}$.}}  
\label{cutflowSS}
\vspace{0.2cm}
%\begin{ruledtabular}
\begin{tabular}{|c|c|c|c|c|}
\hline
 \multicolumn{5}{|c|}{ \quad  $\sqrt{s}=14$ TeV, 300 fb$^{-1}$  }  \cr
 \hline
 cut  &\quad  $WZjj$ \quad &\quad  $WWjj$ QCD \quad &\quad $WWjj$ EW \quad& \quad S ($a_4=0.02$) \quad  \cr
\hline \quad 2 SS leptons  \quad&	4474	&	778	&	1343	&	1289	\cr
\hline \quad $E^{miss}_T > 25$ GeV \quad  &	3705	&	703	&	1225	&	1262	\cr
\hline \quad $\Delta y_{jj} > 2.4$  \quad &	536	&	181	&	746	&	900	\cr
\hline \quad $m_{jj} > 500$ GeV \quad  &	330	&	60	&	678	&	890	\cr
\hline \quad $R_{p_T} > 3.5$  \quad &	6.5	&	0.5	&	17	&	747	\cr
\hline								
\end{tabular}
%\end{ruledtabular}
\end{center}
\end{table}
%%%%%%%%%%%%%%%%%%%%%%%%%
%%%%%%%%%%%%%%%%%%%%%%%%%
%%%%%%%%%%%%%%%%%%%%%%%%%

%%%%%%%%%%%%%%%%%%%%%%%%%%%%%%%%
\subsubsection{Opposite-sign $WW$ channel}
%%%%%%%%%%%%%%%%%%%%%%%%%%%%%%%%

%%%%%%%%%%%%%%%%%%%%%%%
\begin{table}[ht!]
\begin{center}
\caption{\textit{Cutflow (number of events for each process cut by cut) for the OS channel for CM energy $\sqrt{s}= 14$ TeV and luminosity 300 fb$^{-1}$.}}  
\label{cutflowOS}
\vspace{0.2cm}
%\begin{ruledtabular}
\begin{tabular}{|c|c|c|c|c|}
\hline
 \multicolumn{5}{|c|}{ \quad  $\sqrt{s}=14$ TeV, 300 fb$^{-1}$  }  \cr
 \hline
 cut  & \quad $t\bar t$ \quad &\quad  $WWjj$ QCD \quad &\quad $WWjj$ EW \quad& \quad S ($a_5=0.02$) \quad  \cr
\hline
 \quad 2 OS leptons  \quad&1975270 & 68884 & 3221 & 498\cr
\hline
 \quad $E^{miss}_T > 25$ GeV \quad  & 1791100 & 61494 & 2927 & 488 \cr
 \hline
 \quad $m_{jj} > 500$ GeV \quad  & 109885 & 6761 & 1569 & 380 \cr
\hline
 \quad $\Delta y_{jj} > 2.4$  \quad & 78144 & 4543 & 1369 & 394 \cr
\hline
 \quad $R_{p_T} > 3.5$  \quad &  1461 & 114 & 44 & 287\cr
\hline
\quad$m_T^{WW} > 800$ GeV\quad & 504 & 40 & 19 & 231 \cr
\hline
\quad $\Delta \Phi_{\ell \ell} > 2.25$ \quad & 453  & 34 & 19 & 231 \cr
\hline
\quad $b$-tag veto \quad & 353 & 34 & 19 & 227 \cr
\hline
\quad $N$ jets $< 3$ \quad & 21 & 14 & 11 & 148 \cr
\hline
\end{tabular}
%\end{ruledtabular}
\end{center}
\end{table}
%%%%%%%%%%%%%%%%%%%%%%%%%

%%%%%%%%%%%%%%%%%%%%%%%%%
\begin{figure}[!htb!]
 \includegraphics[width=0.32\linewidth]{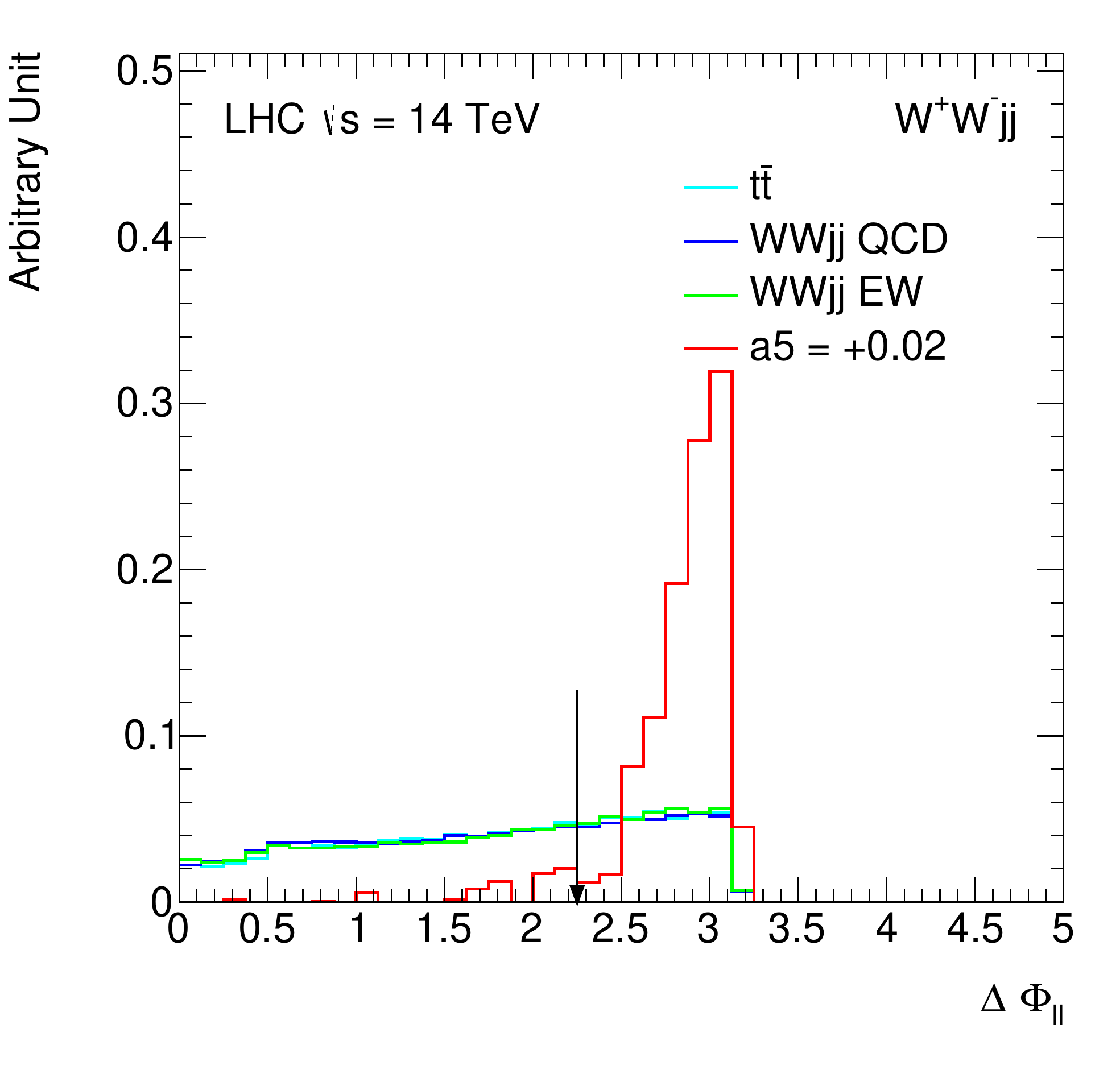}
 \includegraphics[width=0.32\linewidth]{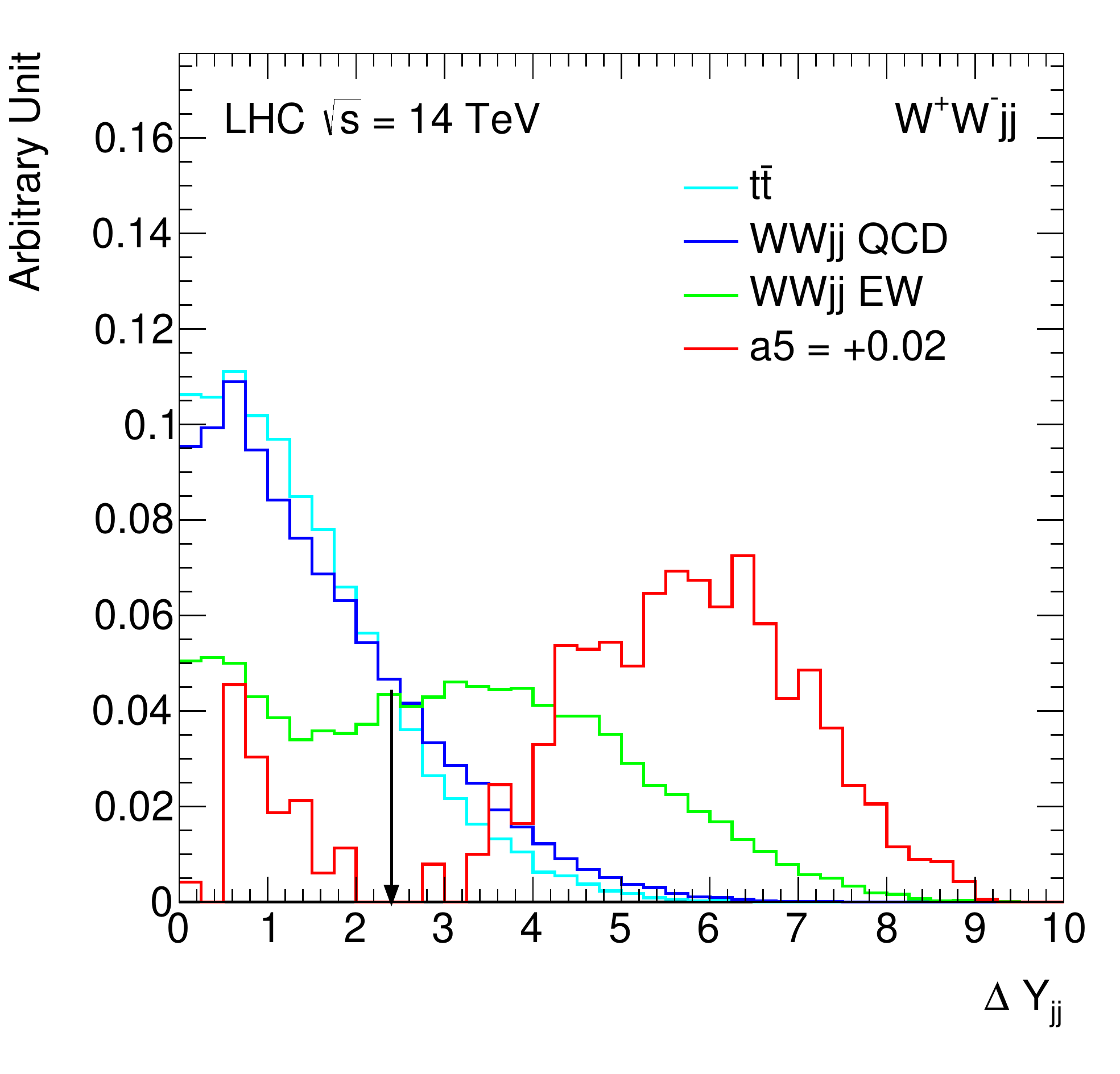}
  \includegraphics[width=0.32\linewidth]{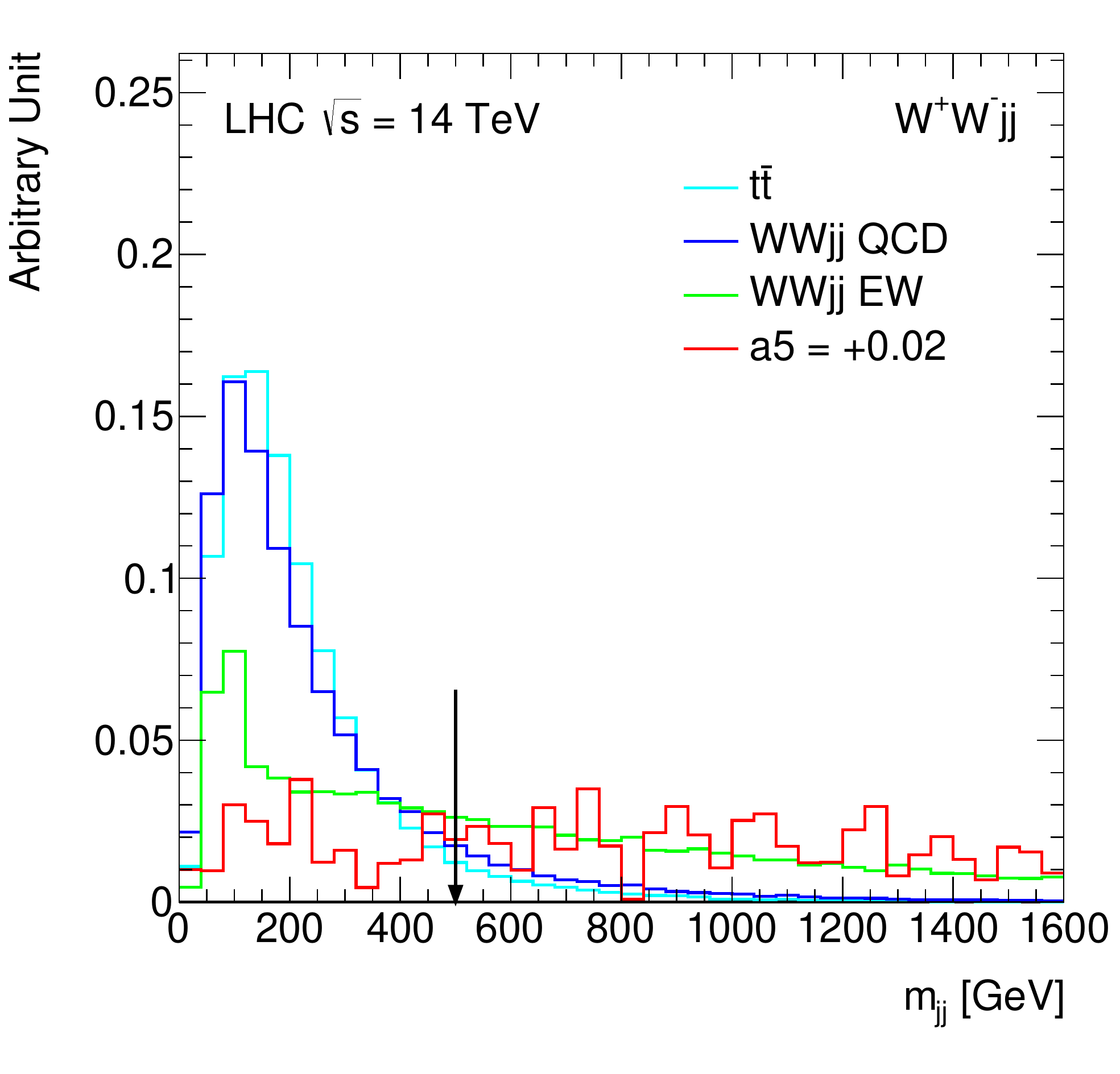}
   \includegraphics[width=0.32\linewidth]{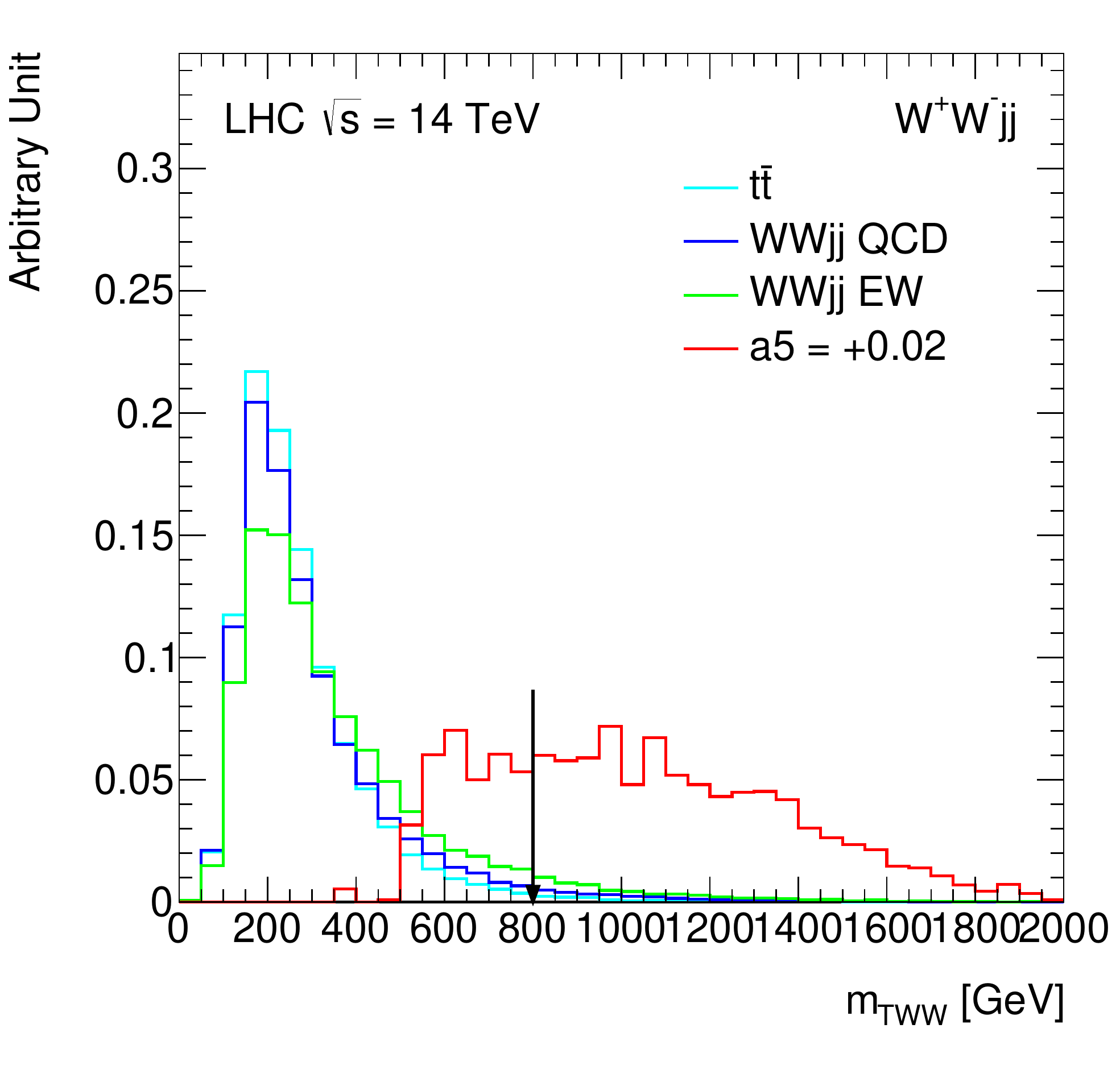}
  \includegraphics[width=0.32\linewidth]{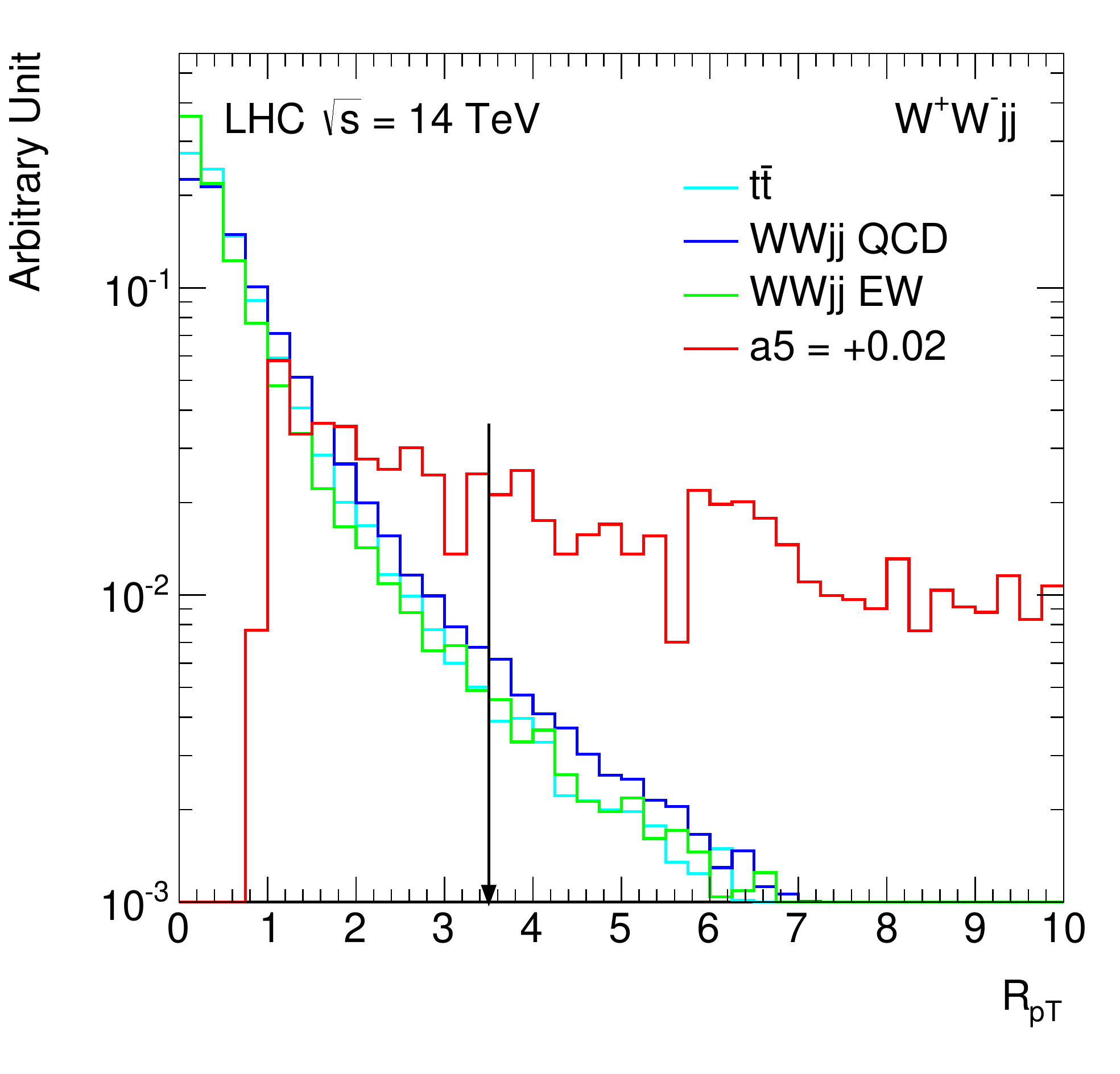}
\caption{\small \textit{Position of the cut selection for the three variables $\Delta \Phi_{\ell\ell}$, $\Delta y_{jj}$, $m_{jj}$, $m^{WW}_T$ and $R_{p_T}$ in the OS channel.}}\label{fig:shapeOS}
\end{figure}
%%%%%%%%%%%%%%%%%%%%%%%%%

The opposite-sign decay channel is less clean because of the large reducible background coming from $t\bar t$ pair production. For this channel in the process $pp\rightarrow l^\pm \nu_{ l} l^\mp  \nu_{ l} j j$ we use the following selection cuts:
\begin{itemize}
\item two opposite-sign leptons with $p_T^{l^\pm}>20$ GeV and $|\eta_{l^\pm}| <  2.5$;
\item missing transverse energy $E_T^{\, miss} >$ 25 GeV\;
\item the two highest $p_T$ jets with an invariant mass $m_{jj} > 500$ GeV;
\item two and only two jets ($p_T^j>25$ GeV and  $|\eta_j|<4.5$) with relative rapidity $|\Delta y_{jj}| > 2.4$;
\item   $R_{p_T} > 3.5$;
\item invariant transverse mass $m^{WW}_T  > 800$ GeV;
\item angular separation between the leptons in the transverse plane $|\Delta\Phi_{ll}|>2.25$;
\item $b$-quark veto ({\it i.e.} no jets tagged by the $b$-tagging algorithm implemented in {\tt Delphes}).
\end{itemize}

The invariant tranverse mass in the cuts above is defined as 
\be
m^{WW}_T = \sqrt{\left( \sqrt{ ( p_T^{ll})^2 + m_{ll}^2 } + \sqrt{ (E_T^{miss})^2 + m_{ll}^2 } \right)^2 - ( \vec p_T^{\;ll} + \vec p_T^{\;miss} )^2}\, ,
\ee
where $\vec p_T^{\;miss}$ is the missing transverse momentum vector, 
$\vec p_T^{\;ll}$ is the transverse momentum of the di-lepton pair and $m_{ll}$ its mass.

Table~\ref{cutflowOS} shows the effect of the various selection cuts on the number of surviving events in OS channel. Fig.~\ref{fig:shapeOS} shows the position of the cut selection for the variables  $\Delta \Phi_{\ell\ell}$, $\Delta y_{jj}$, $m_{jj}$, $m_{\ell\ell}$ and $R_{p_T}$  for this channel.

%%%%%%%%%%%%%%%%%%%%%%%%%%%%%%%%
\subsection{Statistical analysis}\label{sec:Statistics}
%%%%%%%%%%%%%%%%%%%%%%%%%%%%%%%%
In the following we will compute the expected discovery significance and the expected exclusion limits for the coefficients of the effective lagrangian in \eq{L0} and \eq{L1}. 

For a given set of selection cuts, we define the signal $S$ as the enhancement in the number of $WWjj$ events---obtained for certain fixed values of the coefficients $a$, $a_2$, $a_3$, $a_4$ and $a_5$---over the SM prediction (obtained for $a=1$, $a_2=a_3=a_4=a_5=0$)
\be 
S={\cal N}_{\mbox{\tiny ev}}(pp\to WWjj)\Big|_{a, a_2, a_3, a_4, a_5}-{\cal N}_{\mbox{\tiny ev}}(pp\to WWjj)\Big|_{a=1, a_2=a_3=a_4=a_5=0}\,.
\ee
The background $B$ is given by the number of events predicted by the SM 
\be 
B={\cal N}_{\mbox{\tiny ev}}(pp\to WWjj)\Big|_{a=1, a_2=a_3=a_4=a_5=0} + {\cal N}_{\mbox{\tiny ev}}(pp\to t\bar t/ WZjj)\Big|_{a=1, a_2=a_3=a_4=a_5=0} \,.
\ee

The expected number of signal events $S$ is compared with the number of background events $B$ using Poisson statistics without considering any systematic uncertainty. The Poisson probability density function is generalized to non-integer event numbers through the use of the Gamma function.  

%%%%%%%%%%%%%%%%%%%%%55
\subsubsection{Discovery significance and exclusione limits}
%%%%%%%%%%%%%%%%%%%%%%%%%
\begin{figure}[!htb!]
\minipage{0.5\textwidth}
  \includegraphics[width=.9\linewidth]{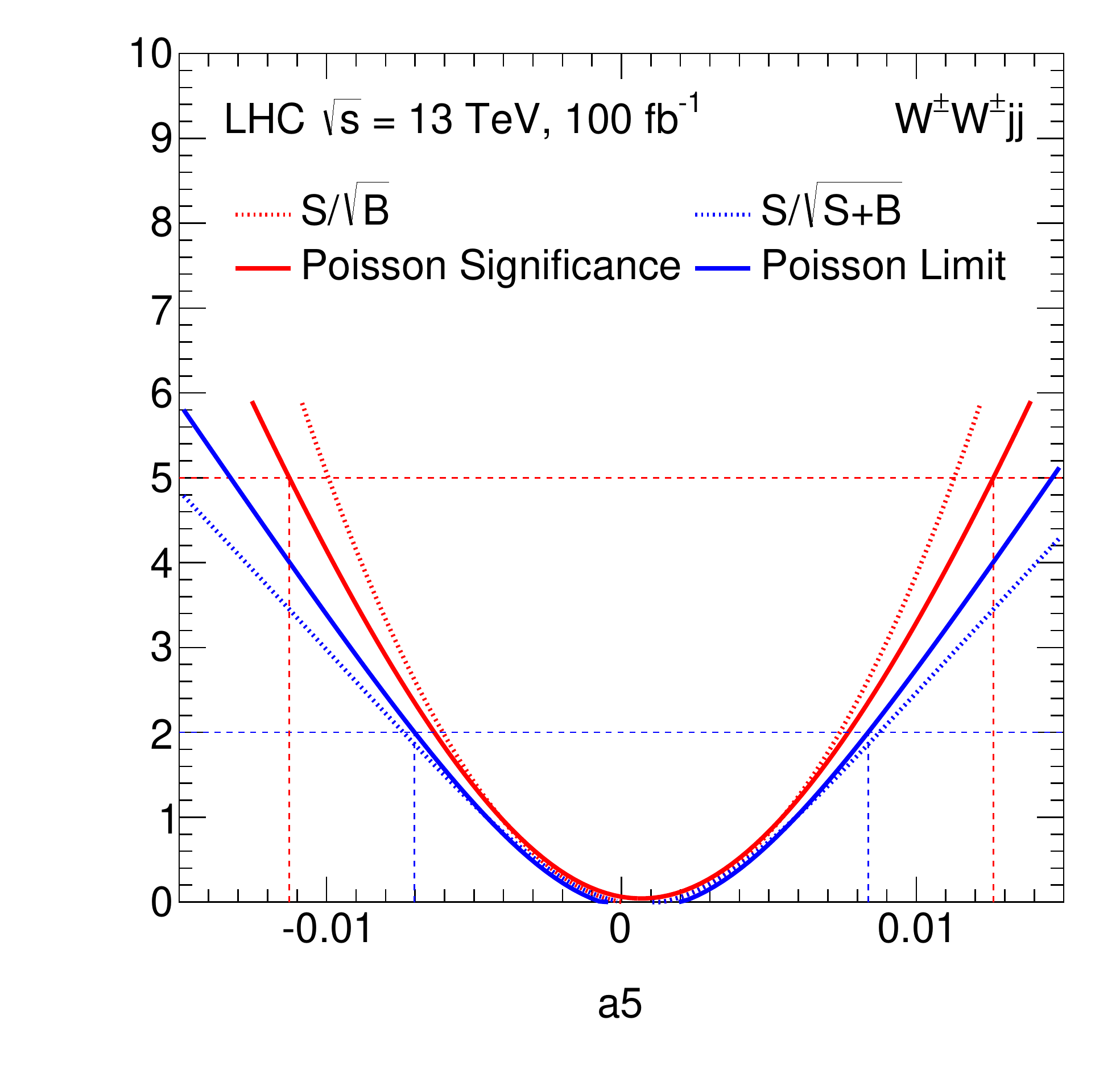}
\endminipage\hfill
\minipage{0.5\textwidth}
  \includegraphics[width=.9\linewidth]{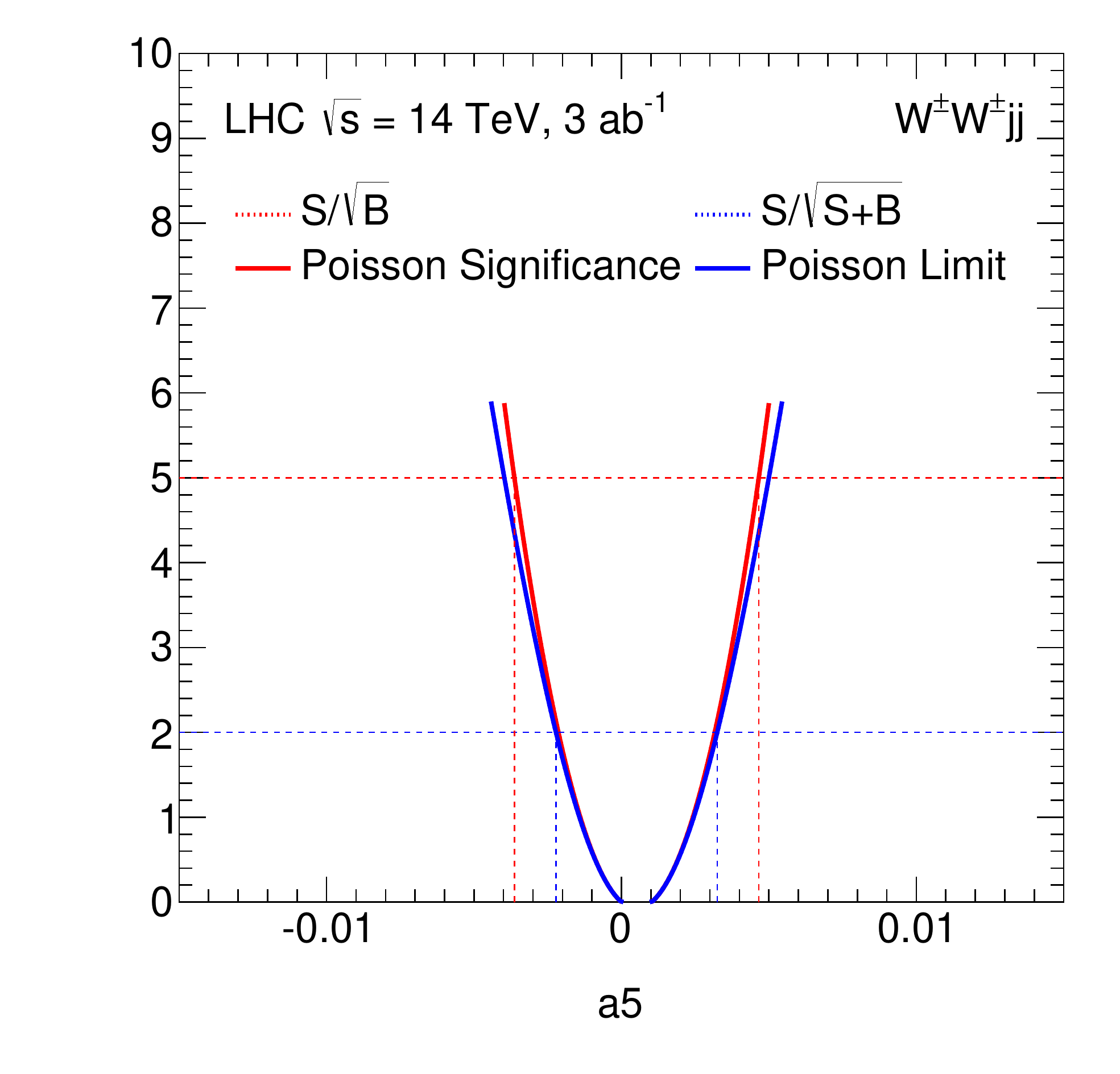}
\endminipage\\
\caption{\small \textit{$\Delta \chi^2$ plot for the coefficient $a_5$ using the Poisson distribution and the simplified formulas. The plot on the right shows the presence of a discrepancy at
low luminosity ($\sqrt{s}=13$ TeV, luminosity 100 fb$^{-1}$). The plot on the right shows that there is no discrepancy at higher luminosity ($\sqrt{s}=14$ TeV, luminosity 3 ab$^{-1}$) where it is impossible to discriminate the continuous from  the dashed lines.}}\label{fig:plot1D}
\end{figure}

For each set of values of the effective couplings, the expected discovery significance is obtained by computing the probability of observing a number of events greater or equal to $S+B$ assuming the background-only hypothesis. This probability is then translated into a number of Gaussian standard deviations: three (five) standard deviations is considered as benchmark for an observation (discovery).
On the other hand, the expected exclusion limits are obtained by computing the probability of observing a number of events less or equal to $B$ assuming the signal-plus-background hypothesis. The specific choice of the parameters is considered excluded at 95\% (99\%) CL if this probability is less or equal than 5\% (1\%).

Notice that, for large values of $B$, the Poisson distribution can be very well approximated by a Gaussian function. In this case the significance (expressed in terms of number of standard deviations) can be computed simply as $S/\sqrt{B}$. In the same limit we can say that a set of parameters is excluded at 95\% (99\%) CL if the quantity $S/\sqrt{S+B}>2$ ($>3$).

The difference between using the exact Poisson distribution and the approximated formulas above can be gauged in Fig.~\ref{fig:plot1D} 
where the $\chi^2$ test is run for the two possibilities. 
As one can see by inspection, while for the case at $\sqrt{s}$ = 13 TeV and luminosity 100 fb$^{-1}$ the difference cannot be ignored, 
there is no difference for the higher energy and luminosity case. We  employ in all cases the Poisson probability distribution.

%%%%%%%%%%%%5
\subsubsection{Systematic uncertainties}
%%%%%%%%%%%%%
\begin{figure}[!htb!]
\minipage{0.5\textwidth}
  \includegraphics[width=.9\linewidth]{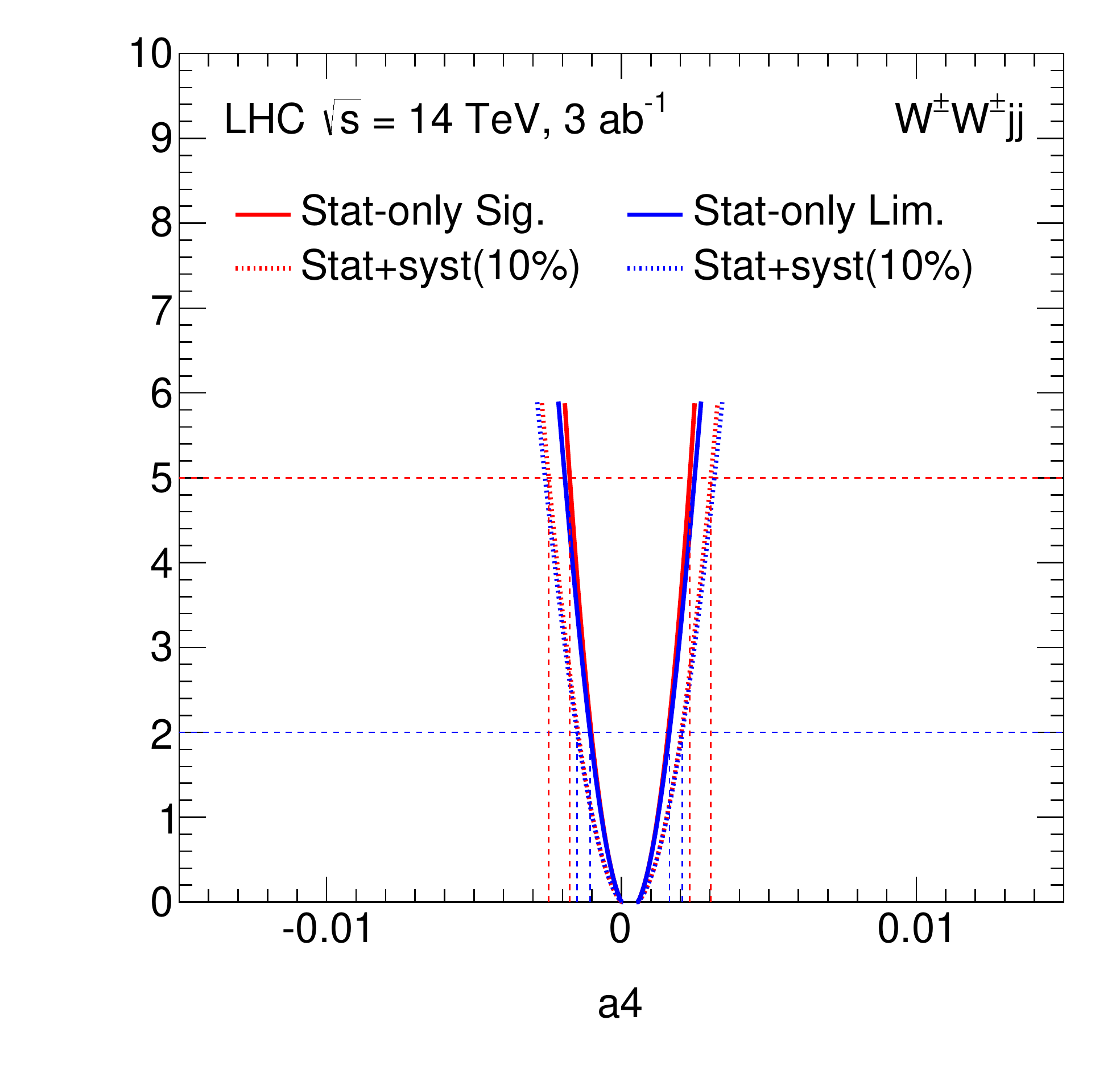}
\endminipage\hfill
\minipage{0.5\textwidth}
  \includegraphics[width=.9\linewidth]{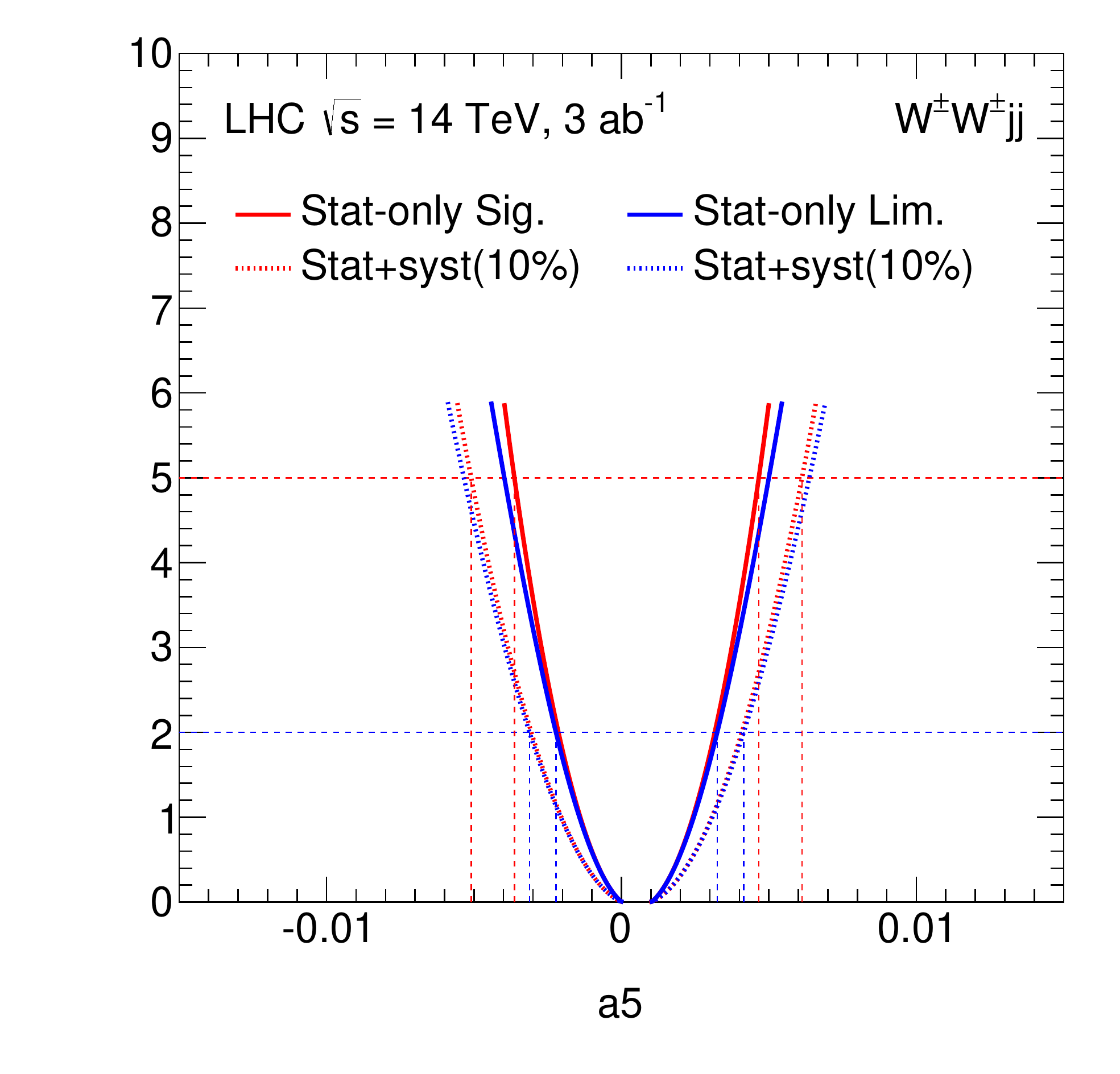}
\endminipage\\
\caption{\small \textit{$\Delta \chi^2$ plot for the coefficient $a_4$ (left) and $a_5$ (right)  with and without 10\% of systematic uncertainty for CM energy
$\sqrt{s}=14$ TeV and luminosity 3 ab$^{-1}$.}}\label{fig:plot1D_syst14}
\end{figure}

\begin{figure}[!htb!]
\minipage{0.5\textwidth}
  \includegraphics[width=.9\linewidth]{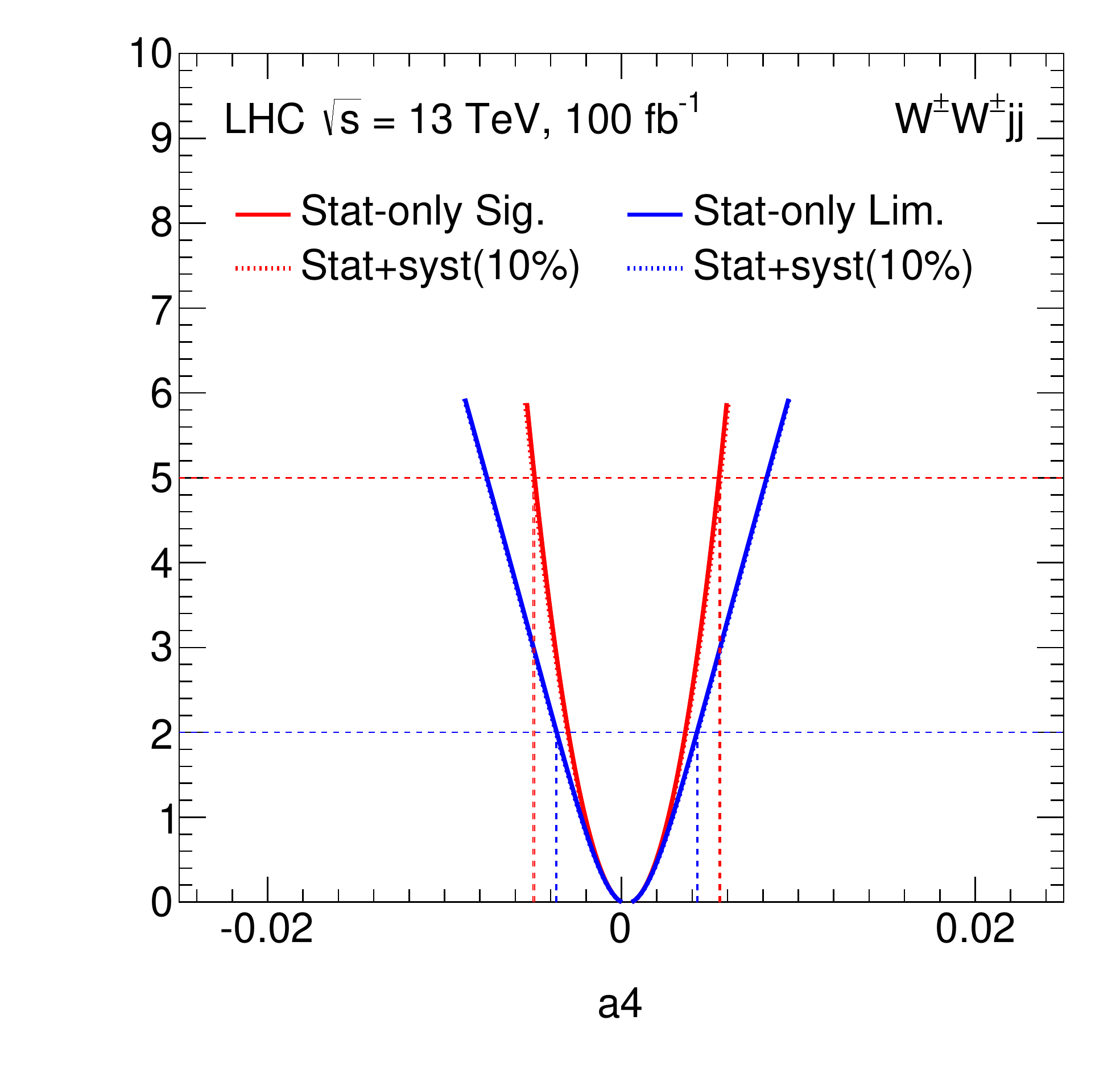}
\endminipage\hfill
\minipage{0.5\textwidth}
  \includegraphics[width=.9\linewidth]{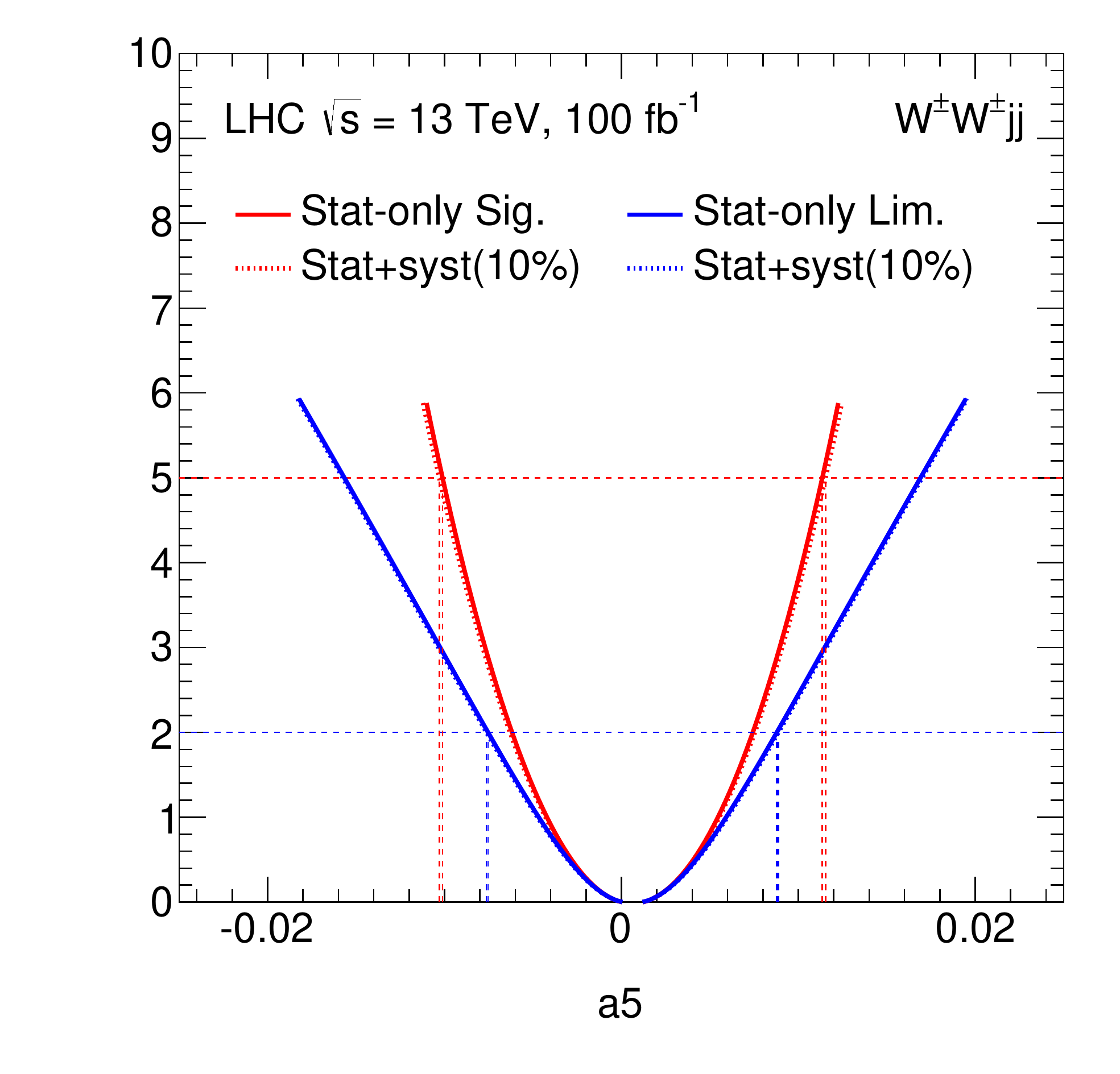}
\endminipage\\
\caption{\small \textit{$\Delta \chi^2$ plot for the coefficient $a_4$ (left) and $a_5$ (right)  with and without 10\% of systematic uncertainty for CM energy
$\sqrt{s}=13$ TeV and luminosity 100 fb$^{-1}$.}}\label{fig:plot1D_syst13}
\end{figure}

%%%%%%%%%%%%%%%%%%%%%%%
%   SYSTEMATICS       %
%%%%%%%%%%%%%%%%%%%%%%%
\begin{table}[ht!]
\begin{center}
\caption{\textit{Upper limits (at  95\% CL)   for the effective lagrangian coefficients $a_4$ and $a_5$, for two  representative CM energies and  luminosities, from the channel $W^{\pm}W^{\pm}jj$.  Comparison with and without the inclusion of a systematic error of 10\%. }}
\label{irr_bckg}
\vspace{0.2cm}
%\begin{ruledtabular}
\begin{tabular}{c|c|c|c|c|}
\cline{2-5} 
 &\multicolumn{2}{|c|}{\quad  $\sqrt{s}=13$ TeV, 100 fb$^{-1}$   \quad }  & \multicolumn{2}{|c|}{ \quad  $\sqrt{s}=14$ TeV, 3 ab$^{-1}$\quad }  \cr
 \cline{2-5}
  \cline{2-5}
 & \qquad  w/out \qquad  & \qquad  w/  \qquad  & \qquad  w/out \qquad  & \qquad w/ \qquad   \cr
\cline{2-5}
\hline
\multicolumn{1}{|c|}{\quad $a_4$ \quad} & 0.0043   & 0.0043  & 0.0016 & 0.0021 \cr
\hline
\multicolumn{1}{|c|}{\quad $a_5$ \quad }& 0.0088   & 0.0089  & 0.0032 & 0.0041 \cr
\hline
\end{tabular}
%\end{ruledtabular}
\end{center}
\end{table}
%%%%%%%%%%%%%%%%%%%%%%%%%

All the results reported in the following are obtained neglecting any systematic uncertainty on the prediction for the number of signal and background events ($S$ and $B$)
because such uncertainties are  mostly related to the experimental techniques used to extract the results. 
To have a feeling of the size of their effect on the results, 
we have included a non-zero systematic uncertainty on $B$ and compared the limits and the significance with the case without systematics.
This comparison is done considering the simplified statistics treatment described above---that is, 
by considering the formulas $S/\sqrt{B}$ and $S/\sqrt{S+B}$.
These two expressions are generalised to the case with non-zero systematic uncertainty  as $S/\sqrt{ B + \delta^2 \cdot B^2}$ and $S/\sqrt{ S + B + \delta^2 \cdot B^2}$ respectively, 
where $\delta$ indicate the relative systematic uncertainty on the expected number of background events $B$.

Table~\ref{irr_bckg} (and the corresponding  plots  in Figs.~\ref{fig:plot1D_syst14}  and ~\ref{fig:plot1D_syst13}) show the result of this comparison, performed considering two   benchmark CM energy and luminosity scenarios for the two coefficients $a_4$ and $a_5$
and a relative systematic uncertainty on $B$ of 10\%. The  smaller statistical error in the case of CM energy $\sqrt{s}=14$ TeV and luminosity 3 ab$^{-1}$ makes the systematic error---assumed to remain the same---more important. 

As expected, the effect is rather important, especially for large values of integrated luminosity where the Gaussian error is smaller, 
and one should bear that in mind. 
Of course, an eventual reduction of such a systematic uncertainty, for instance down at 5\%, would  proportionally reduce the effect, 
and, depending on the size of this uncertainty in a real experiment, 
selection cuts could be further tightened to minimise its impact.

%%%%%%%%%%%%%%%%%%%%%%%%%%%%%%%%
\subsection{Unitarization} \label{sec:Unitarity}
%%%%%%%%%%%%%%%%%%%%%%%%%%%%%%%% 

%%%%%%%%%%%%%%%%%%%
\begin{figure}[h!]
\begin{center}
\includegraphics[width=4in]{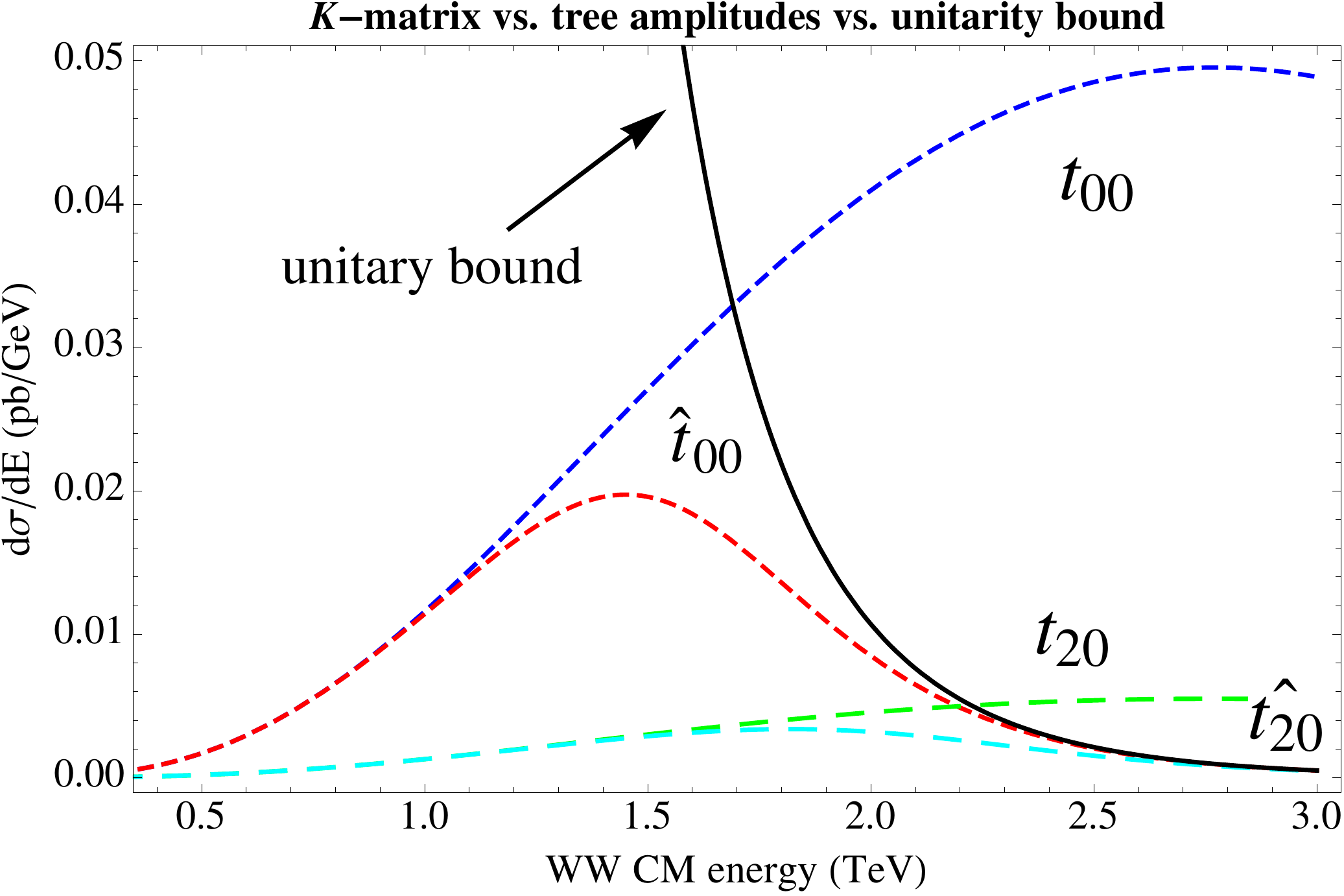}
\caption{\small  \textit{Cross sections for the scattering of longitudinal $W$ bosons as a function of the CM energy. In green (blue) the contribution of the partial wave  $t_{20}$ ($t_{00}$) for  $a_4=a_5=0.001$; in  cyan (red) the same result  after unitarization by $K$-matrix. The continuos black line marks the loss of unitarity.
\label{fig_unitarity}}}
\end{center}
\end{figure}
%%%%%%%%%%%%%%%%%%%%%%%%%% 

For values of the coefficients $a$, $a_2$, $a_3$, $a_4$ and $a_5$ which are different from the SM ones, the computation of the cross section $\sigma(p p \to WWjj)$ obtained using the lagrangian in \eq{L0} and \eq{L1} cannot be trusted because of possible unitarity violation that can arise at the level of some hard scattering diagrams, in particular the ones that involve longitudinal $W$ bosons.
In this case, the cross section of the process  $W_LW_L\to W_LW_L$ breaks unitarity at energies larger than the TeV (the exact violation energy depends on the specific values of the coefficients). 

This breakdown in unitarity can be understood by looking at the longitudinal $W$ bosons scattering amplitudes in the same- and opposite-sign channel---computed using the equivalence theorem in the isospin limit---can be written in terms of isospin amplitudes $A_I(s,t)$ as 
\be
A(W_L^\pm W_L^\pm \rightarrow W_L^\pm W_L^\pm) = A_2\, \quad \mbox{and} \quad A(W_L^\pm W_L^\mp \rightarrow W_L^\pm W_L^\mp) = \frac{1}{3} A_0 +  \frac{1}{2} A_1 +\frac{1}{6} A_2 \, .
\ee
The amplitudes $A_I(s,t)$ can be expanded in terms of partial waves $t_{IJ}(s)$ as 
\be
A_I(s,t) = 32 \pi  \sum^\infty_{J=0} (2J+1)  P_J (\cos\theta)  t_{IJ} (s) \, , \label{transform}
\ee
where
\be 
t_{IJ} (s)= \frac{1}{64\pi}\int_{-1}^{1}d\cos\theta\, A_I(s,t)  P_J (\cos\theta)\,.
\ee
In our case, at tree level, neglecting partial waves higher than the leading $J=0$ wave, we have 
\bea
t_{00} &=&  \frac{s}{16\,  \pi v^2}  (1-a^2  + 3 g^{\prime 2} a_2 + 12 g^2 a_3) + \frac{ s^2}{12 \, \pi v^4} \left[ 11\, a_5 + 7\, a_4 - 2 g^{\prime 2}a_2^2  + 16 g^2 a_3^2\right] \\
 t_{20} & = &  -\frac{s}{32 \, \pi v^2}  (1-a^2 - 6 g^{\prime 2}a_2  +12  g^2 a_3)  + \frac{ s^2}{6 \,\pi v^4} \left[  a_5 +2\,  a_4 - g^{\prime 2}a_2^2   -4  g^2 a_3^2 \right]  \, . 
\eea
The isospin amplitudes $A_I(s,t)$ can then be re-obtained from the partial waves computed above by means of \eq{transform}. In the approximation of neglecting partial waves higher than $J=0$, we have very simple relations:
\be
A_0(s,t) = 32 \pi\,  t_{00} \, ,\quad A_1(s,t) =0\quad \mbox{and} \quad A_2(s,t) = 32 \pi \,  t_{20} \, . \label{amps}
\ee
An example of such unitarity violation is shown in Fig.~\ref{fig_unitarity} where---for values of $a_4=a_5=0.001$---it occurs around 1.5 and 2 TeV  for, respectively, the isospin $I=0$ and $I=2$ component.

The amplitudes in \eq{amps} violate unitarity and we interpret them as an
incomplete approximation to the true amplitudes. One can deal with this problem either by  cutting off the collection of events at a given value of the CM energy  or by implementing an unitarization procedure. 

As an example of the latter, let us look for unitary matrix elements that provides a non-perturbative completion. By inspection of the amplitudes we see that the SS $WW$ channel can only contain double-charged $I=2$ resonances in the $s$-channel, the first two being of spin 0 and 2. We assume that these states are sufficiently heavy to be outside the energy reach of LHC. By extension,  we assume that no resonance is present within the LHC energy range also in the opposite-sign $WW$ channel. Therefore, the most appropriated unitarization procedure for our case in which we do not expect resonances is the $K$-matrix prescription~\cite{Gupta:1949rh}. The $K$-matrix ansatz consists in using the optical theorem
\be
\Im t_{IJ} (s) = | t_{IJ}(s)|^2
\ee
in order to impose the following condition on the unitarized partial wave $\hat t_{IJ}(s)$
\be
\Im \frac{1}{\hat t_{IJ}(s)} = -1 \, .
\ee
The $K$-matrix unitarized partial wave is then defined to be 
\be
\hat t_{IJ}(s) = \frac{t_{IJ}(s)}{1 - i t_{IJ}(s)}\, , \label{k}
\ee
where $t_{IJ}(s)$ is the tree-level partial wave amplitude. The quantity $\hat t_{IJ}(s,t)$ satisfies by construction the optical theorem
and is supposed to represent a re-summation of the higher order terms whose contribution restore unitarity. The result of this unitarization is shown in Fig.~\ref{fig_unitarity} and compared to the tree-level result.

If we define the  rescaling factor for the  SS $WW$ events
\be 
r_{++}(s,a_3,a_4,a_5)=\frac{|\hat t_{20}|^2}{|t_{20}|^2}\, ,
\ee
we can use it to re-weight the events that survive after having applied all the selection cuts, in order to obtain a result that satisfies the unitarity bound.  This procedure is reliable if the events that survive after the selection cuts are dominated by the production of longitudinal polarized $W$.

%%%%%%%%%%%%%%%%%%%%%%%
\begin{table}[ht!]
\begin{center}
\caption{\textit{Comparison of upper exclusion limits (at 95\% and 99\% CL) and discovery significance (at $3$ and $5 \sigma$)  for the effective lagrangian coefficients $a_4$ and $a_5$, for CM energy $\sqrt{s}= 14$ TeV and luminosity 300 fb$^{-1}$, using the $K$-matrix and the sharp cut off unitarization procedures. Values  obtained by  using the SS  $WW$ channel.}}  
\label{comp}
\vspace{0.2cm}
%\begin{ruledtabular}
\begin{tabular}{c|c|c|c|c|}
\cline{2-5} 
 &\multicolumn{4}{|c|}{ \quad  $\sqrt{s}=14$ TeV, 300 fb$^{-1}$ \quad }  \cr
 \cline{2-5}
 &\multicolumn{2}{|c|}{\quad$K$-matrix  \quad }  & \multicolumn{2}{|c|}{ \quad sharp cut off ($E_{WW}<2$ TeV) \quad }  \cr
 \cline{2-5}
  \cline{2-5}
 & \qquad  95\% (99\%) \qquad  & \qquad  3$\sigma$ (5$\sigma$) \qquad  & \qquad  95\% (99\%) \qquad  & \qquad  3$\sigma$ (5$\sigma$) \qquad   \cr
\cline{2-5}				
\hline \multicolumn{1}{|c|}{\quad $a_4$ \quad}	&  0.0028 (0.0038) 	& 0.0035 (0.0053)	& 0.0027 (0.0034)	& 0.0032 (0.0041)	\cr
\hline \multicolumn{1}{|c|}{\quad $a_5$ \quad}	&  0.0053 (0.0072) 	& 0.0066 (0.0107)	& 0.0055 (0.0068)	& 0.0064 (0.0084)	\cr
%\hline					
%\hline \multicolumn{1}{|c|}{\quad $a_3$ \quad}	& -0.052 / 0.102 (-0.07 / 0.139)	& -0.065 / 0.127 (-0.095 / 0.206)	& -0.05 / 0.073 (-0.065 / 0.089)	& -0.061 / 0.084 (-0.084 / 0.108)	\cr
%\hline \multicolumn{1}{|c|}{\quad $a$ \quad}	& 0.64 / 1.37 (0.48 / 1.43)	& 0.53 / 1.41 (0.22 / 1.5)	& 0.64 / 1.37 (0.47 / 1.43)	& 0.52 / 1.42 (0.22 / 1.5)	\cr
\hline					
\end{tabular}
%\end{ruledtabular}
\end{center}
\end{table}
%%%%%%%%%%%%%%%%%%%%%%%%%

The $K$-matrix {\em ansatz} and the cut off in energy are two possible procedures to deal with the violation of unitarity. Table~\ref{comp} shows that the two procedures (for an appropriate choice of cut off) are substantially equivalent. Their differences quantify the dependence on  the unitarization procedure of the limits.

Because it is more difficult to define a rescaling for the OS channel as done above for the SS channel, and because of the additional assumptions entering the $K$-matrix procedure, we  follow the simplest procedure and introduce a sharp cut off in the data collection so as to make the amplitudes unitarity. 

The cut off must be chosen to be less than $4 \pi v$, the limit for the chiral lagrangian expansion, and below the range in which the  growth becames too fast. We take $m_{WW} < 1.25$ TeV for the SS channel and $< 2$ TeV for the OS channel. It can be shown that for these values, as in Table~\ref{comp}, differences between the two unitarization procedures are minimal.

%%%%%%%%%%%%%%%%%%%%%%%%%%%%%%%%
\section{Results} 
\label{sec:r}
%%%%%%%%%%%%%%%%%%%%%%%%%%%%%%%%

As discussed in section~\ref{sec:mc}, we have generated events in which the coefficients of the effective lagrangians in \eq{L0} and \eq{L1} of section~\ref{sec:Notation}, parameterising deviations from the SM, were allowed to vary. We consider  only the coefficients $a$, $a_2$, $a_3$, $a_4$ and $a_5$  because the coefficient $a_1$ is already severely constrained by LEP data, as discussed in section~\ref{sec:Bounds},  and we assume it vanishing  in our analysis. The coefficients $a_4$ and $a_5$,  according to our discussion in section~\ref{sec:coef}, are the leading and most important ones.  They should be searched first. Once they have been constrained, the simulation for the coefficients $a_2$, $a_3$ and  $a$ can be carried out after setting $a_4$ and $a_5$ equal to zero.

We report in Tables \ref{limits13SS}-\ref{limits14OS} the results in terms of exclusion limits (95 and 99\% CL)  and discovery significance  (3 and $5\sigma$)---as discussed in section \ref{sec:Statistics}---for the benchmark luminosities of 100 and 300 fb$^{-1}$  (at CM energy of  $\sqrt{s}=$  13 TeV) and 300 fb$^{-1}$ and 3 ab$^{-1}$ (at   $\sqrt{s}=$  14 TeV). All coefficients are here varied one at the time.

As it can be seen from  Tables \ref{limits13SS}-\ref{limits14OS}, the OS channel does not provide stronger limits for any of the coefficients and the SS channel is sufficient by itself in setting the most stringent constraints. 
 
\begin{figure}[ht!]
\begin{center}
\includegraphics[width=4.5in]{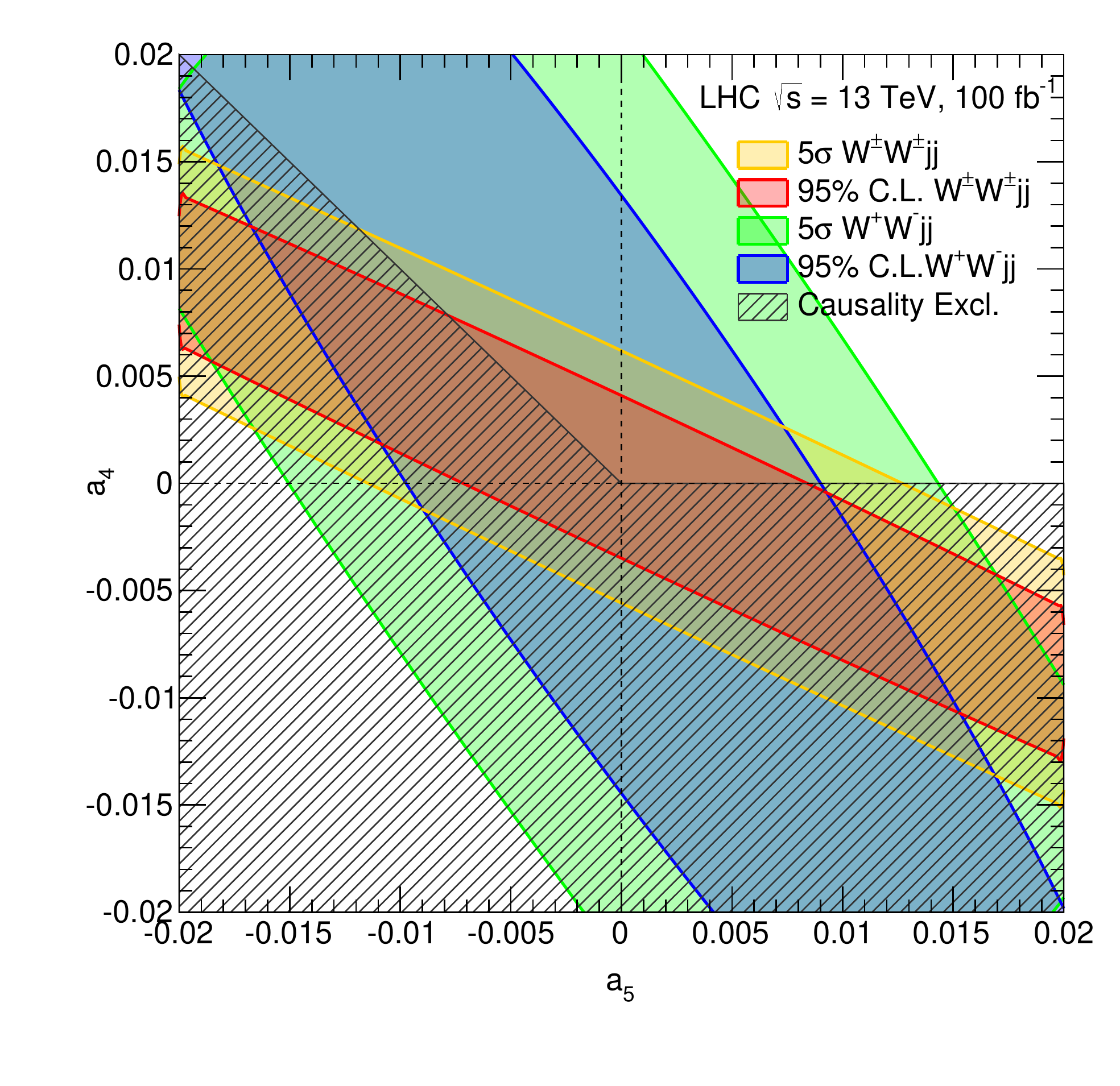}
\includegraphics[width=4.5in]{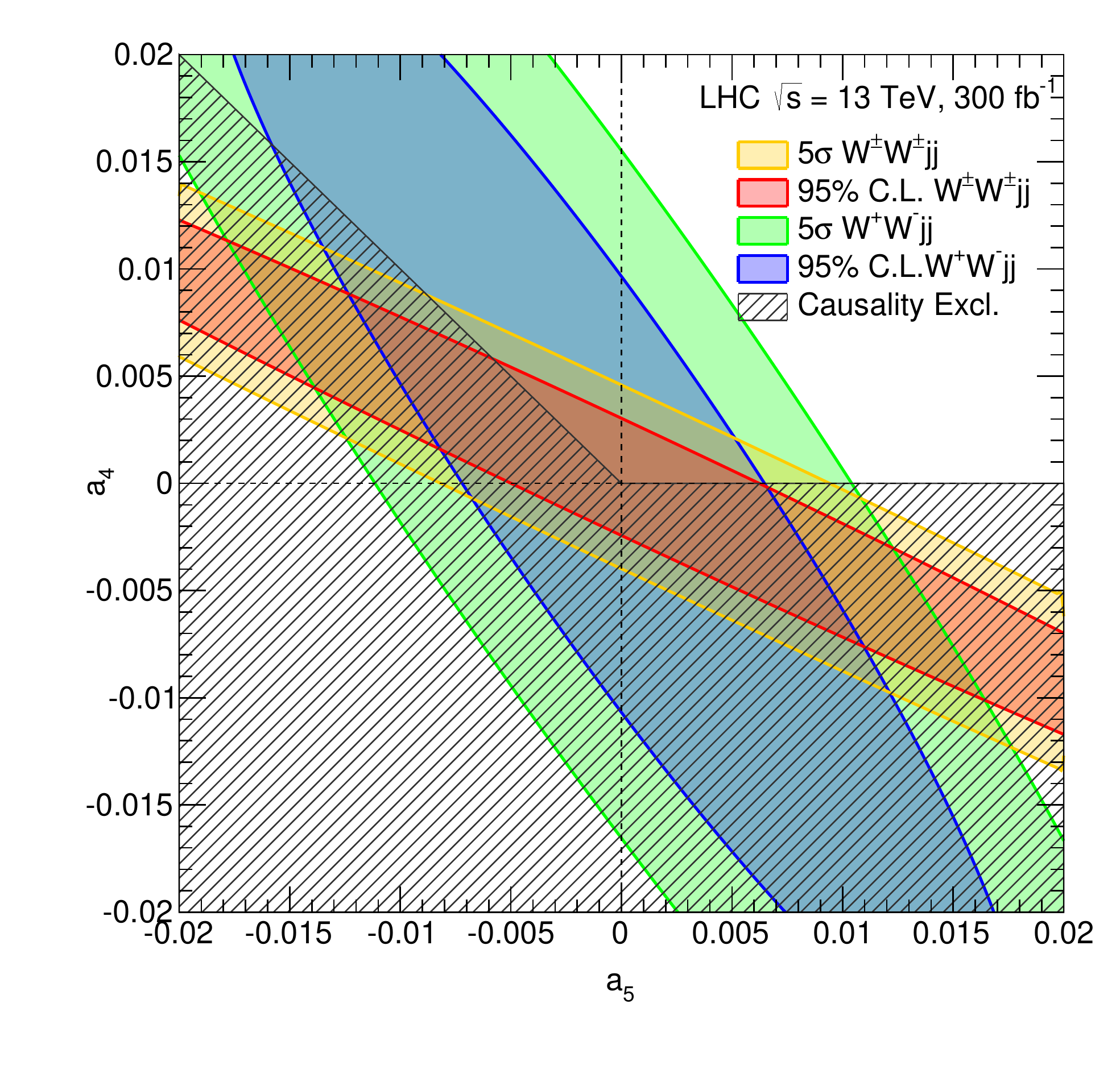}
\caption{\small  \textit{Exclusion limits (at 95\% CL) and discovery significance (5$\sigma$)  for the effective lagrangian coefficients $a_4$ and  $a_5$ at CM energy $\sqrt{s}=$13 TeV from the same-sign (in yellow/orange) and opposite-sign (in light green/blue) channels.  Hatched in grey the area where causality would be violated.}
\label{13}}
\end{center}
\end{figure}

\begin{figure}[ht!]
\begin{center}
\includegraphics[width=4.5in]{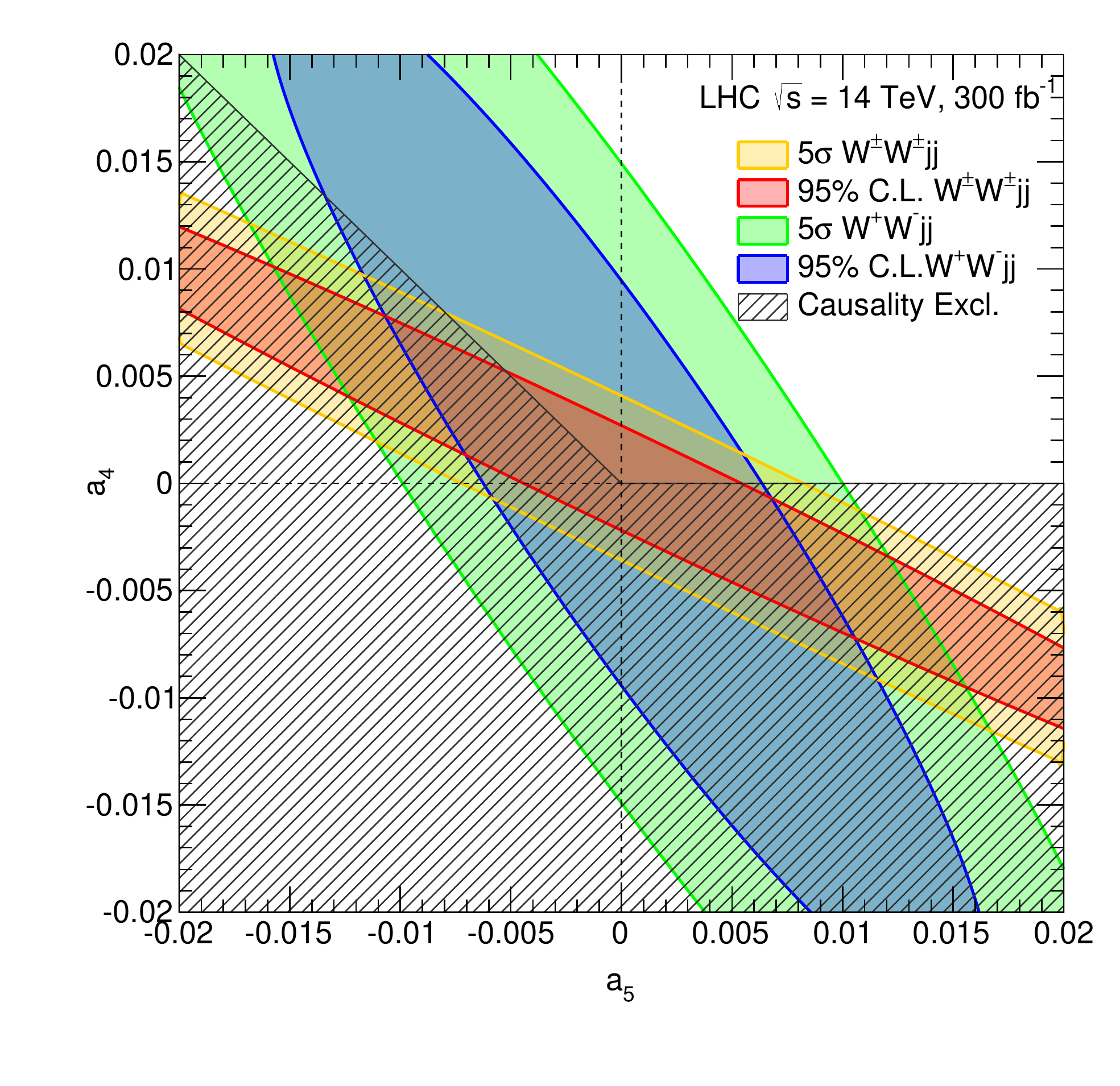}
\includegraphics[width=4.5in]{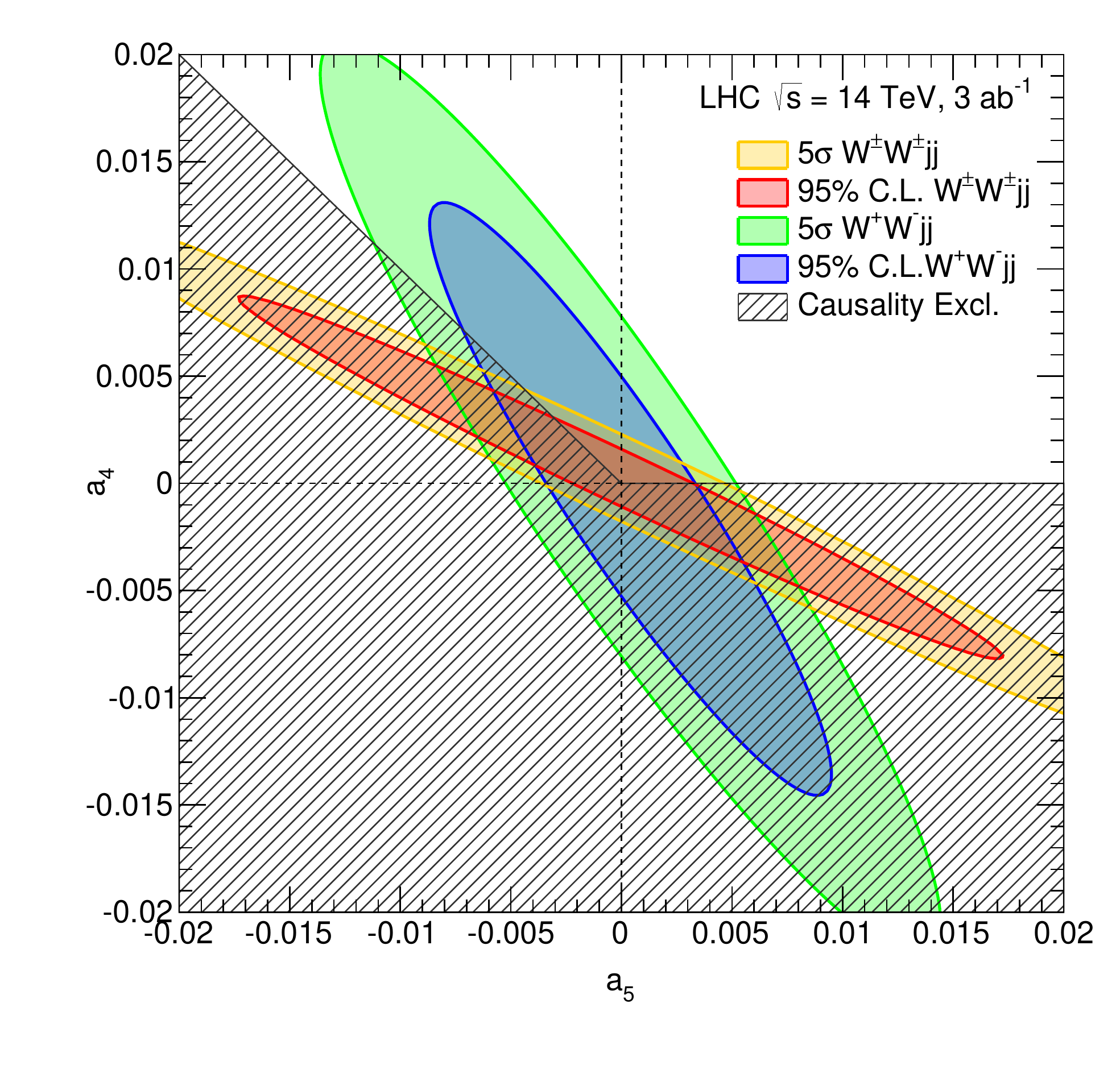}
\caption{\small  \textit{Same as in Fig.~\ref{13}  for CM energy $\sqrt{s}=$ 14 TeV.}
\label{14}}
\end{center}
\end{figure}

Figs.~\ref{13} and \ref{14} show the exclusion limits (95\% CL) and discovery significance ($5\,\sigma$) for the coefficients $a_4$ and $a_5$ obtained from the SS and OS $WW$ channels for, respectively  CM energy $\sqrt{s} =13$ and 14 TeV and the benchmark luminosities. The coefficients $a_4$ and $a_5$ are now varied simultaneously.

%%%%%%%%%%%%%%%%%%%%%%%
\begin{table}[ht!]
\begin{center}
\caption{\textit{Exclusion limits (at 95\% and 99\% CL) and discovery significance (at 3 and 5$\sigma$)  for the effective lagrangian coefficients $a_5$, $a_4$, $a_3$ , $a_2$ and $a$ for CM energy $\sqrt{s}= 13$ TeV and two benchmark  luminosities for LHC run 2. Values obtained by varying the coefficients one at the time. 
All limits are obtained from the $W^{\pm}W^{\pm}jj$ SS channel. }}
\label{limits13SS}
\vspace{0.2cm}
%\begin{ruledtabular}
\begin{tabular}{c|c|c|c|c|}
\cline{2-5} 
 &\multicolumn{4}{|c|}{ \quad  $\sqrt{s} =13$ TeV \quad   ($W^{\pm}W^{\pm}jj$ SS channel) \quad }  \cr
 \cline{2-5}
 &\multicolumn{2}{|c|}{\quad 100 fb$^{-1}$   \quad }  & \multicolumn{2}{|c|}{ \quad 300 fb$^{-1}$\quad }  \cr
 \cline{2-5}
 & \qquad  95\% (99\%) \qquad  & \qquad  3$\sigma$ (5$\sigma$) \qquad  & \qquad  95\% (99\%) \qquad  & \qquad  3$\sigma$ (5$\sigma$) \qquad   \cr
\cline{2-5}					
\hline				
 \multicolumn{1}{|c|}{\quad $a_5$ \quad}	&	$ \begin{matrix} +0.0084\; (+0.0105) \\ -0.007\; (-0.0092) \end{matrix}$	&	$ \begin{matrix} +0.0095\; (+0.0126) \\ -0.0082\; (-0.0113) \end{matrix}$	&	$ \begin{matrix} +0.0062\; (+0.0077) \\ -0.0049\; (-0.0063) \end{matrix}$	&	$ \begin{matrix} +0.0072\; (+0.0094) \\ -0.0059\; (-0.008) \end{matrix}$	\cr
 \hline									
 \multicolumn{1}{|c|}{\quad $a_4$ \quad}	&	$ \begin{matrix} +0.0041\; (+0.0052) \\ -0.0035\; (-0.0046) \end{matrix}$	&	$ \begin{matrix} +0.0047\; (+0.0062) \\ -0.004\; (-0.0056) \end{matrix}$	&	$ \begin{matrix} +0.003\; (+0.0037) \\ -0.0024\; (-0.0031) \end{matrix}$	&	$ \begin{matrix} +0.0035\; (+0.0046) \\ -0.0029\; (-0.004) \end{matrix}$	\cr
 \hline									
 \hline									
 \multicolumn{1}{|c|}{\quad $a_3$ \quad}	&	$ \begin{matrix} +0.097\; (+0.121) \\ -0.072\; (-0.096) \end{matrix}$	&	$ \begin{matrix} +0.109\; (+0.143) \\ -0.085\; (-0.118) \end{matrix}$	&	$ \begin{matrix} +0.074\; (+0.089) \\ -0.049\; (-0.065) \end{matrix}$	&	$ \begin{matrix} +0.085\; (+0.108) \\ -0.060\; (-0.083) \end{matrix}$	\cr
 \hline									
 \multicolumn{1}{|c|}{\quad $a_2$ \quad}	&	$ \begin{matrix} +1.63\; (+2.03) \\ -1.21\; (-1.61) \end{matrix}$	&	$ \begin{matrix} +1.84\; (+2.41) \\ -1.42\; (-1.99) \end{matrix}$	&	$ \begin{matrix} +1.24\; (+1.5) \\ -0.82\; (-1.09) \end{matrix}$	&	$ \begin{matrix} +1.42\; (+1.82) \\ -1.01\; (-1.4) \end{matrix}$	\cr
 \hline									
 \hline									
 \multicolumn{1}{|c|}{\quad $a$ \quad}	&	$ \begin{matrix} +1.52\; (+1.6) \\ 0.17\; (-0.44) \end{matrix}$	&	$ \begin{matrix} +1.56\; (+1.68) \\ -0.11\; (-1.57) \end{matrix}$	&	$ \begin{matrix} +1.43\; (+1.49) \\ 0.54\; (0.31) \end{matrix}$	&	$ \begin{matrix} +1.47\; (+1.56) \\ 0.39\; (-0.07) \end{matrix}$	\cr
 \hline									
\end{tabular}
%\end{ruledtabular}
\end{center}
\end{table}
%%%%%%%%%%%%%%%%%%%%%%%%%

%%%%%%%%%%%%%%%%%%%%%%%
\begin{table}[ht!]
\begin{center}
\caption{\textit{Exclusion limits (at 95\% and 99\% CL) and discovery significance (at 3 and 5$\sigma$)  for the effective lagrangian coefficients $a_5$, $a_4$, $a_3$ , $a_2$ and $a$  for CM energy $\sqrt{s}= 14$ TeV and two  benchmark  luminosities for LHC run 3.  Values obtained by varying the coefficients one at the time. 
All  limits are obtained from the $W^{\pm}W^{\pm}jj$ SS channel.}}
\label{limits14SS}
\vspace{0.2cm}
%\begin{ruledtabular}
\begin{tabular}{c|c|c|c|c|}
\cline{2-5} 
 &\multicolumn{4}{|c|}{ \quad  $\sqrt{s}=14$ TeV \quad   ($W^{\pm}W^{\pm}jj$ SS channel) \quad }  \cr
 \cline{2-5}
 &\multicolumn{2}{|c|}{\quad 300 fb$^{-1}$   \quad }  & \multicolumn{2}{|c|}{ \quad 3 ab$^{-1}$\quad }  \cr
 \cline{2-5}
 & \qquad  95\% (99\%) \qquad  & \qquad  3$\sigma$ (5$\sigma$) \qquad  & \qquad  95\% (99\%) \qquad  & \qquad  3$\sigma$ (5$\sigma$) \qquad   \cr
\cline{2-5}
\hline				
 \multicolumn{1}{|c|}{\quad $a_5$ \quad}	&	$ \begin{matrix} +0.0055\; (+0.0068) \\ -0.0045\; (-0.0058) \end{matrix}$	&	$ \begin{matrix} +0.0064\; (+0.0084) \\ -0.0054\; (-0.0074) \end{matrix}$	&	$ \begin{matrix} +0.0032\; (+0.0039) \\ -0.0022\; (-0.0029) \end{matrix}$	&	$ \begin{matrix} +0.0037\; (+0.0047) \\ -0.0027\; (-0.0036) \end{matrix}$	\cr
 \hline									
 \multicolumn{1}{|c|}{\quad $a_4$ \quad}	&	$ \begin{matrix} +0.0027\; (+0.0034) \\ -0.0022\; (-0.0028) \end{matrix}$	&	$ \begin{matrix} +0.0032\; (+0.0041) \\ -0.0026\; (-0.0036) \end{matrix}$	&	$ \begin{matrix} +0.0016\; (+0.0019) \\ -0.0011\; (-0.0014) \end{matrix}$	&	$ \begin{matrix} +0.0019\; (+0.0023) \\ -0.0013\; (-0.0018) \end{matrix}$	\cr
 \hline									
 \hline									
 \multicolumn{1}{|c|}{\quad $a_3$ \quad}	&	$ \begin{matrix} +0.073\; (+0.089) \\ -0.050\; (-0.065) \end{matrix}$	&	$ \begin{matrix} +0.084\; (+0.108) \\ -0.061\; (-0.084) \end{matrix}$	&	$ \begin{matrix} +0.046\; (+0.054) \\ -0.023\; (-0.030) \end{matrix}$	&	$ \begin{matrix} +0.052\; (+0.063) \\ -0.028\; (-0.039) \end{matrix}$	\cr
 \hline									
 \multicolumn{1}{|c|}{\quad $a_2$ \quad}	&	$ \begin{matrix} +1.14\; (+1.37) \\ -0.70\; (-0.93) \end{matrix}$	&	$ \begin{matrix} +1.30\; (+1.64) \\ -0.86\; (-1.21) \end{matrix}$	&	$ \begin{matrix} +0.75\; (+0.86) \\ -0.31\; (-0.42) \end{matrix}$	&	$ \begin{matrix} +0.83\; (+0.99) \\ -0.39\; (-0.55) \end{matrix}$	\cr
 \hline									
 \hline									
 \multicolumn{1}{|c|}{\quad $a$ \quad}	&	$ \begin{matrix} +1.37\; (+1.43) \\ 0.64\; (0.47) \end{matrix}$	&	$ \begin{matrix} +1.42\; (+1.5) \\ 0.52\; (0.22) \end{matrix}$	&	$ \begin{matrix} +1.27\; (+1.3) \\ 0.86\; (0.81) \end{matrix}$	&	$ \begin{matrix} +1.29\; (+1.33) \\ 0.82\; (0.73) \end{matrix}$	\cr
 \hline									
\end{tabular}
%\end{ruledtabular}
\end{center}
\end{table}
%%%%%%%%%%%%%%%%%%%%%%%%%

%%%%%%%%%%%%%%%%%%%%%%%
\begin{table}[ht!]
\begin{center}
\caption{\textit{Exclusion limits (at 95\% and 99\% CL) and discovery significance (at 3 and 5$\sigma$)  for the effective lagrangian coefficients $a_5$, $a_4$, $a_3$ , $a_2$ and $a$  for CM energy $\sqrt{s}= 13$ TeV and two benchmark  luminosities for LHC run 2. Values obtained by varying the coefficients one at the time. 
All limits are obtained from the $W^{\pm}W^{\mp}jj$ OS channel. }}
\label{limits13OS}
\vspace{0.2cm}
%\begin{ruledtabular}
\begin{tabular}{c|c|c|c|c|}
\cline{2-5} 
 &\multicolumn{4}{|c|}{ \quad  $\sqrt{s} =13$ TeV \quad  ($W^{\pm}W^{\mp}jj$ OS channel) \quad }  \cr
 \cline{2-5}
 &\multicolumn{2}{|c|}{\quad 100 fb$^{-1}$   \quad }  & \multicolumn{2}{|c|}{ \quad 300 fb$^{-1}$\quad }  \cr
 \cline{2-5}
 & \qquad  95\% (99\%) \qquad  & \qquad  3$\sigma$ (5$\sigma$) \qquad  & \qquad  95\% (99\%) \qquad  & \qquad  3$\sigma$ (5$\sigma$) \qquad   \cr
\cline{2-5}
\hline				
 \multicolumn{1}{|c|}{\quad $a_5$ \quad}	&	$ \begin{matrix} +0.0089\; (+0.0114) \\ -0.0095\; (-0.012) \end{matrix}$	&	$ \begin{matrix} +0.0105\; (+0.0141) \\ -0.011\; (-0.0147) \end{matrix}$	&	$ \begin{matrix} +0.0064\; (+0.0081) \\ -0.007\; (-0.0087) \end{matrix}$	&	$ \begin{matrix} +0.0077\; (+0.0103) \\ -0.0083\; (-0.0109) \end{matrix}$	\cr
 \hline									
 \multicolumn{1}{|c|}{\quad $a_4$ \quad}	&	$ \begin{matrix} +0.0141\; (+0.0179) \\ -0.014\; (-0.0178) \end{matrix}$	&	$ \begin{matrix} +0.0165\; (+0.0221) \\ -0.0164\; (-0.022) \end{matrix}$	&	$ \begin{matrix} +0.0103\; (+0.0129) \\ -0.0102\; (-0.0128) \end{matrix}$	&	$ \begin{matrix} +0.0123\; (+0.0162) \\ -0.0122\; (-0.0161) \end{matrix}$	\cr
 \hline									
 \hline									
 \multicolumn{1}{|c|}{\quad $a_3$ \quad}	&	$ \begin{matrix} +0.198\; (+0.245) \\ -0.149\; (-0.195) \end{matrix}$	&	$ \begin{matrix} +0.227\; (+0.295) \\ -0.178\; (-0.246) \end{matrix}$	&	$ \begin{matrix} +0.152\; (+0.183) \\ -0.103\; (-0.134) \end{matrix}$	&	$ \begin{matrix} +0.176\; (+0.224) \\ -0.127\; (-0.174) \end{matrix}$	\cr
 \hline									
 \multicolumn{1}{|c|}{\quad $a_2$ \quad}	&	$ \begin{matrix} +1.17\; (+1.48) \\ -1.07\; (-1.37) \end{matrix}$	&	$ \begin{matrix} +1.36\; (+1.81) \\ -1.26\; (-1.70) \end{matrix}$	&	$ \begin{matrix} +0.87\; (+1.07) \\ -0.77\; (-0.97) \end{matrix}$	&	$ \begin{matrix} +1.03\; (+1.34) \\ -0.92\; (-1.24) \end{matrix}$	\cr
 \hline									
 \hline									
 \multicolumn{1}{|c|}{\quad $a$ \quad}	&	$ \begin{matrix} +1.83\; (+2.08) \\ -0.41\; (-0.65) \end{matrix}$	&	$ \begin{matrix} +1.99\; (+2.35) \\ -0.56\; (-0.92) \end{matrix}$	&	$ \begin{matrix} +1.58\; (+1.75) \\ -0.19\; (-0.34) \end{matrix}$	&	$ \begin{matrix} +1.71\; (+1.97) \\ -0.30\; (-0.54) \end{matrix}$	\cr
 \hline									
\end{tabular}
%\end{ruledtabular}
\end{center}
\end{table}
%%%%%%%%%%%%%%%%%%%%%%%%%

%%%%%%%%%%%%%%%%%%%%%%%
\begin{table}[ht!]
\begin{center}
\caption{\textit{Exclusion limits (at 95\% and 99\% CL) and discovery significance (at 3 and 5$\sigma$)  for the effective lagrangian coefficients $a_5$, $a_4$, $a_3$ , $a_2$ and $a$  for CM energy $\sqrt{s}= 14$ TeV and two  benchmark  luminosities for LHC run 3.  Values obtained by varying the coefficients one at the time. 
All  limits are obtained from the $W^{\pm}W^{\mp}jj$ OS channel.}}
\label{limits14OS}
\vspace{0.2cm}
%\begin{ruledtabular}
\begin{tabular}{c|c|c|c|c|}
\cline{2-5} 
 &\multicolumn{4}{|c|}{ \quad  $\sqrt{s}=14$ TeV \quad  ($W^{\pm}W^{\mp}jj$ OS channel)\quad }  \cr
 \cline{2-5}
 &\multicolumn{2}{|c|}{\quad 300 fb$^{-1}$   \quad }  & \multicolumn{2}{|c|}{ \quad 3 ab$^{-1}$\quad }  \cr
 \cline{2-5}
  \cline{2-5}
 & \qquad  95\% (99\%) \qquad  & \qquad  3$\sigma$ (5$\sigma$) \qquad  & \qquad  95\% (99\%) \qquad  & \qquad  3$\sigma$ (5$\sigma$) \qquad   \cr
\cline{2-5}
\hline					
 \multicolumn{1}{|c|}{\quad $a_5$ \quad}	&	$ \begin{matrix} +0.0061\; (+0.0077) \\ -0.0062\; (-0.0077) \end{matrix}$	&	$ \begin{matrix} +0.0073\; (+0.0097) \\ -0.0074\; (-0.0098) \end{matrix}$	&	$ \begin{matrix} +0.0033\; (+0.0041) \\ -0.0034\; (-0.0042) \end{matrix}$	&	$ \begin{matrix} +0.004\; (+0.0052) \\ -0.0041\; (-0.0052) \end{matrix}$	\cr
 \hline									
 \multicolumn{1}{|c|}{\quad $a_4$ \quad}	&	$ \begin{matrix} +0.0084\; (+0.0107) \\ -0.0097\; (-0.012) \end{matrix}$	&	$ \begin{matrix} +0.0102\; (+0.0136) \\ -0.0115\; (-0.0149) \end{matrix}$	&	$ \begin{matrix} +0.0043\; (+0.0055) \\ -0.0056\; (-0.0068) \end{matrix}$	&	$ \begin{matrix} +0.0053\; (+0.007) \\ -0.0066\; (-0.0083) \end{matrix}$	\cr
 \hline									
 \hline									
 \multicolumn{1}{|c|}{\quad $a_3$ \quad}	&	$ \begin{matrix} +0.134\; (+0.165) \\ -0.115\; (-0.146) \end{matrix}$	&	$ \begin{matrix} +0.158\; (+0.205) \\ -0.140\; (-0.187) \end{matrix}$	&	$ \begin{matrix} +0.077\; (+0.094) \\ -0.059\; (-0.075) \end{matrix}$	&	$ \begin{matrix} +0.091\; (+0.114) \\ -0.072\; (-0.096) \end{matrix}$	\cr
 \hline									
 \multicolumn{1}{|c|}{\quad $a_2$ \quad}	&	$ \begin{matrix} +0.75\; (+0.93) \\ -0.71\; (-0.89) \end{matrix}$	&	$ \begin{matrix} +0.89\; (+1.17) \\ -0.85\; (-1.13) \end{matrix}$	&	$ \begin{matrix} +0.42\; (+0.52) \\ -0.38\; (-0.47) \end{matrix}$	&	$ \begin{matrix} +0.50\; (+0.64) \\ -0.45\; (-0.59) \end{matrix}$	\cr
 \hline									
 \hline									
 \multicolumn{1}{|c|}{\quad $a$ \quad}	&	$ \begin{matrix} +1.56\; (+1.67) \\ -0.50\; (-1.35) \end{matrix}$	&	$ \begin{matrix} +1.64\; (+1.81) \\ -1.15\; (-2.74) \end{matrix}$	&	$ \begin{matrix} +1.36\; (+1.41) \\ 0.55\; (0.31) \end{matrix}$	&	$ \begin{matrix} +1.40\; (+1.49) \\ 0.36\; (-0.06) \end{matrix}$	\cr
 \hline									
\end{tabular}
%\end{ruledtabular}
\end{center}
\end{table}
%%%%%%%%%%%%%%%%%%%%%%%%%

%%%%%%%%%%%%%%%%%%%
%%%%%%%%%%%%%%%%%%%%%%%%%% 

%%%%%%%%%%%%%%%%%%%
%%%%%%%%%%%%%%%%%%%%%%%%%% 

\FloatBarrier
%\clearpage

%%%%%%%%%%%%%%%%%%%%%%%%%%%%%%%%
\section{Discussion} 
%%%%%%%%%%%%%%%%%%%%%%%%%%%%%%%%

While the presence of resonances is the most dramatic signal for a strongly interacting sector, they may be too heavy or broad to be clearly seen at the LHC.
The discovery of a non-vanishing coefficient of the effective lagrangian in \eq{L1}, introduced in section~\ref{sec:Notation}, is a more systematic way to search for the presence of the strongly interacting sector behind the breaking of the EW symmetry. In addition, exclusion limits provide an indirect indication about the energy scale of the masses of those resonances that are expected from such new interactions.

The identification of the most appropriated selection cuts is crucial but it is now well understood that---in addition to the central jet veto necessary to remove the QCD 
background---the control of the large EW background can be achieved by means of a selection on the transverse momenta of the jets and  final leptons.

We have shown that a significant improvement in both discovery significance and exclusion limits for the chiral effective lagrangian coefficients $a_4$ and $a_5$ can be expected from the current and the next run of the LHC. Already at CM energy $\sqrt{s}=13$ TeV and a luminosity of 100 fb$^{-1}$ the limits will  reach the permil precision thus coming within range of the values expected by purely dimensional analysis.
These results can be obtained by studying the SS $WW\rightarrow WW$ channel alone.

The determination of the coefficient  $a_3$ within VBS---the best limits for which come at the moment from precision measurements---will become competitive already at the LHC run 2 when a luminosity of 300 fb$^{-1}$ will be available.  The coefficient $a_2$ gives rise to smaller deviations in VBS and is determined with less precision;  its constraints will be competitive with those from TGC data at LEP only when higher luminosities  become available.

Finally, the  coefficient $a$---controlling the coupling of the Higgs to the vector bosons in \eq{L0} in section~\ref{sec:Notation}---remains best determined in the decay processes of the Higgs boson. Only at future LHC runs a comparable limit will be available from VBS.

While VBS remains our best laboratory to study EW symmetry breaking, the presence of  systematic errors  hard to reduce and even estimate will eventually limit the final precision of the measurements that can be achieved at the LHC. The same is true for the study of the Higgs boson decays and the complementary determination of the coefficient $a$, as defined in \eq{L0} in section~\ref{sec:Notation}.

 \begin{acknowledgments}
We thank Francesco Riva for discussions. MF is associated to SISSA.   The work of AT 
is supported by the S\~ao Paulo Research Foundation (FAPESP) under grants 2011/11973-4 and 2013/02404-1.
The work of MP and AU is supported by the ERC Advanced Grant n.\ 267985, ``Electroweak Symmetry Breaking, Flavour and Dark Matter: One Solution for Three Mysteries" (DaMeSyFla). 
\end{acknowledgments}

%%%%%%%%%%%%%%%%%%%%%%%%%%%%%%%%%%%%%%


\begin{thebibliography}{99}
  
\bibitem{Chatrchyan:2012ufa} 
  G.~Aad {\it et al.}  [ATLAS Collaboration],
  %``Observation of a new particle in the search for the Standard Model Higgs boson with the ATLAS detector at the LHC,''
  Phys.\ Lett.\ B {\bf 716}, 1 (2012)
  [\hhref{arXiv:1207.7214} [hep-ex]].;
   S.~Chatrchyan {\it et al.}  [CMS Collaboration],
  %``Observation of a new boson at a mass of 125 GeV with the CMS experiment at the LHC,''
  Phys.\ Lett.\ B {\bf 716}, 30 (2012)
  [\hhref{arXiv:1207.7235} [hep-ex]].
  
\bibitem{Chanowitz:1985hj} 
  M.~S.~Chanowitz and M.~K.~Gaillard,
  %``The TeV Physics of Strongly Interacting W's and Z's,''
  Nucl.\ Phys.\ B {\bf 261}, 379 (1985).
    
\bibitem{Appelquist:1980vg} 
  T.~Appelquist and C.~W.~Bernard,
  %``Strongly Interacting Higgs Bosons,''
  Phys.\ Rev.\ D {\bf 22}, 200 (1980); 
  A.~C.~Longhitano,
  %``Heavy Higgs Bosons in the Weinberg-Salam Model,''
  Phys.\ Rev.\ D {\bf 22}, 1166 (1980);
  T.~Appelquist and G.~H.~Wu,
  %``The Electroweak chiral Lagrangian and new precision measurements,''
  Phys.\ Rev.\ D {\bf 48}, 3235 (1993)
  [\hhref{hep-ph/9304240}].
  
  \bibitem{Espriu:2012ih} 
  D.~Espriu and B.~Yencho,
  %``Longitudinal WW scattering in light of the ?Higgs boson? discovery,''
  Phys.\ Rev.\ D {\bf 87}, no. 5, 055017 (2013)
  [\hhref{arXiv:1212.4158} [hep-ph]];
  D.~Espriu and F.~Mescia,
  %``Unitarity and causality constraints in composite Higgs models,''
  Phys.\ Rev.\ D {\bf 90}, no. 1, 015035 (2014)
  [\hhref{arXiv:1403.7386} [hep-ph]].
  
  \bibitem{Delgado:2013loa} 
  R.~L.~Delgado, A.~Dobado and F.~J.~Llanes-Estrada,
  %``Light ?Higgs?, yet strong interactions,''
  J.\ Phys.\ G {\bf 41}, 025002 (2014)
  [\hhref{arXiv:1308.1629} [hep-ph]];
  R.~L.~Delgado, A.~Dobado and F.~J.~Llanes-Estrada,
  %``A Strongly Interacting Electroweak Symmetry Breaking Sector with a Higgs-like light scalar,''
  \hhref{arXiv:1412.3277} [hep-ph];
   R.~L.~Delgado, A.~Dobado and F.~J.~Llanes-Estrada,
  %``Unitarity, analyticity, dispersion relations, and resonances in strongly interacting $W_LW_L$, $Z_LZ_L$, and hh scattering,''
  Phys.\ Rev.\ D {\bf 91}, no. 7, 075017 (2015)
  [\hhref{arXiv:1502.04841} [hep-ph]].

  \bibitem{Doroba:2012pd} 
  K.~Doroba, J.~Kalinowski, J.~Kuczmarski, S.~Pokorski, J.~Rosiek, M.~Szleper and S.~Tkaczyk,
  %``The $W_L W_L$ Scattering at the LHC: Improving the Selection Criteria,''
  Phys.\ Rev.\ D {\bf 86}, 036011 (2012)
  [\hhref{arXiv:1201.2768} [hep-ph]].

\bibitem{duncan} 
  D.~A.~Dicus and R.~Vega,
  %``$W W$ Production From $P P$ Collisions,''
  Phys.\ Rev.\ Lett.\  {\bf 57}, 1110 (1986);
  M.~J.~Duncan, G.~L.~Kane and W.~W.~Repko,
  %``W W Physics at Future Colliders,''
  Nucl.\ Phys.\ B {\bf 272}, 517 (1986).
  
\bibitem{Barger:1990py} 
  V.~D.~Barger, K.~m.~Cheung, T.~Han and R.~J.~N.~Phillips,
  %``Strong $W^{+} W^{+}$ scattering signals at $p p$ supercolliders,''
  Phys.\ Rev.\ D {\bf 42}, 3052 (1990);
  J.~Bagger, V.~D.~Barger, K.~m.~Cheung, J.~F.~Gunion, T.~Han, G.~A.~Ladinsky, R.~Rosenfeld and C.-P.~Yuan,
  %``CERN LHC analysis of the strongly interacting W W system: Gold plated modes,''
  Phys.\ Rev.\ D {\bf 52}, 3878 (1995)
  [\hhref{hep-ph/9504426}] 
  
 \bibitem{Bagger:1995mk} 
  J.~Bagger, V.~D.~Barger, K.~m.~Cheung, J.~F.~Gunion, T.~Han, G.~A.~Ladinsky, R.~Rosenfeld and C.-P.~Yuan,
  %``CERN LHC analysis of the strongly interacting W W system: Gold plated modes,''
  Phys.\ Rev.\ D {\bf 52}, 3878 (1995)
  [\hhref{hep-ph/9504426}].
 
\bibitem{Butterworth:2002tt} 
 J.~M.~Butterworth, B.~E.~Cox and J.~R.~Forshaw,
  %``$W W$ scattering at the CERN LHC,''
  Phys.\ Rev.\ D {\bf 65}, 096014 (2002)
  [\hhref{hep-ph/0201098}].
  

  
\bibitem{Englert:2008tn} 
A.~Alboteanu, W.~Kilian and J.~Reuter,
  %``Resonances and Unitarity in Weak Boson Scattering at the LHC,''
  JHEP {\bf 0811}, 010 (2008)
  [\hhref{arXiv:0806.4145} [hep-ph];
  T.~Han, D.~Krohn, L.~T.~Wang and W.~Zhu,
  %``New Physics Signals in Longitudinal Gauge Boson Scattering at the LHC,''
  JHEP {\bf 1003}, 082 (2010)
  [\hhref{arXiv:0911.3656} [hep-ph]].
  C.~Englert, B.~Jager, M.~Worek and D.~Zeppenfeld,
  %``Observing Strongly Interacting Vector Boson Systems at the CERN Large Hadron Collider,''
  Phys.\ Rev.\ D {\bf 80}, 035027 (2009);
   A.~Freitas and J.~S.~Gainer,
  %``High Energy WW Scattering at the LHC with the Matrix Element Method,''
  Phys.\ Rev.\ D {\bf 88}, no. 1, 017302 (2013)
  [\hhref{arXiv:1212.3598}];
  W.~Kilian, T.~Ohl, J.~Reuter and M.~Sekulla,
  %``High-Energy Vector Boson Scattering after the Higgs Discovery,''
  Phys.\ Rev.\ D {\bf 91}, 096007 (2015)
  [\hhref{arXiv:1408.6207} [hep-ph]].
  
  \bibitem{Accomando:2005hz} 
  E.~Accomando, A.~Ballestrero, S.~Bolognesi, E.~Maina and C.~Mariotti,
  %``Boson-boson scattering and Higgs production at the LHC from a six fermion point of view: Four jets + l nu processes at O( alpha(em)**6 ),''
  JHEP {\bf 0603}, 093 (2006)
  [\hhref{hep-ph/0512219}];  A.~Ballestrero, G.~Bevilacqua and E.~Maina,
  %``A Complete parton level analysis of boson-boson scattering and ElectroWeak Symmetry Breaking in lv + four jets production at the LHC,''
  JHEP {\bf 0905}, 015 (2009)
  [\hhref{arXiv:0812.5084} [hep-ph]]; A.~Ballestrero, G.~Bevilacqua, D.~B.~Franzosi and E.~Maina,
  %``How well can the LHC distinguish between the SM light Higgs scenario, a composite Higgs and the Higgsless case using VV scattering channels?,''
  JHEP {\bf 0911}, 126 (2009)
  [\hhref{arXiv:0909.3838} [hep-ph]].
  
  \bibitem{Belyaev:1998ih} 
  S.~Godfrey,
  %``Quartic gauge boson couplings,''
  AIP Conf.\ Proc.\  {\bf 350}, 209 (1995)
  [\hhref{hep-ph/9505252}];
  A.~S.~Belyaev, O.~J.~P.~Eboli, M.~C.~Gonzalez-Garcia, J.~K.~Mizukoshi, S.~F.~Novaes and I.~Zacharov,
  %``Strongly interacting vector bosons at the CERN LHC: Quartic anomalous couplings,''
  Phys.\ Rev.\ D {\bf 59}, 015022 (1999)
  [\hhref{hep-ph/9805229}].

 
 \bibitem{Dobado:1990jy} 
  A.~Dobado, M.~J.~Herrero and J.~Terron,
  %``The Role of Chiral Lagrangians in Strongly Interacting $W$(l) $W$(l) Signals at $p p$ Supercolliders,''
  Z.\ Phys.\ C {\bf 50}, 205 (1991); A.~Dobado, M.~J.~Herrero, J.~R.~Pelaez, E.~Ruiz Morales and M.~T.~Urdiales,
  %``Learning about the strongly interacting symmetry breaking sector at LHC,''
  Phys.\ Lett.\ B {\bf 352}, 400 (1995)
  [\hhref{hep-ph/9502309}].
  
 \bibitem{Fabbrichesi:2007ad} 
  M.~Fabbrichesi and L.~Vecchi,
  %``Possible experimental signatures at the CERN LHC of strongly interacting electro-weak symmetry breaking,''
  Phys.\ Rev.\ D {\bf 76}, 056002 (2007)
  [\hhref{hep-ph/0703236}].
  
  
  \bibitem{Brivio:2013pma} 
  I.~Brivio, T.~Corbett, O.~J.~P.~\`Eboli, M.~B.~Gavela, J.~Gonzalez-Fraile, M.~C.~Gonzalez-Garcia, L.~Merlo and S.~Rigolin,
  %``Disentangling a dynamical Higgs,''
  JHEP {\bf 1403}, 024 (2014)
  [\hhref{arXiv:1311.1823} [hep-ph]].
  
   \bibitem{Eboli:2006wa} 
  O.~J.~P.~Eboli, M.~C.~Gonzalez-Garcia and J.~K.~Mizukoshi,
  %``p p ---> j j e+- mu+- nu nu and j j e+- mu-+ nu nu at O( alpha(em)**6) and O(alpha(em)**4 alpha(s)**2) for the study of the quartic electroweak gauge boson vertex at CERN LHC,''
  Phys.\ Rev.\ D {\bf 74}, 073005 (2006)  [hep-ph/0606118].
 
    \bibitem{Szleper:2014xxa} 
  M.~Szleper,
  %``The Higgs boson and the physics of $WW$ scattering before and after Higgs discovery,''
  \hhref{arXiv:1412.8367} [hep-ph]. 
  
  \bibitem{Giudice:2007fh} 
  A.~Manohar and H.~Georgi,
  %``Chiral Quarks and the Nonrelativistic Quark Model,''
  Nucl.\ Phys.\ B {\bf 234}, 189 (1984);
    H.~Georgi and L.~Randall,
  %``Flavor Conserving CP Violation in Invisible Axion Models,''
  Nucl.\ Phys.\ B {\bf 276}, 241 (1986);
  G.~F.~Giudice, C.~Grojean, A.~Pomarol and R.~Rattazzi,
  %``The Strongly-Interacting Light Higgs,''
  JHEP {\bf 0706}, 045 (2007)
  [\hhref{hep-ph/0703164}].

   %\cite{Peskin:1990zt,Barbieri:2003pr}
\bibitem{Peskin:1990zt} 
  M.~E.~Peskin and T.~Takeuchi,
  %``A New constraint on a strongly interacting Higgs sector,''
  Phys.\ Rev.\ Lett.\  {\bf 65}, 964 (1990);
  %\cite{Golden:1990ig,Holdom:1990tc,Peskin:1991sw,Altarelli:1990zd,Altarelli:1991fk}
%\bibitem{Golden:1990ig} 
  M.~Golden and L.~Randall,
  %``Radiative Corrections to Electroweak Parameters in Technicolor Theories,''
  Nucl.\ Phys.\ B {\bf 361}, 3 (1991);
    %\cite{Holdom:1990tc,Peskin:1991sw,Altarelli:1990zd,Altarelli:1991fk}
%\bibitem{Holdom:1990tc} 
  B.~Holdom and J.~Terning,
  %``Large corrections to electroweak parameters in technicolor theories,''
  Phys.\ Lett.\ B {\bf 247}, 88 (1990).;
  %\cite{Peskin:1991sw,Altarelli:1990zd,Altarelli:1991fk}
%\bibitem{Peskin:1991sw} 
  M.~E.~Peskin and T.~Takeuchi,
  %``Estimation of oblique electroweak corrections,''
  Phys.\ Rev.\ D {\bf 46}, 381 (1992);
%\cite{Altarelli:1990zd,Altarelli:1991fk}
%\bibitem{Altarelli:1990zd} 
  G.~Altarelli and R.~Barbieri,
  %``Vacuum polarization effects of new physics on electroweak processes,''
  Phys.\ Lett.\ B {\bf 253}, 161 (1991);
%\cite{Altarelli:1991fk}
%\bibitem{Altarelli:1991fk} 
  G.~Altarelli, R.~Barbieri and S.~Jadach,
  %``Toward a model independent analysis of electroweak data,''
  Nucl.\ Phys.\ B {\bf 369}, 3 (1992)
  [Erratum-ibid.\ B {\bf 376}, 444 (1992)].

\bibitem{Barbieri:2003pr} 
  R.~Barbieri, A.~Pomarol and R.~Rattazzi,
  %``Weakly coupled Higgsless theories and precision electroweak tests,''
  Phys.\ Lett.\ B {\bf 591}, 141 (2004)
  [\hhref{hep-ph/0310285}]; 
  %\cite{Barbieri:2003pr,Barbieri:2004qk}
%\bibitem{Barbieri:2004qk} 
  R.~Barbieri, A.~Pomarol, R.~Rattazzi and A.~Strumia,
  %``Electroweak symmetry breaking after LEP-1 and LEP-2,''
  Nucl.\ Phys.\ B {\bf 703}, 127 (2004)
  [\hhref{hep-ph/0405040}].

 \bibitem{Hagiwara:1992eh} 
  K.~Hagiwara, S.~Ishihara, R.~Szalapski and D.~Zeppenfeld,
  %``Low-energy constraints on electroweak three gauge boson couplings,''
  Phys.\ Lett.\ B {\bf 283}, 353 (1992).
  
  %\cite{Falkowski:2014tna}
\bibitem{Falkowski:2014tna} 
See, e.g.:  A.~Falkowski and F.~Riva,
  %``Model-independent precision constraints on dimension-6 operators,''
  JHEP {\bf 1502}, 039 (2015)
  [\hhref{arXiv:1411.0669} [hep-ph]].
 
\bibitem{Elias-Miro:2013gya} 
  J.~Elias-Miró, J.~R.~Espinosa, E.~Masso and A.~Pomarol,
  %``Renormalization of dimension-six operators relevant for the Higgs decays $h\rightarrow \gamma\gamma,\gamma Z$,''
  JHEP {\bf 1308}, 033 (2013)
  [\hhref{arXiv:1302.5661} [hep-ph]].
   
  %\cite{Grzadkowski:2010es}
\bibitem{Grzadkowski:2010es} 
W.~Buchmuller and D.~Wyler,
  %``Effective Lagrangian Analysis of New Interactions and Flavor Conservation,''
  Nucl.\ Phys.\ B {\bf 268}, 621 (1986);
  B.~Grzadkowski, M.~Iskrzynski, M.~Misiak and J.~Rosiek,
  %``Dimension-Six Terms in the Standard Model Lagrangian,''
  JHEP {\bf 1010}, 085 (2010)
  [\hhref{arXiv:1008.4884} [hep-ph]].

 \bibitem{Baak:2013fwa} M.~Baak, A.~Blondel, A.~Bodek, R.~Caputo, T.~Corbett, C.~Degrande, O.~Eboli and J.~Erler {\it et al.},
  %``Working Group Report: Precision Study of Electroweak Interactions,''
  \hhref{arXiv:1310.6708} [hep-ph].
  
  %\cite{Falkowski:2013dza}
\bibitem{Falkowski:2013dza} 
  A.~Falkowski, F.~Riva and A.~Urbano,
  %``Higgs at last,''
  JHEP {\bf 1311}, 111 (2013)
  [\hhref{arXiv:1303.1812} [hep-ph]].
  
    \bibitem{Lombardo:2013daa} 
 S.~Schael {\it et al.} [ALEPH Collaboration],
  %``Improved measurement of the triple gauge-boson couplings gamma W W and Z W W in e+ e- collisions,''
  Phys.\ Lett.\ B {\bf 614}, 7 (2005);
  P.~Achard {\it et al.} [L3 Collaboration],
  %``Measurement of triple gauge boson couplings of the $W$ boson at LEP,''
  Phys.\ Lett.\ B {\bf 586}, 151 (2004)
  [hep-ex/0402036];
   G.~Abbiendi {\it et al.} [OPAL Collaboration],
  %``Measurement of charged current triple gauge boson couplings using $W$ pairs at LEP,''
  Eur.\ Phys.\ J.\ C {\bf 33}, 463 (2004)
  [hep-ex/0308067].
    
  \bibitem{Pham:1985cr} 
  T.~N.~Pham and T.~N.~Truong,
  %``Evaluation of the Derivative Quartic Terms of the Meson Chiral Lagrangian From Forward Dispersion Relation,''
  Phys.\ Rev.\ D {\bf 31}, 3027 (1985); 
  A.~Adams, N.~Arkani-Hamed, S.~Dubovsky, A.~Nicolis and R.~Rattazzi,
  %``Causality, analyticity and an IR obstruction to UV completion,''
  JHEP {\bf 0610}, 014 (2006)
  [\hhref{hep-th/0602178}];
  J.~Distler, B.~Grinstein, R.~A.~Porto and I.~Z.~Rothstein,
  %``Falsifying Models of New Physics via WW Scattering,''
  Phys.\ Rev.\ Lett.\  {\bf 98}, 041601 (2007)
  [\hhref{hep-ph/0604255}];  
  L.~Vecchi,
  %``Causal versus analytic constraints on anomalous quartic gauge couplings,''
  JHEP {\bf 0711}, 054 (2007)
  [\hhref{arXiv:0704.1900} [hep-ph]].
  
  \bibitem{Aad:2014zda} 
  G.~Aad {\it et al.}  [ATLAS Collaboration],
  %``Evidence for Electroweak Production of $W^{\pm}W^{\pm}jj$ in $pp$ Collisions at $\sqrt{s}=8$ TeV with the ATLAS Detector,''
  Phys.\ Rev.\ Lett.\  {\bf 113}, no. 14, 141803 (2014)
  [\hhref{arXiv:1405.6241} [hep-ex]]; CMS Collaboration [CMS Collaboration],
  %``Vector boson scattering in a final state with two jets and two same-sign leptons,''
  CMS-PAS-SMP-13-015.

\bibitem{Degrande:2013yda} 
  C.~Degrande, J.~L.~Holzbauer, S.-C.~Hsu, A.~V.~Kotwal, S.~Li, M.~Marx, O.~Mattelaer and J.~Metcalfe {\it et al.},
  %``Studies of Vector Boson Scattering And Triboson Production with DELPHES Parametrized Fast Simulation for Snowmass 2013,''
  \hhref{arXiv:1309.7452} [physics.comp-ph].
  
 \bibitem{Ellis:2013lra} 
  J.~Ellis and T.~You,
  %``Updated Global Analysis of Higgs Couplings,''
  JHEP {\bf 1306}, 103 (2013)
  [\hhref{arXiv:1303.3879} [hep-ph]].
  
\bibitem{CMS:2013xfa} 
 See, for example: [CMS Collaboration],
  %``Projected Performance of an Upgraded CMS Detector at the LHC and HL-LHC: Contribution to the Snowmass Process,''
  arXiv:1307.7135.  
  
 \bibitem{Christensen:2008py}
  N.~D.~Christensen and C.~Duhr,
 % ``FeynRules - Feynman rules made easy,''
  Comput.\ Phys.\ Commun.\  {\bf 180} (2009) 1614
  [\hhref{arXiv:0806.4194} [hep-ph]].
  
  \bibitem{madgraph}
  J. Alwall et. al., 
%  ``MadGraph 5: Going Beyond,''
  JHEP {\bf 1106}, 128 (2011)
  [\hhref{arXiv:1106.052211} [hep-ph]].  
    
\bibitem{Sjostrand:2006za} 
  T.~Sjostrand, S.~Mrenna and P.~Z.~Skands,
  %``PYTHIA 6.4 Physics and Manual,''
  JHEP {\bf 0605}, 026 (2006)
  [hep-ph/0603175].
  
\bibitem{deFavereau:2013fsa} 
  J.~de Favereau {\it et al.} [DELPHES 3 Collaboration],
  %``DELPHES 3, A modular framework for fast simulation of a generic collider experiment,''
  JHEP {\bf 1402}, 057 (2014)
  [arXiv:1307.6346 [hep-ex]].

\bibitem{Gupta:1949rh} 
J.~S.~Schwinger,
  %``Quantum electrodynamics. I A covariant formulation,''
  Phys.\ Rev.\  {\bf 74}, 1439 (1948);
    S.~N.~Gupta,
  %``Theory of longitudinal photons in quantum electrodynamics,''
  Proc.\ Phys.\ Soc.\ A {\bf 63}, 681 (1950).
  
\end{thebibliography}
\end{document}